# Machine Learning Forecasts of Asymmetric Betas Using Firm-Specific Information[*]

Thomas Conlon[1], John Cotter[2], and Iason Kynigakis[3]

**Abstract**

We demonstrate that machine learning methods provide a powerful framework for modelling conditional asymmetric risk. Using a large cross-section of US stocks and a comprehensive set of firm characteristics, we show that allowing for nonlinearities significantly increases the out-of-sample performance across a wide range of asymmetric beta measures and forecasting horizons. Trading frictions, followed by characteristics related to intangibles, momentum and growth, emerge as the most important drivers of future risk dynamics. Reconstructing CAPM beta from forecasts of asymmetric beta components indicates that a more granular decomposition of systematic risk yields a more accurate representation of market beta. We also find that incorporating conditional beta forecasts into discounted cash flow models that account for the term structure of betas enhances equity valuation accuracy. Finally, we show that the statistical outperformance of conditional betas translates into economically significant benefits for market-neutral portfolio investors.

[*] The authors gratefully acknowledge the support of Science Foundation Ireland under Grants Numbers 16/SPP/3347, 13/RC/2106 P2, 17/SPP/5447, 18/CRT/6224 and of the Irish Research Council under grant number GOIPD/2022/721. We also thank Ben Charoenwong and Marcel Prokopczuk for helpful comments and suggestions.

[1] Michael Smurfit Graduate Business School, University College Dublin, Dublin, Ireland, conlon.thomas@ucd.ie
[2] Michael Smurfit Graduate Business School, University College Dublin, Dublin, Ireland, john.cotter@ucd.ie
[3] Michael Smurfit Graduate Business School, University College Dublin, Dublin, Ireland, iason.kynigakis@ucd.ie

# 1. Introduction

Beta from the Capital Asset Pricing Model (CAPM) of Sharpe (1964), Lintner (1965), and Mossin (1966) is one of the most important measures in asset pricing and is widely used for capital budgeting and asset allocation decisions. However, there exists extensive literature that finds regular market betas perform poorly in the context for which they were developed, such as for explaining the cross-sectional variation of stock returns (see, e.g., Fama and French, 1992, 2008; Lewellen and Nagel, 2006; Lewellen, 2015). Early studies by Roy (1952) and Markowitz (1959) suggest that investors perceive risk differently during periods of low and high market returns. In addition, Campbell et al. (2001), indicate that firm-specific betas are difficult to estimate and are unstable over time. As corporate managers and investors require reliable predictions of betas, these issues accentuate the importance of deriving accurate estimates of firm-level betas across different market regimes.

In this study, we forecast asymmetric betas using machine learning methods and firm-specific characteristics. Focusing on asymmetric risk measures offers several key benefits, as asymmetric betas more directly capture how risk varies across different market conditions by distinguishing between downside losses and upside gains. Employing machine learning methods provides two key advantages. First, by allowing for nonlinearities these models can capture complex relationships between firm characteristics and asymmetric risk measures that linear models may overlook. Second, methods that perform variable selection alleviate the issue of overfitting in a high-dimensional predictor space by focusing on the most relevant firm characteristics and shrinking or excluding those that contribute little to predictive accuracy. Together, these features make it possible to generate more accurate and stable forecasts of asymmetric betas. Moreover, this framework allows us to uncover the nonlinear relationships that govern time-varying risk exposures and to identify the important characteristics that explain asymmetric betas. By shifting attention from regular market beta to asymmetric risk measures, and by leveraging high-dimensional data through machine learning, we deliver new insights into the drivers of systematic risk and offer more accurate tools to the financial-decision making process of capital budgeting and portfolio risk management.

Specifically, we forecast the downside and upside betas by Ang et al. (2006) and the four semibetas proposed by Bollerslev et al. (2022). The performance evaluation is designed to prevent overfitting, whether from excessive model complexity or backtest bias. We consider a long sample period covering a broad cross-section of US stocks and a comprehensive predictor set comprised of 214 historical beta estimates and firm-specific characteristics from the Chen and Zimmermann (2022) open-source asset pricing database. Furthermore, we evaluate model performance using an out-of-sample rolling window scheme, consistent with Drobetz et al. (2025) and tune the hyperparameters based on commonly used values in a time series cross-validation framework, following Gu et al. (2020). Our approach, in line with Cosemans et al. (2016), also benefits from combining data sampled from different frequencies, since we use daily data to estimate realized asymmetric betas, while monthly data are used for the firm characteristics and to fit the machine learning models. This differs from the MIDAS approaches by Ghysels et al. (2005) and Engle et al. (2013) that use both low- and high-frequency time-series data to model risk directly. We also assess the role of conditional betas in a discounted cash flow analysis when discount rates are estimated using forecasts from both individual and multiple horizons. Lastly, we examine whether the conditional beta forecasts would benefit an investor constructing optimal portfolios based on factor models by reducing exposure to market risk.

The asymmetric risk effects of betas have been widely explored in the literature. Bawa and Lindenberg (1977) suggest an extension of the CAPM with asymmetric downside and upside betas, where risk is measured using the semivariance based on negative or positive market returns, rather than total variance. Ang et al. (2006) measure downside and upside betas over periods when the excess market return is below or above its mean and find that a downside beta version of the CAPM is better at explaining the cross-sectional variation of stock returns than the regular CAPM. In addition, Lettau et al. (2014) show that accounting for downside risk in the CAPM can better explain the variation in returns across different asset classes. By contrast, Levi and Welch (2020) find that regular market beta is better at explaining the cross-sectional variation in stock returns than downside beta, even during bear markets and crashes.

More recently, Bollerslev et al. (2022), propose a new decomposition of the traditional market beta into four semibetas that depend on the signed covariation between the market and individual asset returns. Consistent with the literature which suggests that investors only care about downside risk, they find that semibetas derived from negative market and negative asset return covariation are better at explaining higher expected returns, while semibetas based on negative market and positive asset return covariation are better at explaining lower expected returns.

Similarly extensive is the literature that focuses on producing more accurate estimates of future market betas. Engle (2016) and Bali et al. (2017) show that dynamic conditional beta based on multivariate GARCH models performs well in the cross-section of stock returns. Levi and Welch (2017), motivated by the need for improved beta estimates by corporate financial managers, propose a linear estimator for forward-looking beta estimates, using only the prevailing empirical beta. Chang et al. (2012) develop an option-implied approach to estimate beta and show that it often outperforms the historical beta in the cross-section. Buss and Vilkov (2012) propose a hybrid methodology that combines option-implied with historical return information, which is found to perform well against other estimators. Hollstein and Prokopczuk (2016) examine the performance of several historical, time-series, and option-implied estimators. They find that including information from option markets can be beneficial, with estimators such as the simple historical beta and the Kalman filter exhibiting smaller errors than those based on multivariate GARCH models. Another strand of literature explores the quality of the beta estimates in different frequencies. Becker et al. (2021) show that a long-memory estimator of beta provides superior forecasting performance than short- or infinite-memory alternatives, while in an international setting, Hollstein (2020), also finds that betas exhibit long-run dependence. Hollstein et al. (2020), show that using intraday data improves the performance of the conditional CAPM, compared to those based on lower frequencies. In contrast, Gilbert et al. (2014), find that betas based on lower frequencies lead to lower pricing errors relative to those estimated using higher frequency data.

A related strand of literature accounts for the time-variation and estimation error in beta by incorporating information based on firm-specific characteristics or aggregate economic variables to the estimates of individual security betas (e.g., Shanken, 1990; Ferson and Schadt, 1996; Jagannathan and Wang, 1996). Commonly used variables include the yield or volatility of the one-month Treasury Bill, the term and default spreads, the dividend yield or a function of consumption, asset wealth, and labour income (Lettau and Ludvigson, 2001). Guo et al. (2017) consider several conditioning variables and uncover that the earnings-price ratio, inflation, and the unemployment rate are the best predictors for the beta of the value premium. Hollstein et al. (2019) assume that beta is a linear function of state variables and find that models that depend on economic information perform worse than the simple historical beta.

Cosemans et al. (2016), propose a conditional beta model that shrinks market beta estimates toward an economically informative prior unique to each firm. Their approach yields superior out-of-sample benefits and outperforms commonly used estimators of regular market beta. Kim et al. (2021) and Kelly et al. (2021) indicate that using firm-specific characteristics can also help improve the estimates of time-varying market betas. More recently, Drobetz et al. (2025), adopt the additive prediction error model by Gu et al. (2020), to model firm-specific realized beta using a large set of firm-level fundamental variables and a wide range of machine learning models. They find that tree-based approaches perform best, with historical betas, turnover and size signals being key predictors.

We complement this literature by estimating conditional asymmetric betas across horizons using a large set of firm-specific characteristics and a range of machine learning models with flexible functional forms that can be nonlinear or perform variable selection. While prior studies have primarily focused on modelling the regular market beta, we extend this framework to asymmetric risk measures that capture how systematic risk varies across market conditions. This enables us to evaluate the extent to which machine learning techniques enhance the estimation of downside and upside betas, and the four semibetas. Furthermore, this approach allows us to assess not only the predictive gains from modelling asymmetry and nonlinearity in risk exposures, but also to uncover the firm-level characteristics that drive their dynamics. Our study bridges the literature on modelling asymmetric risk with machine learning in finance, providing new evidence on the structure, predictability and economic value of firm-level conditional betas.

Our main result is that conditional asymmetric betas estimated using machine learning techniques and a large set of firm-specific characteristics deliver statistically significant outperformance relative to the historical benchmark. Out-of-sample performance is strongest at short horizons and declines monotonically as the forecasting horizon increases. Overall, nonlinear models exhibit higher predictive accuracy than linear models, with the difference becoming more pronounced at longer horizons, indicating the importance of nonlinearities in forecasting asymmetric risk. Forecast combinations of nonlinear models deliver the highest accuracy across the majority of beta measures and horizons, while random forests consistently yields competitive performance among individual models.

For example, in the case of downside beta, a combination of nonlinear models yields an out-of-sample $R^2$ value of approximately 46% that declines to 24% as the forecasting horizon increases from one to twelve months. Similarly, for upside beta, the corresponding values range from 18% to 45%. For concordant semibetas, the predictive accuracy of the forecast combination across horizons is between 25% and 46% for the negative semibeta and from 23% to 42% for the positive semibeta. In contrast, discordant semibetas become more challenging to predict at longer horizons. However, nonlinear methods exhibit the strongest performance consistently outperforming other conditional estimators.

Examining the evolution of model performance over time we find that conditional beta forecasts generate persistent gains relative to the benchmark with models exhibiting a consistent upward trend in cumulative performance across asymmetric beta measures throughout the out-of-sample period, with this pattern being more prominent at shorter horizons. For longer forecasting horizons the differences in performance across models becomes more pronounced, with combinations of nonlinear models and random forests delivering the strongest predictive gains, especially in the aftermath of the global financial crisis. While downside and upside betas display similar dynamics across horizons, performance for semibetas is more heterogeneous, with nonlinear models generally providing the most consistent improvements, particularly for the discordant semibetas in the second half of the sample.

A cross-sectional analysis of forecast errors for the asymmetric betas shows that beta estimators tend to overpredict low-beta stocks and underpredict high-beta stocks, with this bias increasing monotonically across quintile portfolios sorted on beta. We find that machine learning models reduce the magnitude of this bias relative to the benchmark, indicating more stable forecasts. At short horizons, conditional betas produce smaller forecast errors than the benchmark across all quintile portfolios. At long horizons, the predictive gains of conditional models diminish, however, they continue to outperform the benchmark primarily in the middle and upper quintile portfolios.

We also perform a variable importance analysis to identify the predictor groups that contribute most to model performance over the out-of-sample period. Across all beta measures, variables related to trading frictions emerge as the most influential predictors, indicating the importance of characteristics such as historical beta estimators in capturing variation in asymmetric risk. The dominance of this group of characteristics is stronger for linear models, where average variable importance is highly concentrated with values ranging from 80% to 95%, while for nonlinear models it is more balanced with values in the range of 31% to 78%. The more diversified predictor set of the nonlinear models includes characteristics from intangibles, momentum and the value-versus-growth groups. Overall, variable importance is more dispersed across predictor groups for the discordant semibetas relative to other asymmetric beta measures, with these patterns remaining consistent across forecasting horizons.

Building on Bollerslev et al. (2022), we also explore the predictive accuracy of regular market beta reconstructed from asymmetric beta components. We find that combining machine learning-based forecasts from downside and upside betas or from the four semibetas yields statistically significant predictive gains relative to the benchmark, highlighting the value of a more granular decomposition of systematic risk. Overall, forecast combinations of nonlinear models or tree-based methods provide the most accurate representation of CAPM beta, with this outperformance becoming more evident at longer horizons. For example, forecast combinations of nonlinear models attain $R^2$ in the range of 12% to 37%, depending on specification and forecasting horizon. Over time, the model performance diminishes as horizon increases and the combination of nonlinear forecasts shows consistent predictive gains compared to that of linear models, which are stronger for the second half of the out-of-sample period. In the cross-section, conditional estimators reduce the bias of overpredicting low-beta stocks and underpredicting high-beta stocks relative to the benchmark, leading to more stable forecasts especially in the lower and middle parts of the beta distribution.

Given that the vast majority of corporate managers continue to rely on the CAPM for estimating the cost of equity (Graham and Harvey, 2001; Graham, 2022), we examine whether the predictive gains from conditional beta forecasts translate into improved equity valuation within a discounted cash flow framework. We find that valuation accuracy varies with both the forecasting horizon and the underlying economic assumptions. While single-horizon beta forecasts offer improved valuation accuracy at short horizons, their predictive power decays at longer horizons, and under low growth and high market risk premium scenarios. In contrast, allowing discount rates to vary across horizons by using multi-horizon conditional beta forecasts leads to substantial gains in valuation accuracy relative to single-horizon approaches. This outperformance is more pronounced for forecast combinations of nonlinear models and the semibeta decomposition, where in high-growth scenarios the out-of-sample $R^2$ exceeds 60%. These findings highlight the economic value of adopting a more granular, multi-horizon approach to estimating systematic risk, indicating that accounting for the term structure of conditional betas improves equity valuation and capital budgeting decisions.

Lastly, we investigate whether the predictive gains for asymmetric beta forecasts based on machine learning models translate into significant economic value in asset allocation decisions. We demonstrate that conditional beta estimates offer considerable benefits for investors who rely on factor models to construct optimal portfolios, by imposing a factor structure on the covariance matrix and forming market-neutral portfolios. We find that portfolios based on conditional betas deliver improved hedging performance relative to the benchmark. The distribution of ex-post portfolio betas is tightly centered around zero, while benchmark portfolios exhibit significant deviations from market neutrality across horizons. Portfolios based on forecast combinations of nonlinear models exhibit the best performance at short horizons, with their performance persisting at longer horizons. Overall, these results indicate that incorporating conditional beta forecasts based on a large number of characteristics into portfolio construction leads to improved hedging of market risk.

The remainder of this study is organized as follows. Section 2 describes the methodology used to estimate asymmetric betas and the machine learning models. Section 3 provides details on the data, sample splitting and hyperparameter tuning. Section 4 analyses the empirical results of the out-of-sample statistical and economic evaluation, while Section 5 concludes.

## 2. Asymmetric Beta Estimation

In this Section we initially describe the estimation of realized downside and upside betas and the four semibetas, followed by a discussion of modelling realized betas using machine learning.

## 2.1. Realized Asymmetric Betas

The CAPM indicates that the expected return on an asset in excess of the risk-free rate is related to market beta, through the following linear equation:

$$E(r_i) = \beta_i E(r_m), \quad (1)$$

where $E(r_i)$ and $E(r_m)$ denote the expected excess return on the asset $i$ and the market portfolio respectively, and $\beta_i$ is the market beta, for $i = 1, ..., N$, with $N$ being the number of assets. The market beta for asset $i$ is then defined as:

$$\beta_i = \frac{Cov(r_i, r_m)}{Var(r_m)}. \quad (2)$$

According to the CAPM, market beta is constant across periods of high and low market returns. Bawa and Lindenberg (1977) propose an extension of the CAPM by specifying downside and upside betas, to account for the asymmetric treatment of risk. We follow Ang et al. (2006) and compute downside (upside) beta over periods when the excess market return is below (above) a specific cutoff point.[4] Specifically, the CAPM equation during down and up markets is defined as:

$$E(r_i|r_m < 0) = \beta_i^- E(r_m|r_m < 0) \quad \text{and} \quad E(r_i|r_m > 0) = \beta_i^+ E(r_m|r_m > 0), \quad (3)$$

and is conditional that market excess returns being negative or positive, while $\beta_i^-$ and $\beta_i^+$ denote the downside and upside beta respectively. The downside and upside betas can be defined as

$$\beta_i^- = \frac{Cov(r_i, r_m^-)}{Var(r_m^-)} \quad \text{and} \quad \beta_i^+ = \frac{Cov(r_i, r_m^+)}{Var(r_m^+)}, \quad (4)$$

where $r_m^-$ and $r_m^+$ denote the market returns in down and up markets respectively.

Bollerslev et al. (2022) propose a four-way decomposition of regular market beta into four semibetas. This decomposition relies on the semicovariance concepts by Bollerslev et al. (2020). Each semibeta uses a different component of the total covariation $Cov(r_i, r_m)$, with semicovariance components being defined by both asset and market returns being negative, $\mathcal{N}$, both returns being positive, $\mathcal{P}$, mixed sign with negative market return, $\mathcal{M}^-$ and mixed sign with positive market return, $\mathcal{M}^+$. The four semibetas are defined as:

$$\beta_i^{\mathcal{N}} = \frac{Cov(r_i^-, r_m^-)}{Var(r_m)}, \quad \beta_i^{\mathcal{P}} = \frac{Cov(r_i^+, r_m^+)}{Var(r_m)}, \quad \beta_i^{\mathcal{M}^-} = \frac{Cov(r_i^+, r_m^-)}{Var(r_m)}, \quad \beta_i^{\mathcal{M}^+} = \frac{Cov(r_i^-, r_m^+)}{Var(r_m)}, \quad (5)$$

where $r_i^-$ and $r_i^+$ denote the returns of asset $i$ in down and up markets respectively. The semibetas condition on the covariation of both the signed asset returns and signed market returns, in contrast to downside and upside betas, which condition only on the sign of market returns.

---

[4] The measures of $\beta_i^-$ and $\beta_i^+$ we consider use returns relative to the zero rate of return, as a cutoff point to determine down markets and up markets. Alternative cutoff points, between down markets and up markets, include the realized average market excess return and the risk-free rate of return. Ang et al. (2006) show, for both downside and upside measures of beta, that the estimators based on the three types of cutoff points are highly correlated, and that their results are driven by the emphasis placed on losses versus gains instead of the particular cutoff point used as the benchmark.

In this study we focus on modelling the $h$-month ahead realized beta. Specifically, we estimate betas for horizons of $h = 1, 3, 6, 12$ months. Andersen et al. (2006) show that under weak regularity conditions only, realized beta is a consistent estimator of the underlying beta. Using daily returns over the past $h$ months, realized market beta is derived as:

$$\hat{\beta}_i = \frac{\sum_t r_i r_m}{\sum_t r_m^2}, \quad (6)$$

while downside and upside betas are derived as:

$$\hat{\beta}_i^- = \frac{\sum_t r_i r_m^-}{\sum_t (r_m^-)^2}, \qquad \hat{\beta}_i^+ = \frac{\sum_t r_i r_m^+}{\sum_t (r_m^+)^2}, \quad (7)$$

where $r_m^- = \min(r_m, 0)$ and $r_m^+ = \max(r_m, 0)$ are the signed market returns. The realized semibetas are similarly defined as:

$$\hat{\beta}_i^{\mathcal{N}} = \frac{\sum_t r_i^- r_m^-}{\sum_t r_m^2}, \quad \hat{\beta}_i^{\mathcal{P}} = \frac{\sum_t r_i^+ r_m^+}{\sum_t r_m^2}, \quad \hat{\beta}_i^{\mathcal{M}^-} = \frac{\sum_t r_i^+ r_m^-}{\sum_t r_m^2}, \quad \hat{\beta}_i^{\mathcal{M}^+} = \frac{\sum_t r_i^- r_m^+}{\sum_t r_m^2}, \quad (8)$$

where $r_i^- = \min(r_i, 0)$ and $r_i^+ = \max(r_i, 0)$ are the signed returns of asset $i$.[5]

Following studies such as Cosemans et al. (2016) and Hollstein and Prokopczuk (2016), we use ex-post realized beta to evaluate the accuracy of the ex-ante beta estimates obtained from different estimators. Downside and upside betas in addition to the four semibetas are constructed for the realized beta estimator and modelled using a large set of firm-specific characteristics and machine learning techniques.

---

[5] Since the mixed sign semicovariances are always weakly negative numbers, we follow Bollerslev et al. (2022) and set $\hat{\beta}_i^{\mathcal{M}^-} = -\sum_t r_i^+ r_m^- / \sum_t r_m^2$ and $\hat{\beta}_i^{\mathcal{M}^+} = -\sum_t r_i^- r_m^+ / \sum_t r_m^2$ in our analysis for ease of interpretation.

## 2.2. Conditional Measures of Asymmetric Betas

The estimators of beta described in this section are conditioned on firm-characteristics. Specifically, firm-specific signals are incorporated in models that combine information from multiple predictors using flexible functions.

Following Shanken (1990) and Ferson and Schadt (1996) we parametrize asymmetric beta as a function of firm characteristics in a pooled regression setting similar to that of Cosemans et al. (2016) and Drobetz et al. (2025). Specifically, we estimate each $\{\beta^-, \beta^+, \beta^{\mathcal{N}}, \beta^{\mathcal{P}}, \beta^{\mathcal{M}^-}, \beta^{\mathcal{M}^+}\}$ of individual securities using models that combine information through machine learning methods.

Let $\beta_{i,t+h}^{rl}$ be the realized beta for asset $i$ at month $t+h$ and $x_{i,t}$ be the $P \times 1$ vector of predictors for asset $i$ at month $t$, for $i = 1, \ldots, N$ and $t = 1, \ldots, T$. We describe an asset's realized beta using the additive prediction error model by Gu, et al. (2020):

$$\beta_{i,t+h}^{rl} = E_t(\beta_{i,t+h}^{rl}) + \varepsilon_{i,t+h}, \text{ where } E_t(\beta_{i,t+h}^{rl}) = f(x_{i,t}). \tag{9}$$

In this model the expected beta, $E_t(\beta_{i,t+h}^{rl})$, is connected to the $P$-dimensional vector of the predictors, $x_{i,t}$, through a flexible function, $f(\cdot)$, of the predictors.

A regression framework is adopted to estimate beta, where the goal is to minimize the residual sum of squares. In this case the least squares objective of the panel predictive regression is

$$\operatorname*{argmin}_{\boldsymbol{\theta}} \mathcal{L}(\boldsymbol{\theta}) = \operatorname*{argmin}_{\boldsymbol{\theta}} \left[\frac{1}{NT} \sum_{j=1}^{NT} \left(\beta_j^{rl} - f(x_j; \boldsymbol{\theta})\right)^2\right], \tag{10}$$

where $\mathcal{L}(\cdot)$ is the least squares loss and $f(x_j; \boldsymbol{\theta})$ is a flexible function of the predictors with parameters $\boldsymbol{\theta}$. The out-of-sample forecast of beta $\beta^{rl}$ of asset $i$ at month $t+h$ based on the panel predictive regression and data available through $t$, is given by

$$\hat{\beta}_{i,t+h}^{rl} = \hat{f}(x_{i,t}; \widehat{\boldsymbol{\theta}}), \tag{11}$$

where $\hat{f}$ is the estimate of $f$ based on data through $t$. Because the panel is high dimensional, we use a variety of machine learning algorithms to approximate $f$.

## 2.2.1. Linear Machine Learning Methods

Principal component analysis (PCA) and partial least squares (PLS) reduce the dimension of the predictors, $P$, to a smaller dimension of latent factors $K$. The predictive model for both approaches is:

$$\boldsymbol{\beta}^{rl} = f(\mathbf{X}) = (\mathbf{XW})\boldsymbol{\theta} \qquad (12)$$

where $\boldsymbol{\beta}^{rl}$ is the $NT \times 1$ vector of realized asymmetric betas, $\beta_{i,t+h}^{rl}$, $\mathbf{X}$ is the $NT \times P$ matrix of stacked predictors $x_{i,t}$, $\mathbf{W}$ is the $P \times K$ matrix of weights, with columns $(\mathbf{w}_1, \ldots, \mathbf{w}_K)$ for $K \ll P$ and $\boldsymbol{\theta}$ is the $K \times 1$ coefficient vector. Each $\mathbf{w}_k$, $k = 1, \ldots, K$, is the set of linear combination weights used to create the $k$th latent factor thus, $\mathbf{XW}$ is the dimension reduced version of the original predictors. We treat the number of latent factors as a hyperparameter with $K \in [1,10]$.

PCA can be used to find the first $K$ principal component weight vectors by minimizing:

$$\underset{\mathbf{W}}{\operatorname{argmin}} \|\mathbf{X} - \mathbf{XWW}'\|^2, \quad \text{s.t.} \quad \mathbf{W}'\mathbf{W} = \mathbf{I}_K, \qquad (13)$$

where $\mathbf{I}_K$ is a $K \times K$ identity matrix. The matrix $\mathbf{W}$ can be estimated using singular value decomposition: $\mathbf{X} = \mathbf{UDV}'$, by setting $\mathbf{W} = \mathbf{V}$. The columns of $\mathbf{V} = (\mathbf{v}_1, \ldots, \mathbf{v}_K)$ are the principal components weights. Each weight vector is used to derive the $k^{\text{th}}$ principal component, $\mathbf{Xv}_k$. The variable $\mathbf{Xv}_1$ is the first principal component and has the largest sample variance amongst all linear combinations of the columns of $\mathbf{X}$.

PLS identifies the features in a supervised way, by constructing $K$ linear combinations of $\mathbf{X}$ that have maximum correlation with the target. To find the reduced version of the predictor set, the columns of the weight matrix $\mathbf{W}$ need to be obtained through consecutive optimization problems. The criterion to find the $k^{\text{th}}$ estimated weight vector is

$$\underset{w}{\operatorname{argmax}} [\mathbf{w}'\mathbf{Mw}], \quad \text{s.t.} \quad \mathbf{w}'\mathbf{w} = 1, \quad \mathbf{w}'\boldsymbol{\Sigma}_{\mathbf{XX}}\mathbf{w} = 0, \qquad (14)$$

where $\boldsymbol{\Sigma}_{\mathbf{XX}}$ is the covariance of $\mathbf{X}$ and $\mathbf{M} = \mathbf{X}'\boldsymbol{\beta}^{rl}\boldsymbol{\beta}^{rl\prime}\mathbf{X}$. The version of PLS we employ is SIMPLS proposed by de Jong (1993).

Penalized regression is similar to the simple linear model, in that it considers only the baseline, untransformed predictors, however, it modifies the least squares problem by adding one additional penalty term to the loss function:

$$\underset{\boldsymbol{\theta}}{\text{argmin}}[\mathcal{L}(\boldsymbol{\theta}) + \mathcal{P}(\boldsymbol{\theta}; \cdot)], \text{ where } \mathcal{P}(\boldsymbol{\theta}; \cdot) = \sum_{i=1}^{p} \mathcal{P}(\theta_i; \cdot). \tag{15}$$

There are several choices for the penalty function $\mathcal{P}(\cdot)$. In this study we consider the elastic net (EN) penalty proposed by Zou and Hastie (2005), which is a combination of the ridge and lasso penalties. The elastic net is a convex penalty given by the following function:

$$\mathcal{P}(\theta_i; \lambda; a) = \lambda[a|\theta_i| + (1-a)\theta_i^2], \tag{16}$$

where $\lambda > 0$ is a tuning parameter, which is determined separately and controls the amount of shrinkage and $a$ is a hyperparameter that controls the trade-off between $l_1$ and $l_2$ regularization. The elastic net combines both $l_1$ and $l_2$ terms in the penalty, thus simultaneously performing continuous shrinkage, automatic variable selection and can also select groups of correlated variables. The grid of the hyperparameters for the elastic net penalty are $\lambda \in [10^{-3}, 10^3]$ and $a \in [0, 1]$.

### 2.2.2. Nonlinear Machine Learning Methods

A regression tree is a non-parametric model that is constructed using a recursive binary splitting approach. At first, starting from the bottom of the tree or the root, the predictor set is divided into two distinct and non-overlapping rectangular regions or leaves and then the realized beta is computed as simple average of $\beta^{rl}$ within that region. Then one or both of those regions are split into two more regions and this process continues until certain stopping criteria are met. The predictor variable upon which a branch is based, and the value where the branch is split, is chosen to minimize the forecast error. Specifically, the prediction of a tree, $\mathcal{T}$, with $M$ leaves $R_1, R_2, \dots, R_M$ and maximum depth $D$, is defined as $\mathcal{T}(x_{i,t}, c, M, D) = \sum_{m=1}^{M} c_m I_{\{x_{i,t} \in R_m(D)\}}$, where $R_m(D)$ represents one of the partitions of the predictor space. The score $c$ associated with partition $m$, $c_m$, is the simple average of realized beta within region $R_m$, written as $c_m = 1/TN_m \sum_{x_{i,t} \in R_m} \beta^{rl}$, where $TN_m$ denotes the number of observations in region $m$. At each branch the sorting variable is chosen among the set of predictors and the split value that minimize the following loss function:

$$Q_m(x_{i,t}, c, TN_m) = \frac{1}{TN_m} \sum_{x_{i,t} \in R_m} (\beta^{rl} - c_m)^2. \tag{17}$$

Branching halts when the depth, $D$, of the tree reaches a pre-specified threshold that is tuned adaptively using a validation sample.

The first ensemble approach is based on gradient boosting (GB) by Friedman (2001), which recursively combines a large number of shallow regression trees, known as "weak learners", to form an ensemble of trees with greater stability than a single more complex tree. Gradient boosting is initialized by fitting a shallow tree, then a second tree with the same depth is used to fit the residuals of the previous model and the forecasts of these two trees are added together to form a single ensemble prediction. This procedure is repeated sequentially until the total number of iterations $K$ is reached. Gradient boosting is an additive model that can be expressed as

$$f(x_{i,t}, K) = \sum_{k=1}^{K} f_k(x_{i,t}, c_k, M_k, D, v), \qquad (18)$$

where $K$ is the total number of trees and $f_k$ is given by $f_k(x_{i,t}, v) = f_{k-1}(x_{i,t}) + v\mathcal{T}_k(x_{i,t}, c_k, M_k, D)$. The parameter $v$ controls the learning rate of the boosting procedure that scales the contribution of each tree to the ensemble by a factor of $0 < v < 1$ and prevents the model from overfitting the residuals. We set the range of the hyperparameters for the gradient boosting model to $K \in [1,500]$, $D \in [1, 2]$ and $v \in \{0.1, 0.01\}$. We also consider early stopping, where the training process stops prematurely when the validation error has not decreased for 50 iterations.

The second ensemble method we consider is random forests (RF) by Breiman (2001), which is based on bootstrap aggregating or bagging and combines many noisy but approximately unbiased models to reduce the variance of the estimates. The baseline bagging procedure estimates $\mathcal{T}_1, \mathcal{T}_2, \ldots, \mathcal{T}_B$ based on $B$ different bootstrap samples of the data and then averages their forecasts to obtain a single low-variance model, given by

$$f(x_{i,t}, B) = \frac{1}{B}\sum_{b=1}^{B} \mathcal{T}_b(x_{i,t}, c_b, M_b, D). \qquad (19)$$

Random forests build a set of de-correlated trees by considering only a randomly drawn subset of predictors for splitting at each potential branch. This lowers the average correlation among predictions and further reduces the variance relative to bagging. We set the range of the hyperparameters for random forests to $B = 200$, $D \in [5,10,15,20]$ and the fraction of the predictors considered for each split to the square root of the predictors available in each iteration of the rolling window.

Feed-forward neural networks (FFNN) can be defined as a composition of $h^{(1)}, h^{(2)}, \ldots, h^{(L)}$ nonlinear activation functions for each of the $L$ hidden layers of the network $\mathbf{z}^{(L)} = h^{(L)} \circ \ldots \circ h^{(2)} \circ h^{(1)}(\mathbf{x})$, with $\mathbf{z}^{(l)} = h\left(b_0^{(l)} + \mathbf{w}^{(l)\prime} \mathbf{z}^{(l-1)}\right)$, for $l = 1, 2, \ldots, L$, where $\mathbf{z}^{(l)}$ is the $l$th layer of the network with $m = 1, 2, \ldots, M$ nodes, $\mathbf{w}^{(l)}$ is the matrix of weights and $b_0^{(l)}$ is the bias. For the first hidden layer the input is the matrix of predictors, $\mathbf{z}^{(0)} = \mathbf{x}$, such that $\mathbf{z}^{(1)} = h\left(b_0^{(1)} + \mathbf{w}^{(1)\prime} \mathbf{x}\right)$. The results from each hidden layer are aggregated in the output layer:

$$f(\mathbf{x}) = b_0^{(L+1)} + \mathbf{w}^{(L+1)\prime} \mathbf{z}^{(L)}. \tag{20}$$

The activation function applied to each node can take various forms. We follow the existing literature and use the rectified linear unit (ReLU) defined as $h(x) = \max(0, x)$, which encourages sparsity in the number of active nodes and avoids the vanishing gradient problem (Glorot et al., 2011). The neural network is trained using stochastic gradient descent and specifically the adaptive moment estimation algorithm (Adam) by Kingma and Ba (2015). Adam computes individual adaptive learning rates for the model parameters using estimates of first and second moments of the gradients.

Training a neural network can be challenging due to the large number of parameters to be estimated and the nonconvexity of the objective function. To alleviate those concerns we employ various tools to reduce estimation error. First, we use dropout (Srivastava et al., 2014), which randomly sets a fraction of the hidden units to zero at each iteration, thus preventing the neural network from overfitting. To improve the performance of the model we also implement early stopping, batch normalization and forecast averaging.

In each iteration of the optimization algorithm the parameter estimates are updated so as to reduce prediction errors in the training sample and then the predictive performance of the model for that iteration is evaluated using data from the validation sample. Early stopping is implemented by stopping the training process prematurely when the validation error no longer decreases. Specifically, the optimization process halts when the maximum number of iterations (100) is reached or if the validation error has not improved for 5 consecutive iterations, preventing overfitting and significantly speeding up the training process. In both cases the parameter estimates of the best performing model are retrieved.

Batch normalization (Ioffe and Szegedy, 2015) reduces the variability of the predictors by scaling the input of activations. It was proposed to solve the problem of internal covariate shift in which the distribution of the inputs of the hidden layer change during training, as the parameters of the previous layers change. For each node in each training step, the algorithm cross-sectionally standardizes the output of a previous activation to restore the representation power of the node. This helps to increase the stability of the neural network and increase the speed of training.

Finally, to reduce the prediction variance of the neural network, due to the stochastic nature of the optimization, we adopt an ensemble approach by forming the final prediction from the average of the forecasts from ten models initialized from different random seeds. For the hyperparameters of the neural networks we set the learning rate to $\{0.001, 0.01\}$, the dropout rate to $\{0.1, 0.2, 0.3\}$, while the parameters for Adam are left to the defaults. We also treat the depth of the network as a hyperparameter and consider neural networks with up to three layers. The number of units in each layer depends on the depth of the network (a network with three layers would have 32 units in the first hidden layer, 16 in the second and 8 in the last hidden layer).

### 2.2.3. Forecast Combinations

We also construct forecast combinations of the linear (CLIN) and nonlinear (CNL) machine learning models to combine the predictive power of multiple models (Goyal and Welch, 2008). Following Bates and Granger (1969), we consider a simple combination that equally weights the individual forecasts from each set of models, an approach which Rapach et al. (2010) show can lead to significant benefits in predictive performance. The forecast combination estimate of realized asymmetric beta, $\hat{\beta}_i^{rl,C}$, of asset $i$ at month $t+h$, based on $M$ individual model forecasts, $\hat{\beta}_i^{rl,m}$ is:

$$\hat{\beta}_{i,t+h}^{rl,C} = \frac{1}{M} \sum_{m=1}^{M} \hat{\beta}_{i,t+h}^{rl,m}. \tag{21}$$

Creating ensembles of linear and nonlinear models allows us to directly assess the benefits of accounting for nonlinearities and combine the predictive power of multiple machine learning approaches.

## 3. Data and Sample Splitting

The data set used to construct the asymmetric betas is based on the post-1978 CRSP database. Each period, the sample contains stock returns from common stocks (share codes 10 and 11), trading on the NYSE, the AMEX, and the NASDAQ exchanges (exchange codes 1, 2 or 3), with positive volume, with a stock price greater than $5 and with monthly market capitalization above the 20% NYSE threshold. In addition to stock return data, we use the daily value-weighted return (including distributions) market index from CRSP as a proxy for the market portfolio and the risk-free rate of return from Kenneth French's online data library.[6]

The realized asymmetric betas are estimated for each stock in the sample using daily returns over a specific period. The beta estimates for a particular month are derived using daily returns over four different estimation windows, with $h = 1, 3, 6, 12$ months.[7] A rolling window is used to estimate the monthly realized beta measures, which expands by one (calendar) month at a time until it reaches the end of the sample.

Overall, the predictor set is comprised of 214 firm-specific characteristics. Of those, 207 are from the Open Source Asset Pricing database[8] by Chen and Zimmermann (2022) and the remaining 7 predictors are realized beta estimators of the CAPM beta, the upside and downside betas and the four semibetas based on daily data. We categorise the predictors into six groups from Hou et al. (2020) and Han et al. (2024), with the number of characteristics in each group given in parentheses: Intangibles (59), Investment (34), Momentum (36), Profitability (15), Trading Frictions (35), and Value vs Growth (28). Following Gu et al. (2020), we replace missing values in a firm characteristic with the cross-sectional median of that characteristic at each month for all stocks. Furthermore, we follow Cosemans et al. (2016), and winsorize outliers in all firm characteristics to the 0.5th and 99.5th percentile values of their cross-sectional distributions and standardize all characteristics by subtracting the cross-sectional mean and dividing by the cross-sectional standard deviation in each month. To avoid look-ahead bias we lag all predictors by $h$ months. Further details on the predictor set can be found in Appendix 2.

---

[6] https://mba.tuck.dartmouth.edu/pages/faculty/ken.french/data_library.html
[7] Monthly betas for horizons $h = 1, 3, 6, 12$ are derived for stocks that have at least 15, 50, 100, or 200 daily observations available over the corresponding preceding period.
[8] https://www.openassetpricing.com/

## 3.1. Sample Splitting and Hyperparameter Tuning

Once we obtain the realized measures of beta, the full sample period begins in January 1980 and ends on December 2024, for a total of $T = 540$ monthly observations (or 45 years). To estimate the conditional beta measures, we adopt a rolling window approach. The size of the rolling window used to estimate the parameters of the predictive models is set to $\tau = 120$ monthly observations (or ten years) and moves across the full sample by one year (or 12 monthly observations) at a time. Following Cosemans et al. (2016), a stock is included in a rolling window iteration if its return for the last 36 months of the rolling window is available. Since the asymmetric measures of beta being forecasted are constructed using $h$-months of daily return data, there needs to be an $h$-month gap between the end of the in-sample period of the rolling window, used for estimating the models, and the out-of-sample period, used for constructing the forecasts based on the fitted models and their evaluation with the respective realized asymmetric beta measure. The initial rolling window spans the period from January 1980 to December 1989, leading to an out-of-sample period from January 1990 to December 2024 that is used for the out-of-sample evaluation of the alternative beta estimators.

The machine learning models used to derive the conditional asymmetric beta forecasts rely on hyperparameter tuning. The choice of hyperparameters controls the amount of model complexity and is critical for the performance of the model. We follow Gu et al. (2020) and adopt the validation sample approach, which maintains the temporal ordering of the data and select the optimal set of values for the tuning parameters based on the validation sample. Specifically, in each iteration of the rolling window, the in-sample is split into two disjointed periods, the training subsample consisting of the first nine years, with the last year used as the validation subsample. In the training subsample the model is estimated for several sets of values of the tuning parameters. The second subsample is used to select the optimal set of tuning parameters. Using the parameter estimates for each set of hyperparameters from the training sample, conditional beta forecasts are constructed for the observations in the validation sample. The optimal set of hyperparameters is chosen to minimize the pooled mean squared error (MSE) over the validation subsample. The model is re-estimated using the chosen set of hyperparameters and all in-sample observations in the iteration of the rolling window. [9]

### 3.2. Descriptive Statistics

Table 1 reports the time series averages of the cross-sectional means, medians, standard deviations and correlations of the realized asymmetric betas across different horizons ($h = 1, 3, 6, 12$ months). The downside and upside betas ($\beta^-$, $\beta^+$) are on average close to unity, while the two concordant semibetas ($\beta^{\mathcal{N}}$, $\beta^{\mathcal{P}}$) exhibit on average lower values, and consistent with Bollerslev et al. (2022) we observe that they far exceed the two discordant semibetas ($\beta^{\mathcal{M}^-}$, $\beta^{\mathcal{M}^+}$). The standard deviation of $\beta^-$ and $\beta^+$ is higher than that of the concordant semibetas, which also exceeds that of the discordant semibeta measures. The downside and upside betas are highly correlated with the negative and positive semibetas respectively, while the discordant semibetas are weakly positively correlated with each other and exhibit low positive or negative correlations with the other beta measures. For all estimation windows, the average values remain relatively unchanged, however, the standard deviation decreases and correlations increase as the estimation window increases.

[Insert Table 1 About Here]

Figure 1 presents the unconditional distributions of all beta measures across all months and stocks in the sample. The distribution of realized downside and upside betas is symmetric around unity, while the semibetas are weakly positive and their distributions are right skewed and centered below unity. The distributions of the downside and upside betas, the two concordant betas and those of the two discordant betas are respectively almost the same and their shape persists across horizons.

[Insert Figure 1 About Here]

---

[9] For example, for $h = 12$, the sample splitting for the last iteration of the rolling window is January 2013 to December 2021 (108 months) for the training sample, January 2022 to December 2022 (12 months) for the validation sample. The model is refit for the optimal hyperparameter set using all in-sample observations, January 2013 to December 2022 (120 months). Forecasts are constructed based on the fitted model parameters and compared to the realizations from January 2024 to December 2024 (12 months).

# 4. Empirical Analysis

This section presents the out-of-sample evaluation of the conditional asymmetric beta measures. Specifically, we investigate the predictive ability of the conditional beta estimators to forecast the respective $h$-month ahead ($h = 1, 3, 6, 12$) realized asymmetric beta.

## 4.1. Evaluation of Conditional Asymmetric Betas

Initially, we analyse the forecasts in terms of out-of-sample $R^2$ (see, e.g., Campbell and Thompson, 2008). The $R^2$ statistic measures the proportional reduction in mean squared error for the conditional asymmetric beta forecast relative to the benchmark:

$$R^2 = 1 - \frac{\text{MSE}^m}{\text{MSE}^b},$$

where $\text{MSE}^m$ is the mean squared error of the alternative forecasting model and $\text{MSE}^b$ is the mean squared error of the benchmark model. Specifically, we compare the forecasts from the conditional estimators of asymmetric beta to the historical estimator based on the $h$-lagged realized beta measure for the respective asymmetric beta. A positive out-of-sample $R^2$ value indicates that the conditional beta forecast outperforms the benchmark in terms of MSE, while a negative value signals the opposite.

We consider three versions of out-of-sample $R^2$ that differ based on how the MSE statistic has been calculated. MSE is derived across all stocks throughout the out-of-sample period as

$$\text{Panel MSE}_h = \sum_{j=1}^{NT^{OOS}} w_j \left(\beta_j^{rl} - \hat{\beta}_j\right)^2,$$

$$\text{Time Series MSE}_h = \sum_{i=1}^{N} w_i \frac{1}{T_i^{OOS}} \sum_{t=1}^{T_i^{OOS}} \left(\beta_{i,t}^{rl} - \hat{\beta}_{i,t}\right)^2,$$

$$\text{Cross Section MSE}_h = \frac{1}{T^{OOS}} \sum_{t=1}^{T^{OOS}} \sum_{i=1}^{N_t} w_{i,t} \left(\beta_{i,t}^{rl} - \hat{\beta}_{i,t}\right)^2.$$

where $\beta^{rl}$ is the realized beta, $\hat{\beta}$ denotes the predicted beta, $T_i^{OOS}$ is the total number of out-of-sample observations for firm $i$, $N_t$ is the number of firms in period $t$ of the out-of-sample and $w$ is the weight of each stock. Panel MSE refers to how well a model describes the realized asymmetric beta in a pooled context. Time Series MSE (TS) summarizes predictive performance as the cross-sectional average of performance for each firm $i$. Cross Section MSE (CS) measures the cross-sectional predictive performance as in a Fama-MacBeth context.

To examine the differences in the out-of-sample predictive accuracy between the conditional asymmetric beta estimator and the benchmark we use the Clark and West (CW, 2007) test for equal predictive accuracy. To compare the panel, time-series and cross-sectional average of prediction errors between the forecasting model, $m$, and the benchmark, $b$, we use the following CW test statistics:

$$CW_{m,b,h}^{Q} = \frac{\bar{d}_{m,b}^{Q}}{\hat{\sigma}_{m,b}^{Q}}, \qquad Q \in \{\text{Panel, TS, CS}\},$$

where $\bar{d}_{m,b}$ is the weighted average of the Clark–West error differences and $\hat{\sigma}_{m,b}$ is the heteroskedasticity and autocorrelation consistent (HAC) standard error using the Newey and West (1987) estimator with 4 lags. The CW adjusted term of squared error differences is defined as:

$$d_{i,t} = \left(\hat{e}_{i,t}^{(m)} - \hat{e}_{i,t}^{(b)} + \hat{e}_{i,t}^{(m,b)}\right),$$

where $\hat{e}_{i,t}^{(m)}$ and $\hat{e}_{i,t}^{(b)}$ are the squared prediction errors for the conditional beta forecast and the benchmark respectively, $\hat{e}_{i,t}^{(m,b)}$ is the squared difference between the conditional beta forecast and the benchmark. The weighted average $\bar{d}_{m,b}^{Q}$ for each case is calculated as:

$$\bar{d}_{m,b}^{\text{Panel}} = \sum_{j=1}^{NT^{OOS}} w_j d_j,$$

$$\bar{d}_{m,b}^{\text{TS}} = \sum_{i=1}^{N} w_i \frac{1}{T_i^{OOS}} \sum_{t=1}^{T_i^{OOS}} d_{i,t},$$

$$\bar{d}_{m,b}^{\text{CS}} = \frac{1}{T^{OOS}} \sum_{t=1}^{T^{OOS}} \sum_{i=1}^{N_t} w_{i,t} d_{i,t}.$$

In each case, the HAC estimator is applied to the series that underlies the corresponding mean: pooled observations for the Panel case, aggregating at the firm level for the TS case, and at the month-level for the CS case. For the main results we report predictive performance for $R^2$ and the statistical significance based on the CW test statistics with equal weighting ($w_j = 1/NT^{OOS}$, $w_i = 1/N$, $w_{i,t} = 1/N_t$).

[Insert Table 2 About Here]

The results of the predictive performance evaluation of the asymmetric betas based on panel $R^2$ are reported in Table 2.[10] Several patterns emerge that remain consistent among different asymmetric betas and across forecasting horizons. All models deliver high and statistically significant out-of-sample $R^2$ over the benchmark especially at short horizons, with predictive accuracy declining monotonically as the forecast horizon increases. In general, nonlinear models, exhibit the strongest predictive performance, with the gap in performance between linear and nonlinear methods widening for higher $h$, which suggests that nonlinearities and interactions become increasingly important when forecasting beta at longer horizons. The combination of nonlinear models (CNL) yields the highest forecasting performance across the majority of beta measures and horizons. Among individual models, random forests (RF) consistently achieves the highest $R^2$ across most asymmetric beta measures and horizons, with the predictive gains being particularly pronounced for downside and upside betas and the concordant semibetas.

Comparing across beta measures, downside and upside betas exhibit similar levels of predictability, with downside beta being marginally easier to forecast at longer horizons. Negative semibeta displays strong predictability that is comparable to downside beta. In contrast, predictive performance varies more between semibeta measures. Performance of positive semibeta remains strong at short horizons but deteriorates more quickly compared to negative semibeta as forecasting horizon increases. Furthermore, the positive semibeta is the only case where linear methods, such as the elastic net (EN) or the combination of linear models (CLIN), outperform the nonlinear models. The discordant semibetas present a greater forecasting challenge, with nonlinear machine learning models providing the most consistent gains for these risk measures, however, with generally smaller gains relative to the benchmark compared to the other asymmetric risk measures.

Specifically, for the downside and upside betas and the concordant semibetas the results vary depending on the model and the risk measure being forecast, with $R^2$ values being typically in the range of 39% to 46% for $h = 1$, between 31% and 39% for $h = 3$, from 19% to 32% for $h = 6$, and declining to the range of 10% to 25% for $h = 12$. The mixed-sign semibetas based on negative (positive) market returns, exhibit $R^2$ values in the range of 42% to 44% (36% to 39%) for $h = 1$, from 26% to 33% (20% to 28%) for $h = 3$, from 9% to 20% (-0.4% to 11%) for $h = 6$, and becoming negative for all models and asymmetric risk measures in the case of $h = 12$. Overall, all instances of outperformance of the machine learning models from the benchmark are statistically significant at the 1% level, with ensembles of regression trees and nonlinear methods successfully capturing the dynamics of market asymmetries at longer horizons, which are more challenging to forecast.[11]

---

[10] The results for the time series and cross section $R^2$ are reported in Tables A1 and A2 in Appendix 1 and are consistent with those based on panel $R^2$.

The performance measures reported so far provide a static view of the predictive ability of the machine learning models. To examine a model's performance over time, we construct equal-weighted portfolios of beta forecasts in each month and plot the cumulative difference of forecast errors (CDFE) of the benchmark relative to the conditional beta forecasting models throughout the out-of-sample period for each asymmetric beta measure. The evolution of model performance is depicted in Figures 2 to 5 for horizons $h = 1, 3, 6, 12$ months.

[Insert Figures 2-5 About Here]

Overall, Figure 2 highlights that short term ($h = 1$) conditional beta forecasts not only improve upon the benchmark on average but also generate persistent cumulative gains throughout the out-of-sample period, with the CDFE of all models and across beta measures displaying a similar upward trend. For the 3- and 6-month horizons (Figures 3 and 4) the differences in performance among models become more distinct, as forecasting further into the future becomes a more challenging task. The CDFE pattern for downside and upside betas for $h = 3, 6$ remains similar to when $h = 1$, with forecasts based on RF and the CNL outperforming the remaining models, exhibiting a steeper slope after the global financial crisis. The performance over time for the concordant semibetas, indicates that the combination of nonlinear models and that of linear models yield the best predictive gains for the negative and positive semibetas respectively, while for both discordant semibetas nonlinear models and their combination show clear signs of outperformance, particularly in the second half of the out-of-sample period.

---

[11] The out-of-sample performance evaluation of asymmetric betas is based on equal-weighted measures. In unreported analysis, we examine the out-of-sample $R^2$ and their significance according to the Clark-West test with weights based on the log of market capitalization. The results are qualitatively similar with the equal-weighted findings reported in Table 2, confirming that the predictive gains are not driven by small-cap stocks. These results are available upon request.

For the 12-month horizon (Figure 5), the dynamics for downside and upside betas are broadly consistent with the results of shorter-term horizons, with ensemble methods and the forecast combination of nonlinear models producing the strongest cumulative gains, especially in the aftermath of the global financial crisis. For the negative semibeta, RF and CNL are again the best performing models, while for the positive semibeta, performance gains are flatter until the global financial crisis, after which RF, PLS, EN and the two forecast combinations steadily diverge from the remaining approaches. In contrast, the mixed sign semibetas remain particularly difficult to predict at a 12-month horizon, with cumulative forecast error differences persistently negative across the sample, although an upward trend emerges from approximately 2013 onwards, driven primarily by nonlinear machine learning models.

After examining the predictive accuracy over time, we turn to the cross-sectional predictive ability of the models, by exploring the forecast errors of cross-sectional portfolio sorts. The goal of this analysis is to uncover the pattern of forecast errors of the stocks in the sample with regards to beta. At the end of each month $t$ stocks are sorted into quintile portfolios for each realized asymmetric beta. Portfolio betas are constructed as equal-weighted averages for the different models used to forecast beta and for the one-year ahead realized beta. Forecast errors are defined as the MSE between realized and forecast portfolio betas. To examine the direction of forecast errors, we calculate the fraction of stocks within each quintile portfolio for which the difference between realized and forecast betas is positive. A value above 0.5 would indicate that an alternative beta estimator on average underestimates realized betas, while a value below 0.5 indicates that on average the forecast overestimates realized beta. The results for cross-sectional performance of the forecast combinations of linear and nonlinear models for the asymmetric betas are presented in Figures 7-9.[12]

[Insert Figures 7-9 About Here]

---

[12] The cross-sectional predictive performance of individual models is presented in Figures A1-A4 in Appendix 1.

Focusing on the degree of over- or under-prediction a clear pattern emerges for all conditional estimators in that low-beta portfolios tend to be overpredicted, while high-beta portfolios are systematically underpredicted. This bias is monotonic, as overprediction diminishes across the lower quintiles, while underprediction increases steadily across the upper quintiles. This pattern persists across all asymmetric risk measures and forecasting horizons. More importantly, for 1- and 3-month horizons (Figures 6 and 7) conditional estimators consistently yield smaller errors than the benchmark for all quintile portfolios, suggesting that machine learning forecasts produce more stable estimates across the beta spectrum, mitigating the extreme biases observed in the benchmark in short-term forecasting horizons. For the downside and upside betas the highest errors are observed in the upper and lower quintiles, while for the semibetas the MSEs monotonically increase as portfolios are sorted on higher beta values. On the other hand, for longer forecasting horizons the pattern of the MSEs for downside and upside betas starts to resemble that of the semibetas. For the 6-month horizon (Figure 8) the forecast combinations continue to outperform the benchmark model for the downside and upside betas and the concordant semibetas, however, at a lesser degree. For the discordant semibetas the difference in MSE becomes significantly smaller, with outperformance from the models arising from the upper quintile portfolios. The performance of the models diminishes further at a 12-month horizon (Figure 9) with the alternative models outperforming the benchmark primarily in the middle quintile portfolios for the downside and upside betas and the discordant semibetas, while for the mixed-sign semibetas some evidence of outperformance is observed in the third and fourth quintile.

### 4.2. Variable Importance of Asymmetric Betas

To gain further insight into the drivers leading to the outperformance found for the machine learning models we examine the variable importance of each of the six groups of predictors. The variable importance for each group is computed in each period according to the permutation feature importance and is defined as the change in equal-weighted panel MSE of the variables within that group. The aggregate measure of variable importance is constructed as the time series average throughout the out-of-sample period for each group and then normalized across all groups to sum to 100 for each model. The variable importance for the asymmetric betas for a 1-month horizons presented in Figure 10.[13]

[Insert Figure 10 About Here]

Across all asymmetric beta measures, variables from the trading frictions group stand out as the most influential predictors, confirming their central role in capturing time-variation in risk exposures. This pattern is more pronounced for linear models, where variable importance for this group of characteristics is above 90% for downside and upside betas and the concordant semibetas, and remains above 80% for discordant semibetas. For nonlinear models the importance of trading frictions remains significant, however, average variable importance is more dispersed across asymmetric risk measures. Specifically, gradient boosting (GB) selects predictors from the trading frictions group on average from 42% to 77%, random forests (RF) from 30% to 47% and the neural network (FFNN) from 57% to 75%, depending on the asymmetric beta measure. The dispersion in variable importance is most evident for the discordant semibetas, where nonlinear models rely on a broader set of characteristics relative to other beta measures. In this case, random forests in particular draw on a more diversified predictor set. The additional explanatory power comes primarily from variables in the intangibles group, and to a lesser extent from those in the momentum and value-versus-growth groups, suggesting that these asymmetric risk measures are driven by a richer set of firm-specific characteristics.[14]

## 4.3. Evaluation of Reconstructed CAPM Beta

Bollerslev et al. (2022) demonstrate that the regular market beta can be expressed as a weighted combination of downside and upside betas, with weights determined by the relative variance of negative and positive market returns:

$$\beta_i = \beta_i^- \frac{Var(r_m^-)}{Var(r_m)} + \beta_i^+ \frac{Var(r_m^+)}{Var(r_m)}, \qquad (22)$$

---

[13] The variable importance for greater forecasting horizons exhibits similar patterns to that for $h = 1$ and the results are reported in Figures A5-A7 in Appendix 1.
[14] In unreported analysis, we examine whether the contribution of each predictor group to the predictive performance of the conditional beta models varies over time. The results indicate that the patterns observed on average persist throughout the out-of-sample period, with characteristics from the trading frictions group consistently dominating all other groups across beta measures, forecast horizons, and models. These results are available upon request.

or using the realized downside and upside betas estimated from daily returns, the regular beta can be reconstructed using the following function:

$$\hat{\beta}_i = \hat{\beta}_i^- \frac{\sum_t (r_m^-)^2}{\sum_t r_m^2} + \hat{\beta}_i^+ \frac{\sum_t (r_m^+)^2}{\sum_t r_m^2}. \tag{23}$$

Alternatively, regular market beta can be reconstructed as the sum of the four semibetas, which decompose the signed covariation between asset and market returns:

$$\beta_i = \frac{Cov(r_i^-, r_m^-) + Cov(r_i^+, r_m^+) + Cov(r_i^+, r_m^-) + Cov(r_i^-, r_m^+)}{Var(r_m)}$$

$$= \beta_i^{\mathcal{N}} + \beta_i^{\mathcal{P}} + \beta_i^{\mathcal{M}^-} + \beta_i^{\mathcal{M}^+} \tag{24}$$

or the realized semibetas using the following function:

$$\hat{\beta}_i = \hat{\beta}_i^{\mathcal{N}} + \hat{\beta}_i^{\mathcal{P}} + \hat{\beta}_i^{\mathcal{M}^-} + \hat{\beta}_i^{\mathcal{M}^+}. \tag{25}$$

Building on Bollerslev et al. (2022), who show that regular market beta can be reconstructed from asymmetric components, we evaluate the predictive accuracy of reconstructed CAPM beta forecasts relative to the historical beta benchmark. The CAPM beta is reconstructed under both approaches: first, as a weighted combination of downside and upside beta forecasts, and second, as the sum of semibeta forecasts. The reconstructed betas are assessed on average across forecasting horizons for the full sample, over time, and in the cross-section. Table 3 reports the out-of-sample performance of reconstructed CAPM beta forecasts based on panel $R^2$. [15]

[Insert Table 3 About Here]

Overall, the results show that combining forecasts from downside and upside betas or semibeta components yield statistically significant predictive gains, which indicates that a granular decomposition of systematic risk exposures provides an accurate representation of CAPM beta dynamics. The results further demonstrate that machine learning models substantially enhance the forecasting accuracy of reconstructed CAPM betas, with nonlinear models consistently providing the most robust gains, with their advantage over linear models becoming more pronounced at longer forecasting horizons. All models deliver high and statistically significant out-of-sample $R^2$ at the one-month horizon, with values ranging from 34 to 39% across both reconstruction approaches. Predictive accuracy declines steadily as the forecasting horizon increases from 1 to 12 months, with $R^2$ decreasing to the range of 20%-28% at the 3-month horizon, to 9%-18% at six months and to 3%-13% at twelve months. Across specifications, the performance of CAPM beta reconstructed from semibetas is similar to that based on downside and upside betas. Among individual models, random forests achieves the highest $R^2$ values across both CAPM beta reconstruction approaches and all forecasting horizons. Furthermore, the combination of nonlinear forecasting models (CNL) outperforms consistently the combination comprised of linear models (CLIN).[16]

---

[15] The performance of conditional reconstructed CAPM betas according to the time series and cross section $R^2$ remain consistent with those based on panel $R^2$ and are reported in Tables A3 and A4 in Appendix 1.

To examine the performance of reconstructed CAPM beta over time, we form equal-weighted portfolios of beta forecasts in each month and plot the cumulative difference of forecast errors of the benchmark from the conditional beta estimators. Figure 11 presents the results for the linear and nonlinear forecast combinations across both CAPM specifications for horizons $h = 1, 3, 6, 12$ months.[17]

[Insert Figure 11 About Here]

Overall, the results for performance over time complement the findings from Table 3. Specifically, in that CAPM beta reconstructed from semibetas performs similarly to that based on downside and upside betas, predictive gains diminish as the horizon window increases and that a forecast combination of nonlinear models (CNL) tends to outperform a combination of linear models (CLIN). At the one-month horizon, the upward trend in cumulative performance indicates that conditional CAPM beta forecasts outperform the benchmark throughout the out-of-sample period. At the three-month horizon, the performance for both CNL and CLIN peaks during the dot-com bubble, after which CNL plateaus and CLIN declines until the global financial crisis, before both resuming an upward trend that continues through the end of the sample. A similar pattern emerges at the six-month horizon, although periods of underperformance are more pronounced. At the twelve-month horizon, CLIN performance declines from the dot-com bubble up to the global financial crisis, remains relatively flat until 2013, before increasing until the end of the sample. In contrast CNL, begins to exhibit sustained positive cumulative gains from the global financial crisis onward, highlighting its superior performance at longer horizons.

---

[16] The out-of-sample performance evaluation of reconstructed CAPM betas is based on equal-weighted measures. In unreported analysis, we calculate value-weighted $R^2$ and their significance, with weights based on the log of market capitalization. The results are qualitatively similar to those in Table 3, indicating that the predictive gains are not driven by small-cap stocks. These results are available upon request.

[17] The performance of reconstructed CAPM betas for individual models based on cumulative differences of forecast errors is presented in Figures A8-A11 in Appendix 1.

The cross-sectional predictive performance of the models for reconstructed CAPM beta is evaluated by calculating forecast errors for quintile portfolios sorted on realized CAPM betas. Figure 12 presents the results for the linear and nonlinear forecast combinations across CAPM specifications and forecasting horizons.[18]

[Insert Figure 12 About Here]

Consistent with the findings for asymmetric betas, all conditional estimators display the characteristic pattern of overestimating low-beta portfolios and underestimating high-beta portfolios. However, the magnitude of these errors is smaller relative to the benchmark for the first four quintiles, with the most pronounced gains in predictive accuracy observed for the portfolio based on stocks with the lowest betas. This suggests that the forecasting models are better at capturing variation among low- to medium-beta stocks, where betas may be more stable and easier to predict. Cross-sectional performance is similar when comparing CAPM beta reconstructed from downside and upside betas or from the four semibetas. The distribution of forecast errors also changes across forecasting horizons. At short horizons errors are higher in both the lowest and highest beta quintiles, while at longer horizons errors become relatively flat across the first four quintiles but increase sharply for the highest-beta portfolio, again suggesting that high-beta stocks exhibit less predictable dynamics across horizons.

## 4.4. Equity Valuation and Discounted Cash Flow Analysis

In an extensive survey of US CFOs regarding their practice on capital budgeting, cost of capital, and capital structure conducted by Graham and Harvey (2001), the CAPM was found to be the most widely used approach to estimate cost of equity among 73.5% of the respondents. A more recent survey by Graham (2022) shows that 85% of the surveyed companies continue to rely on the CAPM to estimate their firm's cost of equity.

Given the enduring popularity of CAPM among corporate managers, we employ discounted cash flow models with discount rates based on conditional beta forecasts to evaluate the performance of alternative beta estimators in a share price prediction framework. Our analysis relates to Homel et al. (2023), who examine the performance of a wide range of discounting methods for deriving the value of a project, including discount rates based on factor models, and Augustin and Pukthuanthong (2025), who incorporate the term structure of betas in discounted cash flow analysis. Specifically, we estimate the price of firm $i$ and month $t$ as:

$$PV_{i,t}^h = \sum_{j=1}^{12} \frac{CF_{i,j}}{(1+r_{i,t}^h)^i} + \frac{1}{(1+r_{i,t}^h)^{12}} \times \frac{CF_{i,12}}{r_{i,t}^h - g}, \qquad (26)$$

where $CF_{i,j}$ is the cash flow of firm $i$ in period $j$, $r^h$ is the cost of equity and $g$ is the long-term growth rate. We consider a holding period of one year and a stream of 12 monthly cash flows. For each firm $i$ at month $t$ we estimate the cost of equity using conditional reconstructed CAPM betas for horizon $h$ as:

$$r_{i,t}^h = r_f + \beta_{i,t}^h \times r_m. \qquad (27)$$

---

[18] The cross-sectional performance of reconstructed CAPM betas for individual models based on forecast errors of quintile portfolios sorted on beta is presented in Figures A12-A15 in Appendix 1.

As an alternative we derive the price of firm $i$ and month $t$ by using discount rates that vary according to the conditional beta horizon:

$$PV_{i,t} = \sum_{j=1}^{4} \frac{CF_{i,h_j}}{(1+r_{i,t}^{h_j})^i} + \frac{1}{(1+r_{i,t}^{12})^{12}} \times \frac{CF_{i,12}}{r_{i,t}^{12} - g}, \qquad (28)$$

where $h_j = (1,3,6,12)$ for $j = 1, \ldots 4$, with cash flows for each period derived from monthly cash flows as: $CF_1 = CF_1$, $CF_3 = 2 \times CF_1$, $CF_6 = 3 \times CF_1$ and $CF_{12} = 6 \times CF_1$.

The analysis is performed on firms with positive monthly cash flows, with the cash flow of month $j$, $CF_j$, calculated as the 12-month rolling average of ordinary dividends obtained from CRSP. Discount rates are estimated using a risk-free rate of zero and annual market risk premiums in the range of 8% to 12%, while terminal values are based on annual long-term growth rates from 0% to 2%. Predicted prices are compared to realized CRSP prices based on the out-of-sample panel $R^2$, with the corresponding historical CAPM beta set as the benchmark. Table 4 reports the results for observed price prediction based on intrinsic values derived from discounted cash flow models with cost of equity estimated using conditional betas from forecast combinations.[19]

---

[19] Share price prediction performance of reconstructed CAPM betas for individual models is presented in Tables A5-A9 in Appendix 1.

[Insert Table 4 About Here]

Across Panels A-D, where intrinsic values are determined using horizon-specific betas, the out-of-sample $R^2$ tends to decay as horizon increases. Short-term beta forecasts provide improved valuation results at low growth. For example, when $g = 0\%$ and $r_m = 8\%$ the performance of CLIN based on semibetas starts at approximately 21% and declines to 2% at longer horizons. In contrast, at higher growth rates $R^2$ values are more stable across horizons. For example, when $g = 2\%$ and $r_m = 8\%$ the performance of CNL based on semibetas has a range between 33% and 42%. Performance is stronger at lower market risk premiums, while higher premiums reduce predictive accuracy. For example, $R^2$ becomes negative at longer horizons when both growth is low and the market risk premium is high ($g = 0\%$, $r_m = 12\%$, and $h = 12$). Relative to the variation in performance induced by different long-term growth rates, market risk premiums, and horizons, the differences in $R^2$ are smaller across forecast combination methods and CAPM beta specifications.

Panel E reports the performance of the DCF models that account for the term structure of conditional betas. Combining betas across horizons leads to a substantial improvement in performance relative to using single-horizon forecasts. The predictive gains relative to the benchmark are particularly pronounced at higher growth rates, where $R^2$ exceeds 60% when the discount rate is based on the semibeta decomposition and nonlinear forecast combinations. Short-horizon betas are individually informative, however, constructing cost of equity estimates by combining information across horizons leads to a significant boost in predictive performance, especially at higher growth rates, indicating that accounting for the term structure of betas is beneficial for equity valuation.

### 4.5. Market-Neutral Minimum-Variance Portfolios

We have shown that when machine learning-based estimators of asymmetric beta components are used to reconstruct CAPM beta, they lead to statistically significant improvements in predictive accuracy over the benchmark. In this section, we explore whether these gains benefit portfolio construction by evaluating the performance of optimal portfolios formed using the reconstructed CAPM beta forecasts. This setting is particularly relevant for hedge fund managers implementing market-neutral strategies, where accurate beta estimates are crucial for neutralizing the market risk of a portfolio.

The accuracy of beta estimates becomes important to portfolio management when a factor structure is imposed on the covariance matrix of asset returns. A factor structure leads to more precise estimates of the covariance matrix when the asset universe is large due to the decrease in parameters that need to be estimated. Following Ghysels and Jacquier (2006), Cosemans et al. (2016) and Drobetz et al. (2025), we construct market-neutral minimum-variance portfolios using covariance matrices based on exact factor models for each conditional beta estimator.

The forecast of the covariance matrix of asset returns, $\Sigma_{r,t+h|t}$, implied by the single factor model using the different conditional beta estimates is:

$$\Sigma_{r,t+h|t} = \sigma^2_{m,t+h|t} \boldsymbol{\beta}_{t+h|t} \boldsymbol{\beta}'_{t+h|t} + \Sigma_{e,t+h|t}, \tag{29}$$

where $\boldsymbol{\beta}_{t+h|t}$ is the $N_t \times 1$ vector of beta forecasts, $\sigma^2_{m,t+h|t}$ is the estimate of the variance of the excess market return using data up to time $t$, $\Sigma_{e,t+h|t}$ is a diagonal matrix which contains the residual variances $\sigma^2_{e,i,t+h|t}$. Following Ghysels and Jacquier (2007), the residuals of stock $i$ are derived using the conditional beta estimates as $\hat{e}_{i,t+h} = r_{i,t} - \hat{\beta}_{i,t+h} r_{m,t}$. Both market and idiosyncratic variances are estimated using in-sample daily data over a two-year rolling window.

The forecasts of the factor-based covariance matrix are then used to construct market-neutral minimum variance portfolios, whose weights are obtained by solving the following optimization problem:

$$\underset{\mathbf{w}_{t+h}}{\mathrm{argmin}}\ \mathbf{w}'_{t+h} \hat{\Sigma}_{r,t+h|t} \mathbf{w}_{t+h},$$

$$\text{s.t.} \quad \sum_{i=1}^{N_t} w_{i,t+h} = 1, \quad \sum_{i=1}^{N_t} w_{i,t+h} \hat{\beta}_{i,t+h|t} = 0, \quad -0.3 \leq w_{i,t+h} \leq 0.3, \tag{30}$$

the first constraint ensures full investment, the second enforces market neutrality, and the third imposes bounds on individual portfolio weights.

We construct monthly portfolios using the 500 stocks with the highest market capitalization at each formation period. Given the short investment horizon typical of hedge fund strategies, portfolios are rebalanced monthly, and we derive the realized portfolio beta in the subsequent month. Following Augustin and Pukthuanthong (2025), performance is assessed by comparing the distribution of ex-post realized portfolio betas across models. Specifically, we compare portfolios based on reconstructed

CAPM betas that are derived from either downside and upside betas or the four semibetas to those based on the historical CAPM beta benchmark. Ideally, accurate beta forecasts should produce portfolio beta distributions centered around zero. Figures 13 to 16 present the distributions of realized portfolio betas for the benchmark model and for forecast combinations of linear and nonlinear models across different forecasting horizons.[20]

[Insert Figures 13 to 16 About Here]

Market-neutral minimum-variance portfolios based on reconstructed CAPM betas generate ex-post portfolio betas that are tightly centered around zero. In contrast, portfolios formed using the historical CAPM beta estimator are not truly market neutral, as reflected by the displacement of their beta distributions away from zero and consistently highly positive modes across all horizons. Specifically, the distribution of ex-post portfolio betas based on the historical estimator exhibits modes ranging from 0.40 to 0.66 at short horizons ($h = 1,3$) and declining to 0.26-0.29 at longer horizons ($h = 6,12$). Turning to minimum-variance portfolios based on conditional estimators, the performance varies across horizons. At short horizons, portfolios based on CAPM betas reconstructed from semibetas using nonlinear forecast combinations (CNL) perform best, yielding modes closest to zero, with values 0.01 at $h = 1$ and approximately zero at $h = 3$. At the six-month horizon, the best performance is achieved by portfolios based on CAPM beta reconstructed from downside and upside betas based on CNL forecasts, with a mode of 0.02. At the twelve-month horizon, performance converges across specifications and models, with modes ranging from $-0.07$ to $0.08$. These results indicate that the reconstructed beta estimates lead to more accurate hedging of market risk, and that hedge fund managers would consistently benefit by incorporating beta forecasts based on firm-specific characteristics into the portfolio formation process.

## 5. Conclusion

This study investigates whether machine learning methods and firm-specific characteristics can improve the estimation of asymmetric betas across different horizons. Using a large cross-section of US stocks over more than four decades, we evaluate the predictive performance of conditional downside and upside betas by Ang et al. (2006), as well as the four semibetas proposed by Bollerslev et al. (2022). Across a comprehensive set of models, we find that conditional asymmetric betas based on machine learning exhibit statistically significant predictive gains that translate to economically significant out-of-sample benefits in equity valuation and for investors constructing market-neutral portfolios.

---

[20] Minimum-variance portfolio performance of reconstructed CAPM betas for individual models is presented in Figures A16-A19 in Appendix 1.

Predictive performance is strongest at short horizons and declines monotonically as the forecasting horizon increases. Among asymmetric risk measures, the discordant semibetas are more challenging to predict than downside and upside betas or concordant semibetas. However, nonlinear machine learning models outperform other conditional estimators, with forecast combinations of nonlinear models or random forests providing the highest predictive accuracy across the majority of beta measures and horizons. Beyond model performance, our analysis shows that characteristics from the trading frictions group, which includes historical beta indicators, is the dominant set of predictors, followed by characteristics from intangibles, momentum and growth groups.

These results extend to the reconstruction of the regular CAPM beta from asymmetric beta components. Forecast combinations of nonlinear models provide stronger predictive gains compared to combinations of linear models, indicating that a more granular decomposition of systematic risk exposures provides a richer and more accurate representation of the CAPM beta dynamics. Estimating discount rates based on conditional betas lead to more accurate share price prediction relative to the benchmark, especially when estimating the cost of equity using beta forecasts across multiple horizons. Finally, market-neutral portfolios constructed using conditional betas produce ex-post betas tightly centered around zero, in contrast to those based on historical estimates that exhibit systematic deviations from market neutrality. These findings highlight the value of combining machine learning with firm-level signals to generate more accurate beta forecasts, improve valuation performance and to more effectively hedge market risk.

# References

Andersen, T. G., Bollerslev, T., Diebold, F. X., & Wu, G. (2006). Realized beta: Persistence and predictability. In *Econometric Analysis of Financial and Economic Time Series*. Emerald Group Publishing Limited.

Ang, A., Chen, J., & Xing, Y. (2006). Downside risk. *The Review of Financial Studies*, 19(4), 1191-1239.

Augustin, N., & Pukthuanthong, K. Forecasting the Term Structure of Equity Betas: Implications for Valuation.

Bali, T. G., Engle, R. F., & Tang, Y. (2017). Dynamic conditional beta is alive and well in the cross section of daily stock returns. *Management Science*, 63(11), 3760-3779.

Bates, J. M., & Granger, C. W. (1969). The combination of forecasts. *Journal of the Operational Research Society*, 20(4), 451-468.

Bawa, V. S., & Lindenberg, E. B. (1977). Capital market equilibrium in a mean-lower partial moment framework. *Journal of Financial Economics*, 5(2), 189-200.

Becker, J., Hollstein, F., Prokopczuk, M., & Sibbertsen, P. (2021). The memory of beta. *Journal of Banking & Finance*, 124, 106026.

Bollerslev, T., Li, J., Patton, A. J., & Quaedvlieg, R. (2020). Realized semicovariances. *Econometrica: Journal of the Econometric Society*, 88(4), 1515-1551.

Bollerslev, T., Patton, A. J., & Quaedvlieg, R. (2022). Realized semibetas: Disentangling "good" and "bad" downside risks. *Journal of Financial Economics*, 144(1), 227-246.

Breiman, L. (2001). Random forests. *Machine Learning*, 45(1), 5-32.

Buss, A., & Vilkov, G. (2012). Measuring equity risk with option-implied correlations. *The Review of Financial Studies*, 25(10), 3113-3140.

Campbell, J. Y., & Thompson, S. B. (2008). Predicting excess stock returns out of sample: Can anything beat the historical average? *The Review of Financial Studies*, 21(4), 1509-1531.

Campbell, J. Y., Lettau, M., Malkiel, B. G., & Xu, Y. (2001). Have individual stocks become more volatile? An empirical exploration of idiosyncratic risk. *The Journal of Finance*, 56(1), 1-43.

Chang, B. Y., Christoffersen, P., Jacobs, K., & Vainberg, G. (2012). Option-implied measures of equity risk. *Review of Finance*, 16(2), 385-428.

Chen, A. Y., & Zimmermann, T. (2022). Open Source Cross-Sectional Asset Pricing. *Critical Finance Review*, 11(2), 207-264.

Clark, T. E., & West, K. D. (2007). Approximately normal tests for equal predictive accuracy in nested models. *Journal of Econometrics*, 138(1), 291-311.

Cosemans, M., Frehen, R., Schotman, P. C., & Bauer, R. (2016). Estimating security betas using prior information based on firm fundamentals. *The Review of Financial Studies*, 29(4), 1072-1112.

De Jong, S. (1993). SIMPLS: An alternative approach to partial least squares regression. *Chemometrics and Intelligent Laboratory Systems*, 18(3), 251-263.

Drobetz, W., Hollstein, F., Otto, T., & Prokopczuk, M. (2025). Estimating stock market betas via machine learning. *Journal of Financial and Quantitative Analysis*, 60(3), 1074-1110.

Engle, R. F. (2016). Dynamic conditional beta. *Journal of Financial Econometrics*, 14(4), 643-667.

Engle, R. F., Ghysels, E., & Sohn, B. (2013). Stock market volatility and macroeconomic fundamentals. *Review of Economics and Statistics*, 95(3), 776-797.

Fama, E. F., & French, K. R. (1992). The cross-section of expected stock returns. *The Journal of Finance*, 47(2), 427-465.

Fama, E. F., & French, K. R. (2008). Dissecting anomalies. *The Journal of Finance*, 63(4), 1653-1678.

Ferson, W. E., & Schadt, R. W. (1996). Measuring fund strategy and performance in changing economic conditions. *The Journal of Finance*, 51(2), 425-461.

Friedman, J. H. (2001). Greedy function approximation: a gradient boosting machine. *Annals of Statistics*, 1189-1232.

Ghysels, E., & Jacquier, E. (2006). Market beta dynamics and portfolio efficiency. Available at SSRN 711942.

Ghysels, E., Santa-Clara, P., & Valkanov, R. (2005). There is a risk-return trade-off after all. *Journal of Financial Economics*, 76(3), 509-548.

Gilbert, T., Hrdlicka, C., Kalodimos, J., & Siegel, S. (2014). Daily data is bad for beta: Opacity and frequency-dependent betas. *The Review of Asset Pricing Studies*, 4(1), 78-117.

Glorot, X., Bordes, A., & Bengio, Y. (2011). Deep sparse rectifier neural networks. In *Proceedings of the fourteenth international conference on artificial intelligence and statistics* (pp. 315-323). JMLR Workshop and Conference Proceedings.

Graham, J. R. (2022). Presidential address: Corporate finance and reality. *The Journal of Finance*, 77(4), 1975-2049.

Graham, J. R., & Harvey, C. R. (2001). The theory and practice of corporate finance: Evidence from the field. *Journal of Financial Economics*, 60(2-3), 187-243.

Gu, S., Kelly, B., & Xiu, D. (2020). Empirical asset pricing via machine learning. *The Review of Financial Studies*, 33(5), 2223-2273.

Guo, H., Wu, C., & Yu, Y. (2017). Time-varying beta and the value premium. *Journal of Financial and Quantitative Analysis*, 52(4), 1551-1576.

Han, Y., He, A., Rapach, D. E., & Zhou, G. (2024). Cross-sectional expected returns: new Fama–MacBeth regressions in the era of machine learning. *Review of Finance*, 28(6), 1807-1831.

Hollstein, F. (2020). Estimating beta: The international evidence. *Journal of Banking & Finance*, 121, 105968.

Hollstein, F., & Prokopczuk, M. (2016). Estimating beta. *Journal of Financial and Quantitative Analysis*, 51(4), 1437-1466.

Hollstein, F., Prokopczuk, M., & Simen, C. W. (2019). Estimating beta: Forecast adjustments and the impact of stock characteristics for a broad cross-section. *Journal of Financial Markets*, 44, 91-118.

Hollstein, F., Prokopczuk, M., & Simen, C. W. (2020). The conditional capital asset pricing model revisited: Evidence from high-frequency betas. *Management Science*, 66(6), 2474-2494.

Hommel, N., Landier, A., & Thesmar, D. (2023). Corporate valuation: An empirical comparison of discounting methods (No. w30898). *National Bureau of Economic Research*.

Hou, K., Xue, C., & Zhang, L. (2020). Replicating Anomalies. *The Review of Financial Studies*, 33(5), 2019-2133.

Ioffe, S., & Szegedy, C. (2015). Batch normalization: Accelerating deep network training by reducing internal covariate shift. In *International conference on machine learning* (pp. 448-456).

Jagannathan, R., & Wang, Z. (1996). The conditional CAPM and the cross-section of expected returns. *The Journal of Finance*, 51(1), 3-53.

Kelly, B. T., Moskowitz, T. J., & Pruitt, S. (2021). Understanding momentum and reversal. *Journal of Financial Economics*, 140(3), 726-743.

Kim, S., Korajczyk, R. A., & Neuhierl, A. (2021). Arbitrage portfolios. *The Review of Financial Studies*, 34(6), 2813-2856.

Kingma, D. P., & Ba, J. (2014). Adam: A method for stochastic optimization. *arXiv preprint arXiv:1412.6980*.

Lettau, M., & Ludvigson, S. (2001). Consumption, aggregate wealth, and expected stock returns. *The Journal of Finance*, 56(3), 815-849.

Lettau, M., Maggiori, M., & Weber, M. (2014). Conditional risk premia in currency markets and other asset classes. *Journal of Financial Economics*, 114(2), 197-225.

Levi, Y., & Welch, I. (2017). Best practice for cost-of-capital estimates. *Journal of Financial and Quantitative Analysis*, 52(2), 427-463.

Levi, Y., & Welch, I. (2020). Symmetric and asymmetric market betas and downside risk. *The Review of Financial Studies*, 33(6), 2772-2795.

Lewellen, J. (2015). The Cross-section of Expected Stock Returns. *Critical Finance Review*, 4(1), 1-44.

Lewellen, J., & Nagel, S. (2006). The conditional CAPM does not explain asset-pricing anomalies. *Journal of Financial Economics*, 82(2), 289-314.

Lintner, J. (1965). Security prices, risk, and maximal gains from diversification. *The Journal of Finance*, 20(4), 587-615.

Markowitz, H. M. (1959). *Portfolio selection: efficient diversification of investments*. New York, New York: Wiley.

Mossin, J. (1966). Equilibrium in a capital asset market. *Econometrica: Journal of the Econometric Society*, 768-783.

Newey, W. K., & West, K. D. (1987). A Simple, Positive Semi-Definite, Heteroskedasticity and Autocorrelation. *Econometrica: Journal of the Econometric Society*, 55(3), 703-708.

Rapach, D. E., Strauss, J. K., & Zhou, G. (2010). Out-of-sample equity premium prediction: Combination forecasts and links to the real economy. *The Review of Financial Studies*, 23(2), 821-862.

Roy, A. D. (1952). Safety first and the holding of assets. *Econometrica: Journal of the Econometric Society*, 431-449.

Shanken, J. (1990). Intertemporal asset pricing: An empirical investigation. *Journal of Econometrics*, 45(1-2), 99-120.

Sharpe, W. F. (1964). Capital asset prices: A theory of market equilibrium under conditions of risk. *The Journal of Finance*, 19(3), 425-442.

Srivastava, N., Hinton, G., Krizhevsky, A., Sutskever, I., & Salakhutdinov, R. (2014). Dropout: a simple way to prevent neural networks from overfitting. *The Journal of Machine Learning Research*, 15(1), 1929-1958.

Welch, I., & Goyal, A. (2008). A comprehensive look at the empirical performance of equity premium prediction. *The Review of Financial Studies*, 21(4), 1455-1508.

Zou, H., & Hastie, T. (2005). Regularization and variable selection via the elastic net. *Journal of the Royal Statistical Society: Series B (Statistical Methodology)*, 67(2), 301-320.

# Tables and Figures

Table 1 Descriptive Statistics of Asymmetric Betas for Different Estimation Windows.
This table reports the time series averages of the cross-sectional means, medians, standard deviations and correlations for the monthly realized asymmetric betas across four different estimation windows ($h = 1, 3, 6, 12$ months) constructed using daily returns. The asymmetric risk measures are the downside and upside betas ($\beta^-$, $\beta^+$), the concordant semibetas ($\beta^{\mathcal{N}}$, $\beta^{\mathcal{P}}$), and the discordant semibetas ($\beta^{\mathcal{M}^-}$, $\beta^{\mathcal{M}^+}$). The sample period is from January 1980 to December 2024.

|  | 1-month estimation window | | | | | | 3-month estimation window | | | | | |
|---|---|---|---|---|---|---|---|---|---|---|---|---|
|  | $\beta^-$ | $\beta^+$ | $\beta^{\mathcal{N}}$ | $\beta^{\mathcal{P}}$ | $\beta^{\mathcal{M}^-}$ | $\beta^{\mathcal{M}^+}$ | $\beta^-$ | $\beta^+$ | $\beta^{\mathcal{N}}$ | $\beta^{\mathcal{P}}$ | $\beta^{\mathcal{M}^-}$ | $\beta^{\mathcal{M}^+}$ |
| Mean | 0.95 | 0.98 | 0.58 | 0.73 | 0.15 | 0.19 | 0.96 | 0.98 | 0.60 | 0.68 | 0.13 | 0.17 |
| Median | 0.88 | 0.90 | 0.52 | 0.65 | 0.10 | 0.14 | 0.90 | 0.91 | 0.55 | 0.63 | 0.11 | 0.14 |
| St.Dev. | 1.04 | 0.91 | 0.35 | 0.42 | 0.17 | 0.20 | 0.62 | 0.61 | 0.27 | 0.30 | 0.11 | 0.13 |
|  | Correlations | | | | | | Correlations | | | | | |
| $\beta^-$ | 1 | 0.32 | 0.89 | 0.35 | -0.51 | -0.01 | 1 | 0.56 | 0.92 | 0.60 | -0.34 | 0.01 |
| $\beta^+$ | 0.32 | 1 | 0.35 | 0.90 | -0.05 | -0.46 | 0.56 | 1 | 0.58 | 0.92 | -0.06 | -0.31 |
| $\beta^{\mathcal{N}}$ | 0.89 | 0.35 | 1 | 0.44 | -0.09 | 0.10 | 0.92 | 0.58 | 1 | 0.68 | 0.03 | 0.16 |
| $\beta^{\mathcal{P}}$ | 0.35 | 0.90 | 0.44 | 1 | 0.05 | -0.06 | 0.60 | 0.92 | 0.68 | 1 | 0.10 | 0.07 |
| $\beta^{\mathcal{M}^-}$ | -0.51 | -0.05 | -0.09 | 0.05 | 1 | 0.21 | -0.34 | -0.06 | 0.03 | 0.10 | 1 | 0.38 |
| $\beta^{\mathcal{M}^+}$ | -0.01 | -0.46 | 0.10 | -0.06 | 0.21 | 1 | 0.01 | -0.31 | 0.16 | 0.07 | 0.38 | 1 |
|  | 6-month estimation window | | | | | | 12-month estimation window | | | | | |
|  | $\beta^-$ | $\beta^+$ | $\beta^{\mathcal{N}}$ | $\beta^{\mathcal{P}}$ | $\beta^{\mathcal{M}^-}$ | $\beta^{\mathcal{M}^+}$ | $\beta^-$ | $\beta^+$ | $\beta^{\mathcal{N}}$ | $\beta^{\mathcal{P}}$ | $\beta^{\mathcal{M}^-}$ | $\beta^{\mathcal{M}^+}$ |
| Mean | 0.96 | 0.98 | 0.60 | 0.66 | 0.13 | 0.16 | 0.96 | 0.99 | 0.60 | 0.65 | 0.13 | 0.15 |
| Median | 0.91 | 0.92 | 0.56 | 0.62 | 0.11 | 0.14 | 0.91 | 0.92 | 0.57 | 0.61 | 0.11 | 0.14 |
| St.Dev. | 0.52 | 0.52 | 0.25 | 0.27 | 0.08 | 0.10 | 0.46 | 0.48 | 0.23 | 0.25 | 0.07 | 0.08 |
|  | Correlations | | | | | | Correlations | | | | | |
| $\beta^-$ | 1 | 0.71 | 0.94 | 0.73 | -0.22 | 0.03 | 1 | 0.82 | 0.95 | 0.81 | -0.11 | 0.06 |
| $\beta^+$ | 0.71 | 1 | 0.71 | 0.93 | -0.04 | -0.20 | 0.82 | 1 | 0.80 | 0.94 | 0.00 | -0.09 |
| $\beta^{\mathcal{N}}$ | 0.94 | 0.71 | 1 | 0.80 | 0.11 | 0.21 | 0.95 | 0.80 | 1 | 0.87 | 0.19 | 0.26 |
| $\beta^{\mathcal{P}}$ | 0.73 | 0.93 | 0.80 | 1 | 0.15 | 0.16 | 0.81 | 0.94 | 0.87 | 1 | 0.22 | 0.23 |
| $\beta^{\mathcal{M}^-}$ | -0.22 | -0.04 | 0.11 | 0.15 | 1 | 0.52 | -0.11 | 0.00 | 0.19 | 0.22 | 1 | 0.67 |
| $\beta^{\mathcal{M}^+}$ | 0.03 | -0.20 | 0.21 | 0.16 | 0.52 | 1 | 0.06 | -0.09 | 0.26 | 0.23 | 0.67 | 1 |

Table 2 Predictive Panel Performance of Asymmetric Betas Across Forecasting Horizons.

This table reports the out-of-sample $R^2$ of the alternative prediction models across horizons ($h = 1, 3, 6, 12$ months). The forecasting performance is evaluated using the panel $R^2$ calculated using equal weighting. The benchmark is the $h$-lagged realized beta measure for the respective asymmetric beta. The conditional beta forecasts are based on principal component analysis (pca), partial least squares (pls), elastic net (en), random forests (rf), gradient boosting (gb) and feed-forward neural network (ffnn). Performance is also reported for the equally weighted forecast combinations of the three linear models (clin) and the three non-linear models (cnl). This table also reports the statistical significance of the forecasts according to the Clark and West (2007) test for equal predictive accuracy. The statistical significance of the alternative beta estimators from the benchmark is denoted by: *, **, and *** for significance at the 10%, 5%, and 1% level, respectively. The out-of-sample period is from January 1990 to December 2024.

|  | Downside Beta | | | |  | Upside Beta | | | |
|---|---|---|---|---|---|---|---|---|---|
|  | $h=1$ | $h=3$ | $h=6$ | $h=12$ |  | $h=1$ | $h=3$ | $h=6$ | $h=12$ |
| pca | 45.15*** | 35.74*** | 27.32*** | 17.59*** | pca | 43.45*** | 32.19*** | 19.84*** | 10.68*** |
| pls | 45.84*** | 37.04*** | 29.31*** | 19.89*** | pls | 44.63*** | 34.14*** | 22.29*** | 13.82*** |
| en | 45.89*** | 37.19*** | 29.49*** | 20.62*** | en | 44.65*** | 34.11*** | 22.83*** | 14.81*** |
| clin | 45.87*** | 37.21*** | 29.50*** | 20.63*** | clin | 44.62*** | 34.20*** | 22.67*** | 14.63*** |
| gb | 45.82*** | 37.56*** | 30.13*** | 21.02*** | gb | 44.75*** | 35.53*** | 23.51*** | 14.75*** |
| rf | 45.96*** | 38.85*** | 32.66*** | 24.51*** | rf | 45.00*** | 36.43*** | 25.95*** | 18.65*** |
| ffnn | 46.00*** | 37.29*** | 29.38*** | 19.58*** | ffnn | 44.88*** | 34.32*** | 22.31*** | 13.42*** |
| cnl | 46.35*** | 39.06*** | 32.38*** | 24.15*** | cnl | 45.39*** | 36.68*** | 25.71*** | 18.35*** |
|  | Semibeta - Negative | | | |  | Semibeta - Positive | | | |
|  | $h=1$ | $h=3$ | $h=6$ | $h=12$ |  | $h=1$ | $h=3$ | $h=6$ | $h=12$ |
| pca | 45.10*** | 35.07*** | 25.57*** | 17.70*** | pca | 42.24*** | 37.17*** | 25.09*** | 23.22*** |
| pls | 46.12*** | 36.41*** | 27.14*** | 19.00*** | pls | 43.29*** | 38.45*** | 26.51*** | 24.32*** |
| en | 46.18*** | 36.72*** | 27.55*** | 20.01*** | en | 43.07*** | 38.51*** | 26.07*** | 24.65*** |
| clin | 46.14*** | 36.66*** | 27.55*** | 20.11*** | clin | 43.18*** | 38.54*** | 26.66*** | 25.26*** |
| gb | 43.83*** | 33.94*** | 27.90*** | 22.68*** | gb | 39.12*** | 31.08*** | 19.25*** | 16.16*** |
| rf | 45.67*** | 37.01*** | 30.41*** | 25.09*** | rf | 41.29*** | 36.20*** | 24.11*** | 24.49*** |
| ffnn | 46.29*** | 36.63*** | 27.05*** | 18.79*** | ffnn | 43.47*** | 38.16*** | 23.41*** | 18.02*** |
| cnl | 46.05*** | 37.79*** | 30.79*** | 25.28*** | cnl | 42.55*** | 37.05*** | 24.79*** | 23.34*** |
|  | Semibeta - Mixed Sign, Negative Market Return | | | |  | Semibeta - Mixed Sign, Positive Market Return | | | |
|  | $h=1$ | $h=3$ | $h=6$ | $h=12$ |  | $h=1$ | $h=3$ | $h=6$ | $h=12$ |
| pca | 42.86*** | 26.01*** | 9.79*** | -15.08*** | pca | 36.26*** | 20.69*** | -0.43*** | -18.55*** |
| pls | 44.15*** | 29.09*** | 13.84*** | -9.79*** | pls | 37.54*** | 23.75*** | 4.07*** | -13.09*** |
| en | 44.21*** | 29.40*** | 14.06*** | -9.84*** | en | 37.58*** | 23.64*** | 3.45*** | -14.34*** |
| clin | 44.11*** | 28.99*** | 13.96*** | -9.75*** | clin | 37.48*** | 23.59*** | 3.78*** | -13.64*** |
| gb | 42.18*** | 31.98*** | 17.85*** | -5.26*** | gb | 38.02*** | 23.58*** | 8.92*** | -8.46*** |
| rf | 44.84*** | 32.87*** | 18.25*** | -3.69*** | rf | 38.96*** | 28.31*** | 8.50*** | -5.21*** |
| ffnn | 44.64*** | 30.21*** | 19.07*** | -6.90*** | ffnn | 38.05*** | 26.31*** | 7.52*** | -10.30*** |
| cnl | 44.86*** | 33.32*** | 20.48*** | -1.33*** | cnl | 39.65*** | 28.49*** | 11.29*** | -3.59*** |

Table 3 Predictive Panel Performance of Reconstructed CAPM Betas Across Forecasting Horizons.

This table reports the out-of-sample $R^2$ of the alternative prediction models across horizons ($h = 1, 3, 6, 12$ months). The forecasting performance is evaluated using the panel $R^2$ calculated using equal weighting. The benchmark is the $h$-lagged realized CAPM beta. The conditional beta forecasts are based on principal component analysis (pca), partial least squares (pls), elastic net (en), random forests (rf), gradient boosting (gb) and feed-forward neural network (ffnn). Performance is also reported for the equally weighted forecast combinations of the three linear models (clin) and the three non-linear models (cnl). Estimates of CAPM beta are derived as a combination of forecasts of downside and upside betas, and as a combination of forecasts of four semibetas. This table also reports the statistical significance of the forecasts according to the Clark and West (2007) test for equal predictive accuracy. The statistical significance of the alternative beta estimators from the benchmark is denoted by: *, **, and *** for significance at the 10%, 5%, and 1% level, respectively. The out-of-sample period is from January 1990 to December 2024.

| | CAPM Beta (Downside and Upside Betas) | | | |
|---|---|---|---|---|
| | $h = 1$ | $h = 3$ | $h = 6$ | $h = 12$ |
| pca | 35.13*** | 20.99*** | 9.30*** | 3.13*** |
| pls | 37.22*** | 23.99*** | 12.85*** | 6.66*** |
| en | 37.32*** | 24.07*** | 13.21*** | 7.81*** |
| clin | 37.19*** | 23.98*** | 13.01*** | 7.52*** |
| gb | 37.81*** | 26.44*** | 15.43*** | 9.96*** |
| rf | 38.04*** | 27.80*** | 18.33*** | 13.03*** |
| ffnn | 37.70*** | 24.56*** | 13.27*** | 6.73*** |
| cnl | 38.73*** | 28.01*** | 17.82*** | 12.67*** |
| | CAPM Beta (Semibetas) | | | |
| | $h = 1$ | $h = 3$ | $h = 6$ | $h = 12$ |
| pca | 35.36*** | 20.89*** | 9.71*** | 3.31*** |
| pls | 37.63*** | 24.14*** | 13.04*** | 6.61*** |
| en | 37.56*** | 24.63*** | 13.29*** | 8.08*** |
| clin | 37.47*** | 24.17*** | 13.35*** | 7.81*** |
| gb | 34.56*** | 23.55*** | 16.32*** | 9.84*** |
| rf | 36.89*** | 26.36*** | 16.99*** | 12.50*** |
| ffnn | 37.78*** | 24.33*** | 12.01*** | 4.14*** |
| cnl | 37.67*** | 26.95*** | 17.78*** | 12.45*** |

Table 4 Share Price Prediction for Forecast Combinations.

This table reports the out-of-sample panel $R^2$ for observed prices compared to intrinsic values derived from discounted cash flow models with alternative discount rates. Intrinsic values are calculated using a one-year holding period with monthly cash flows based on 12-month rolling averages of ordinary dividends. Terminal values assume annual long-term growth rates from 0% to 2%. Conditional betas are converted to discount rates using a risk-free rate of zero and market risk premiums from 8% to 12%. Panels A-D report the results for horizon-specific conditional betas ($h = 1,3,6,12$), while Panel E for conditional betas across horizons. The benchmark is the respective $h$-lagged realized CAPM beta. The conditional beta forecasts are based on equally weighted forecast combinations of the three linear models (clin) and the three non-linear models (cnl). Estimates of CAPM beta are derived as a combination of forecasts of downside and upside betas, and as a combination of forecasts of four semibetas. The out-of-sample period is from January 1990 to December 2024.

| Growth Rate | 0% | | | 1% | | | 2% | | |
|---|---|---|---|---|---|---|---|---|---|
| Market Premium | 8% | 10% | 12% | 8% | 10% | 12% | 8% | 10% | 12% |
| A. 1-month horizon betas | | | | | | | | | |
| clin (Downside and Upside Betas) | 20.53 | 11.76 | 6.74 | 31.75 | 19.93 | 11.45 | 20.80 | 23.97 | 15.34 |
| cnl (Downside and Upside Betas) | 20.78 | 11.74 | 6.58 | 34.58 | 20.88 | 11.80 | 18.11 | 26.91 | 17.20 |
| clin (Semibetas) | 20.98 | 12.05 | 6.94 | 33.68 | 20.78 | 12.01 | 25.32 | 26.20 | 16.49 |
| cnl (Semibetas) | 19.98 | 10.86 | 5.72 | 34.79 | 20.29 | 10.97 | 37.82 | 28.22 | 17.30 |
| B. 3-month horizon betas | | | | | | | | | |
| clin (Downside and Upside Betas) | 14.33 | 7.46 | 3.78 | 32.89 | 17.73 | 10.18 | 32.63 | 27.37 | 16.20 |
| cnl (Downside and Upside Betas) | 14.46 | 7.44 | 3.69 | 34.55 | 18.42 | 10.42 | 25.55 | 29.34 | 17.30 |
| clin (Semibetas) | 14.46 | 7.54 | 3.83 | 33.68 | 18.15 | 10.39 | 32.60 | 27.96 | 16.69 |
| cnl (Semibetas) | 13.83 | 6.79 | 3.07 | 34.46 | 17.87 | 9.77 | 33.53 | 30.22 | 16.97 |
| C. 6-month horizon betas | | | | | | | | | |
| clin (Downside and Upside Betas) | 6.97 | 2.89 | 0.89 | 27.77 | 14.02 | 7.08 | 38.44 | 33.41 | 15.74 |
| cnl (Downside and Upside Betas) | 6.78 | 2.63 | 0.61 | 28.67 | 14.22 | 6.98 | 34.40 | 33.01 | 15.95 |
| clin (Semibetas) | 7.03 | 2.93 | 0.92 | 27.97 | 14.17 | 7.15 | 37.79 | 33.21 | 15.93 |
| cnl (Semibetas) | 6.54 | 2.36 | 0.36 | 28.73 | 14.00 | 6.70 | 36.21 | 34.98 | 16.41 |
| D. 12-month horizon betas | | | | | | | | | |
| clin (Downside and Upside Betas) | 1.98 | 0.02 | -0.82 | 23.73 | 11.28 | 5.06 | 40.02 | 29.83 | 15.94 |
| cnl (Downside and Upside Betas) | 1.84 | -0.19 | -1.04 | 24.49 | 11.39 | 4.96 | 40.48 | 30.29 | 16.31 |
| clin (Semibetas) | 2.03 | 0.04 | -0.82 | 23.96 | 11.35 | 5.09 | 39.30 | 29.78 | 16.03 |
| cnl (Semibetas) | 1.64 | -0.39 | -1.24 | 24.53 | 11.23 | 4.76 | 41.93 | 30.83 | 16.34 |
| E. Betas across horizons | | | | | | | | | |
| clin (Downside and Upside Betas) | 5.46 | 2.37 | 0.87 | 35.33 | 18.50 | 10.18 | 60.66 | 48.26 | 28.06 |
| cnl (Downside and Upside Betas) | 5.32 | 2.16 | 0.65 | 35.98 | 18.60 | 10.08 | 62.03 | 48.53 | 28.41 |
| clin (Semibetas) | 5.50 | 2.39 | 0.88 | 35.53 | 18.57 | 10.21 | 60.41 | 48.33 | 28.15 |
| cnl (Semibetas) | 5.13 | 1.96 | 0.46 | 36.01 | 18.46 | 9.89 | 63.26 | 49.21 | 28.43 |

Figure 1 Unconditional Distributions of Asymmetric Betas for Different Estimation Windows.
This figure displays the kernel density estimates of the unconditional distribution of the monthly realized asymmetric betas across four different estimation windows ($h = 1, 3, 6, 12$ months) constructed using daily returns. The sample period is from January 1980 to December 2024.

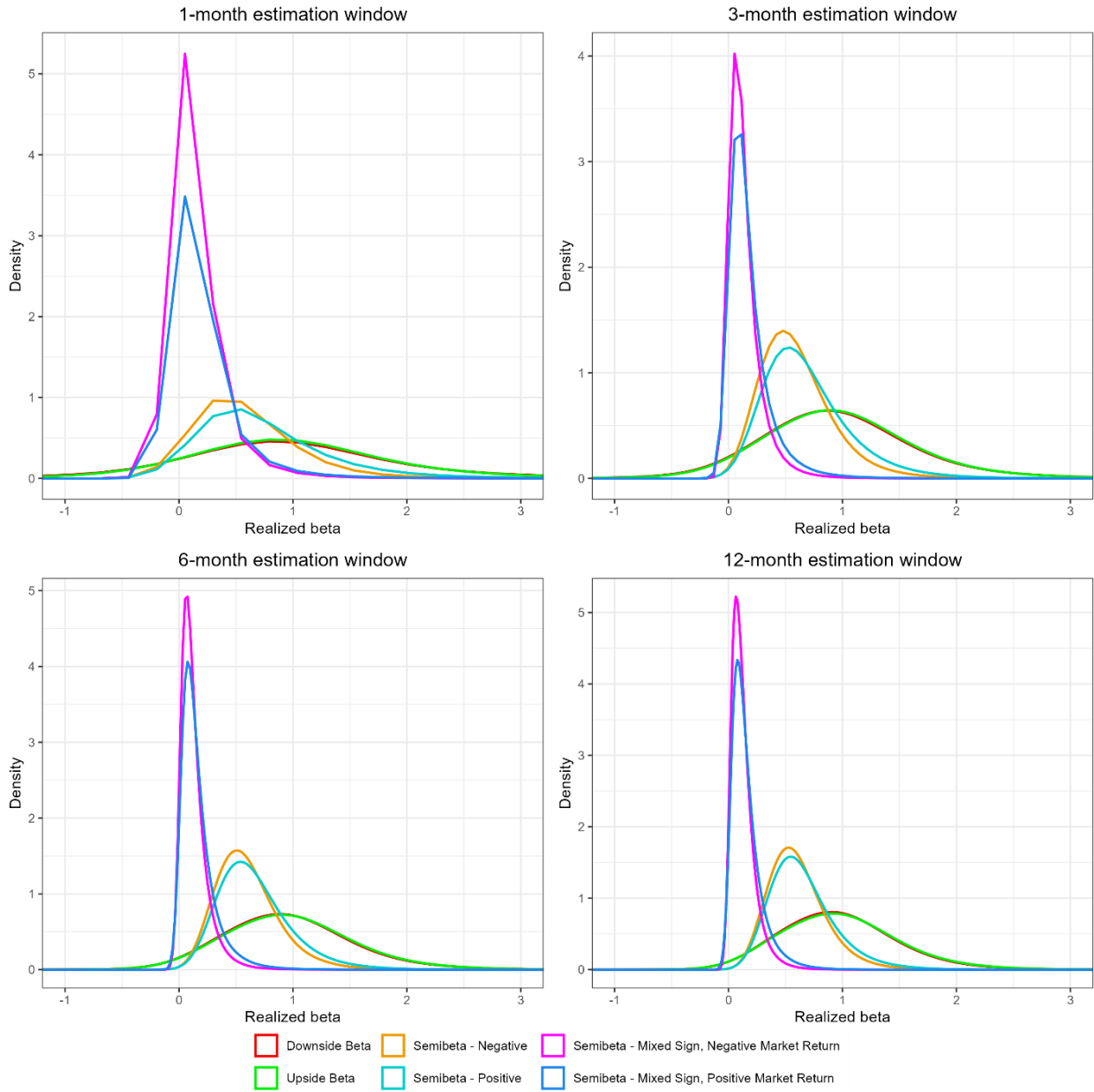

Figure 2 Cumulative Difference of Forecast Errors Over Time: 1-Month Horizon.
This figure displays the cumulative differences in squared forecast errors for the benchmark relative to the conditional beta forecasting models for $h = 1$. The benchmark is the $h$-lagged realized beta measure for the respective asymmetric beta. The conditional beta forecasts are based on principal component analysis (pca), partial least squares (pls), elastic net (en), random forests (rf), gradient boosting (gb) and feed-forward neural network (ffnn). Performance is also reported for the equally weighted forecast combinations of the three linear models (clin) and the three non-linear models (cnl). The shaded regions depict NBER-dated recessions. Higher values indicate improved predictive performance from the benchmark. The out-of-sample period is from January 1990 to December 2024.

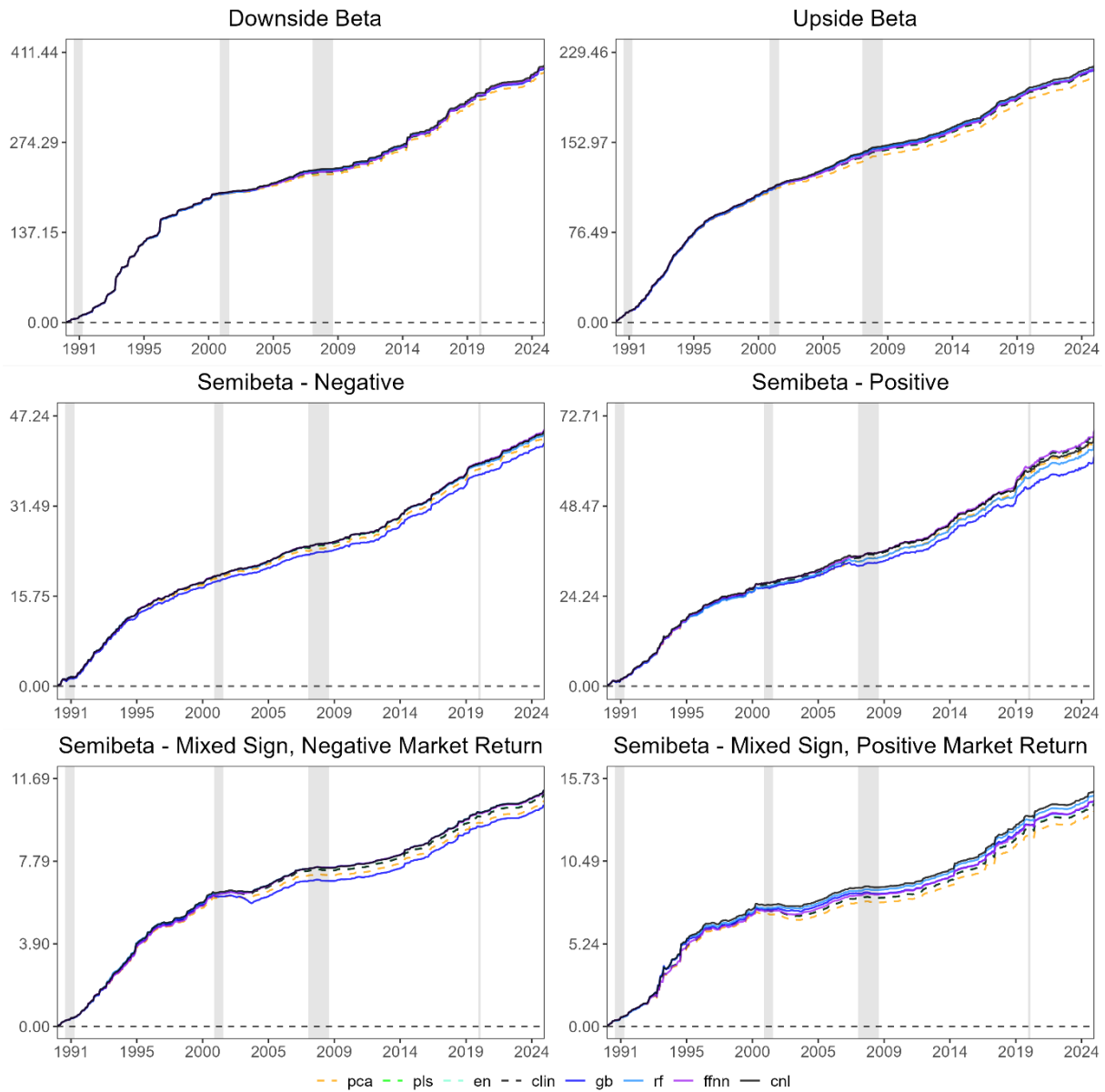

Figure 3 Cumulative Difference of Forecast Errors Over Time: 3-Month Horizon.
This figure displays the cumulative differences in squared forecast errors for the benchmark relative to the conditional beta forecasting models for $h = 3$. The benchmark is the $h$-lagged realized beta measure for the respective asymmetric beta. The conditional beta forecasts are based on principal component analysis (pca), partial least squares (pls), elastic net (en), random forests (rf), gradient boosting (gb) and feed-forward neural network (ffnn). Performance is also reported for the equally weighted forecast combinations of the three linear models (clin) and the three non-linear models (cnl). The shaded regions depict NBER-dated recessions. Higher values indicate improved predictive performance from the benchmark. The out-of-sample period is from January 1990 to December 2024.

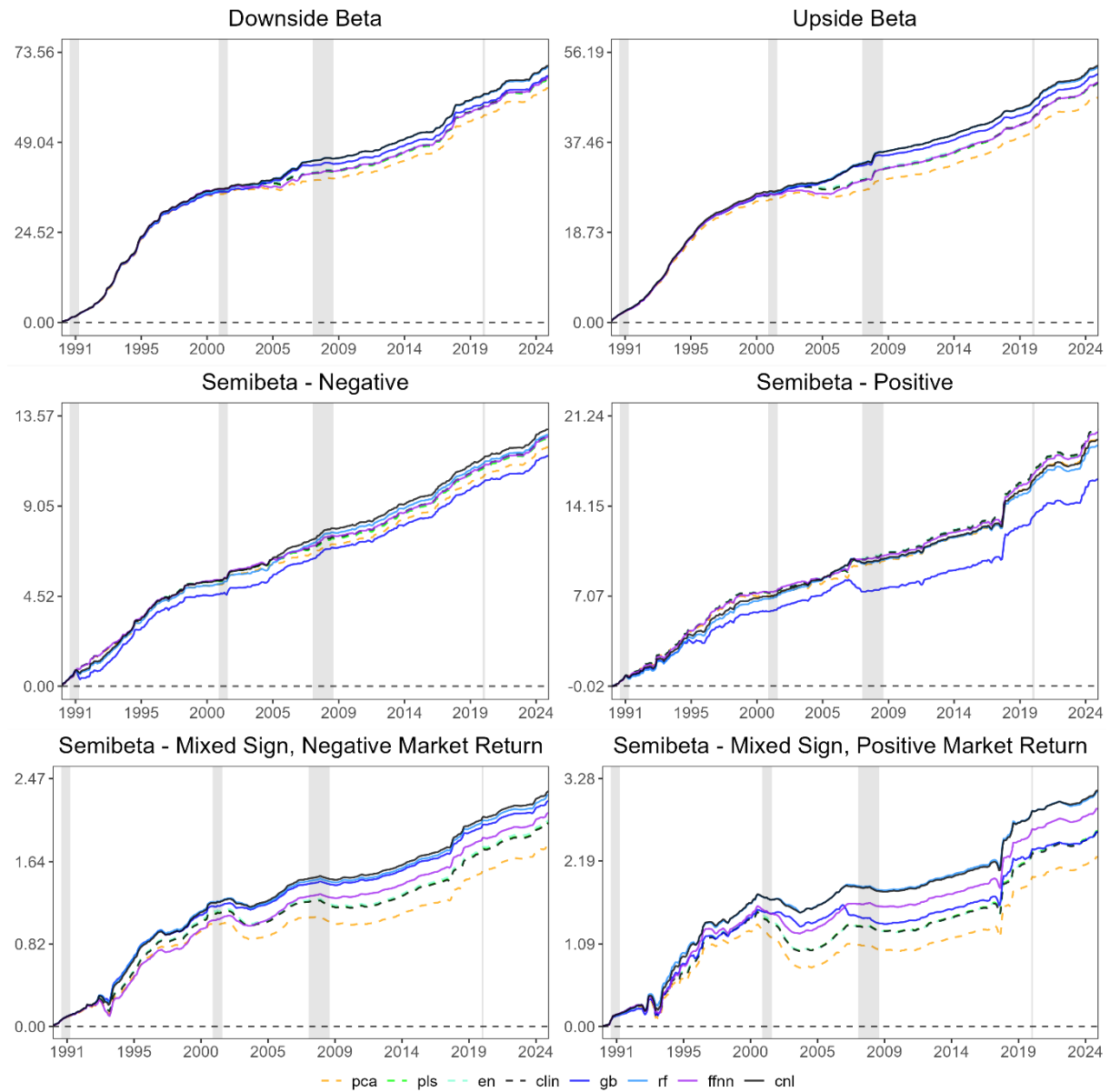

Figure 4 Cumulative Difference of Forecast Errors Over Time: 6-Month Horizon.
This figure displays the cumulative differences in squared forecast errors for the benchmark relative to the conditional beta forecasting models for $h = 6$. The benchmark is the $h$-lagged realized beta measure for the respective asymmetric beta. The conditional beta forecasts are based on principal component analysis (pca), partial least squares (pls), elastic net (en), random forests (rf), gradient boosting (gb) and feed-forward neural network (ffnn). Performance is also reported for the equally weighted forecast combinations of the three linear models (clin) and the three non-linear models (cnl). The shaded regions depict NBER-dated recessions. Higher values indicate improved predictive performance from the benchmark. The out-of-sample period is from January 1990 to December 2024.

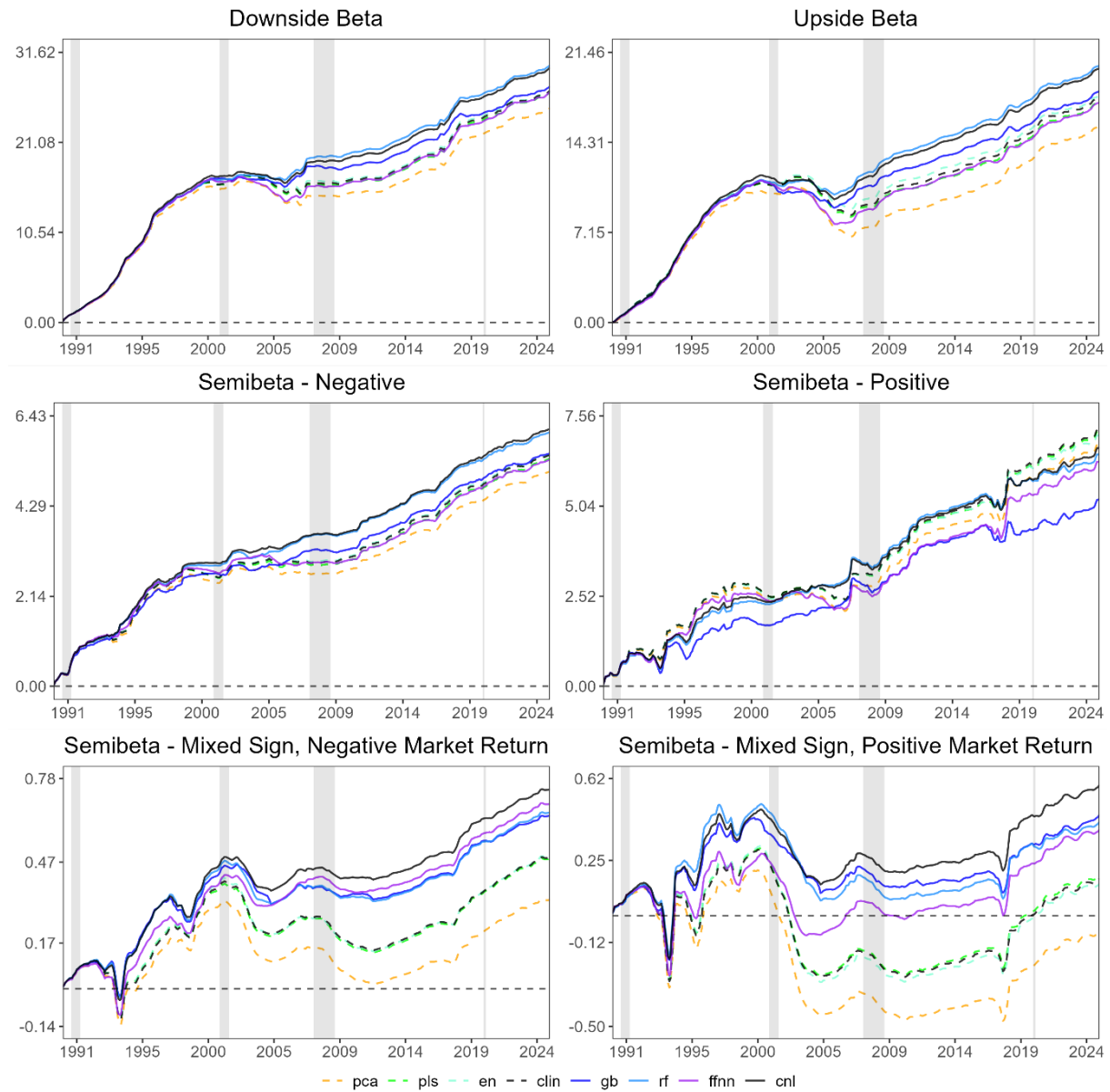

Figure 5 Cumulative Difference of Forecast Errors Over Time: 12-Month Horizon.
This figure displays the cumulative differences in squared forecast errors for the benchmark relative to the conditional beta forecasting models for $h = 12$. The benchmark is the $h$-lagged realized beta measure for the respective asymmetric beta. The conditional beta forecasts are based on principal component analysis (pca), partial least squares (pls), elastic net (en), random forests (rf), gradient boosting (gb) and feed-forward neural network (ffnn). Performance is also reported for the equally weighted forecast combinations of the three linear models (clin) and the three non-linear models (cnl). The shaded regions depict NBER-dated recessions. Higher values indicate improved predictive performance from the benchmark. The out-of-sample period is from January 1990 to December 2024.

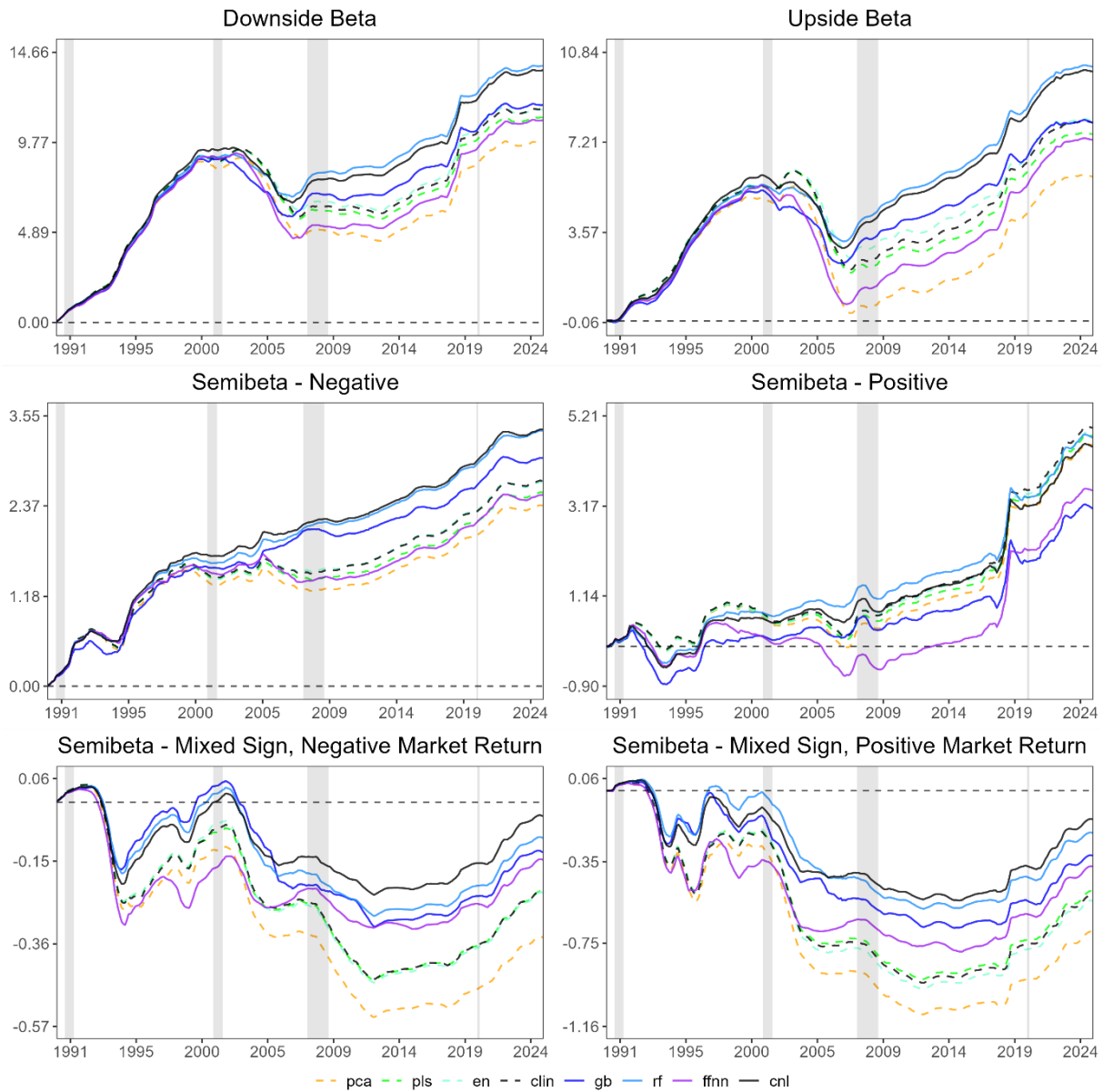

Figure 6 Average Predictive Performance of Quintile Portfolios for Forecast Combinations: 1-Month Horizon.
This figure displays the time series averages of monthly mean squared errors (MSE) for quintile portfolios for $h = 1$. Portfolios are formed monthly by sorting stocks into quintiles according to realised beta values (1-low beta, 5-high beta). The equally weighted MSE is constructed for each quintile, for the benchmark (red bars) and the conditional beta forecasts (blue bars). The figure also reports the fraction of stocks within each portfolio for which the difference between realized and forecast portfolio betas is positive (dots, right-hand axis). The benchmark is the $h$-lagged realized beta measure for the respective asymmetric beta. The conditional beta forecasts are based on the equally weighted forecast combinations of the linear models (clin) and the non-linear models (cnl). The out-of-sample period is from January 1990 to December 2024.

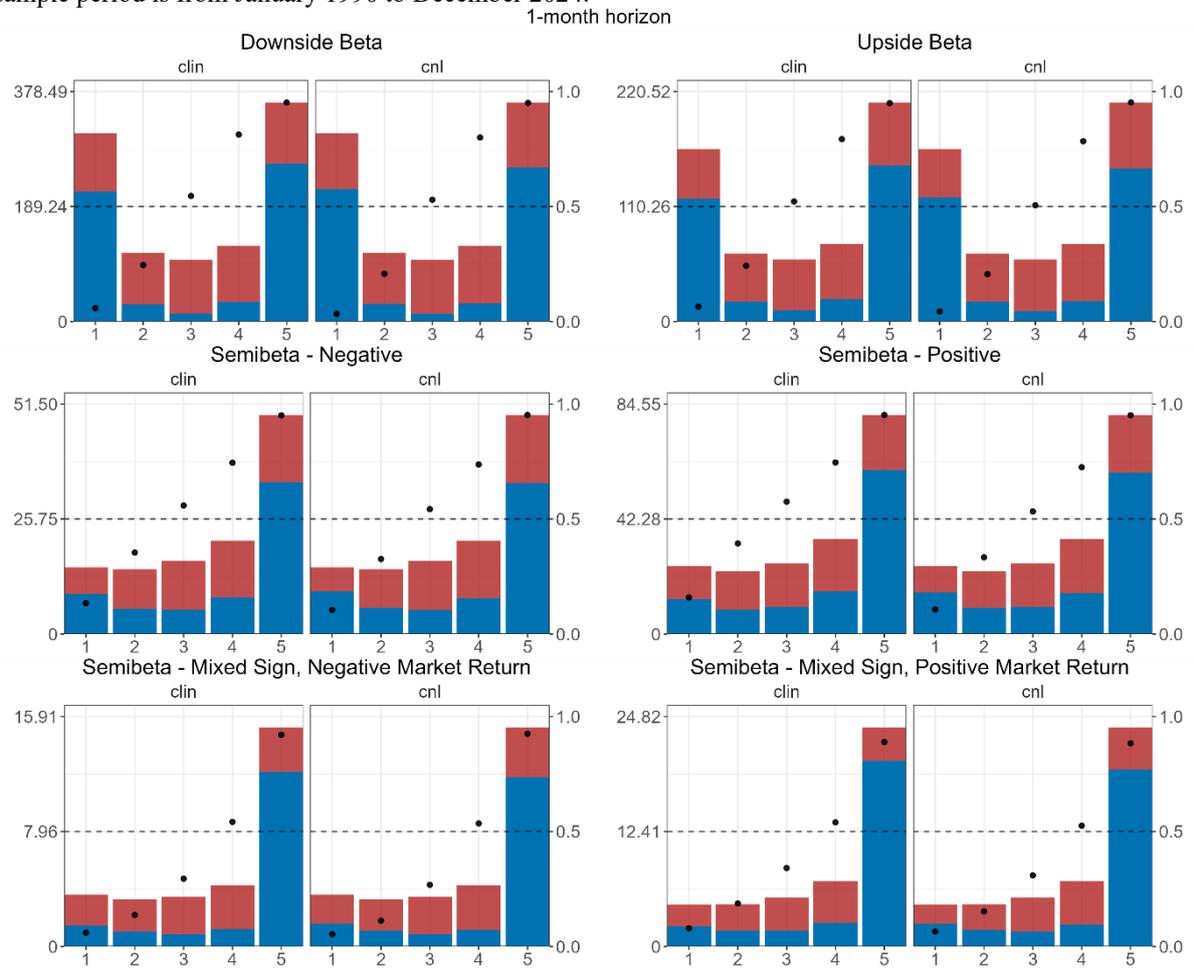

Figure 7 Average Predictive Performance of Quintile Portfolios for Forecast Combinations: 3-Month Horizon. This figure displays the time series averages of monthly mean squared errors (MSE) for quintile portfolios for $h = 3$. Portfolios are formed monthly by sorting stocks into quintiles according to realised beta values (1-low beta, 5-high beta). The equally weighted MSE is constructed for each quintile, for the benchmark (red bars) and the conditional beta forecasts (blue bars). The figure also reports the fraction of stocks within each portfolio for which the difference between realized and forecast portfolio betas is positive (dots, right-hand axis). The benchmark is the $h$-lagged realized beta measure for the respective asymmetric beta. The conditional beta forecasts are based on the equally weighted forecast combinations of the linear models (clin) and the non-linear models (cnl). The out-of-sample period is from January 1990 to December 2024.

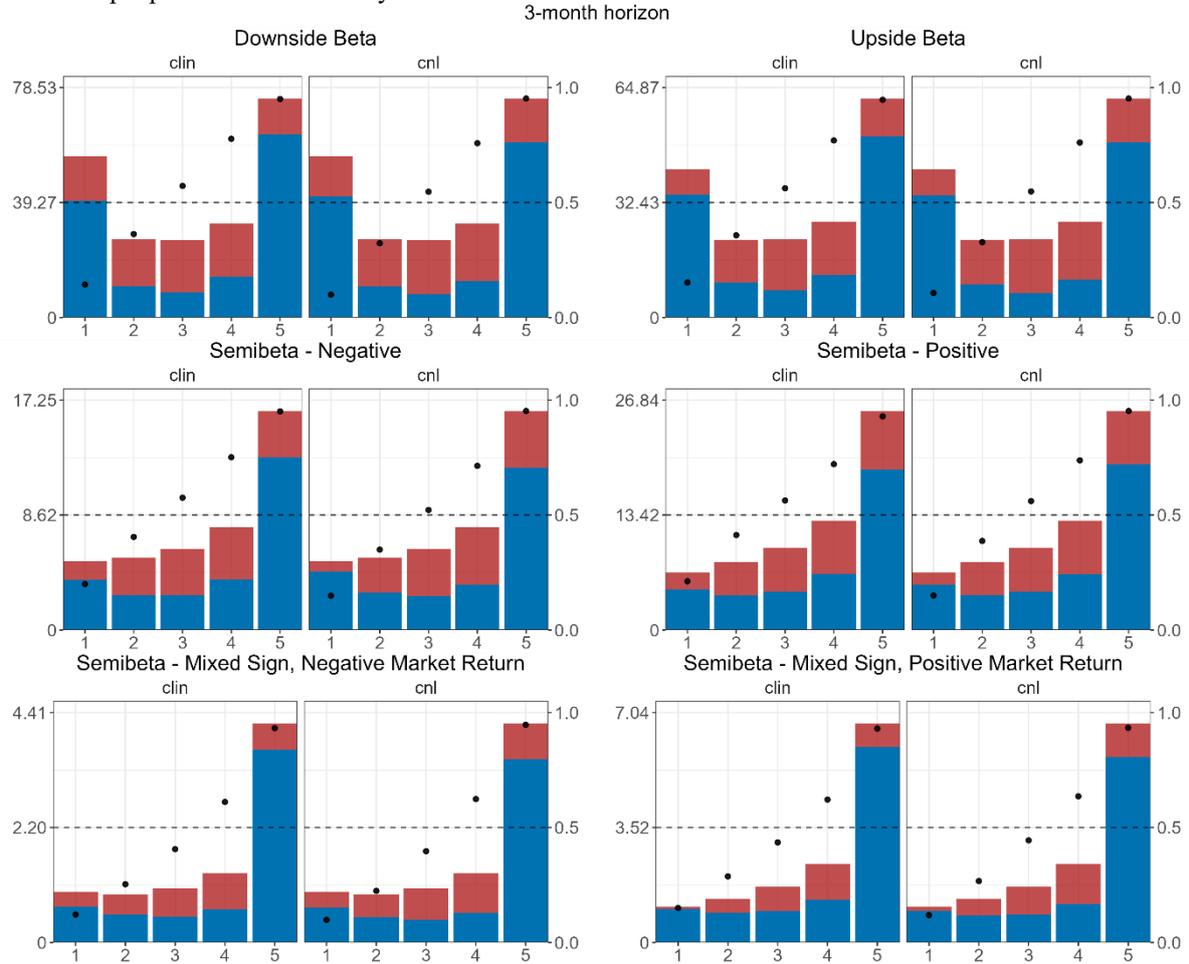

Figure 8 Average Predictive Performance of Quintile Portfolios for Forecast Combinations: 6-Month Horizon. This figure displays the time series averages of monthly mean squared errors (MSE) for quintile portfolios for $h = 6$. Portfolios are formed monthly by sorting stocks into quintiles according to realised beta values (1-low beta, 5-high beta). The equally weighted MSE is constructed for each quintile, for the benchmark (red bars) and the conditional beta forecasts (blue bars). The figure also reports the fraction of stocks within each portfolio for which the difference between realized and forecast portfolio betas is positive (dots, right-hand axis). The benchmark is the $h$-lagged realized beta measure for the respective asymmetric beta. The conditional beta forecasts are based on the equally weighted forecast combinations of the linear models (clin) and the non-linear models (cnl). The out-of-sample period is from January 1990 to December 2024.

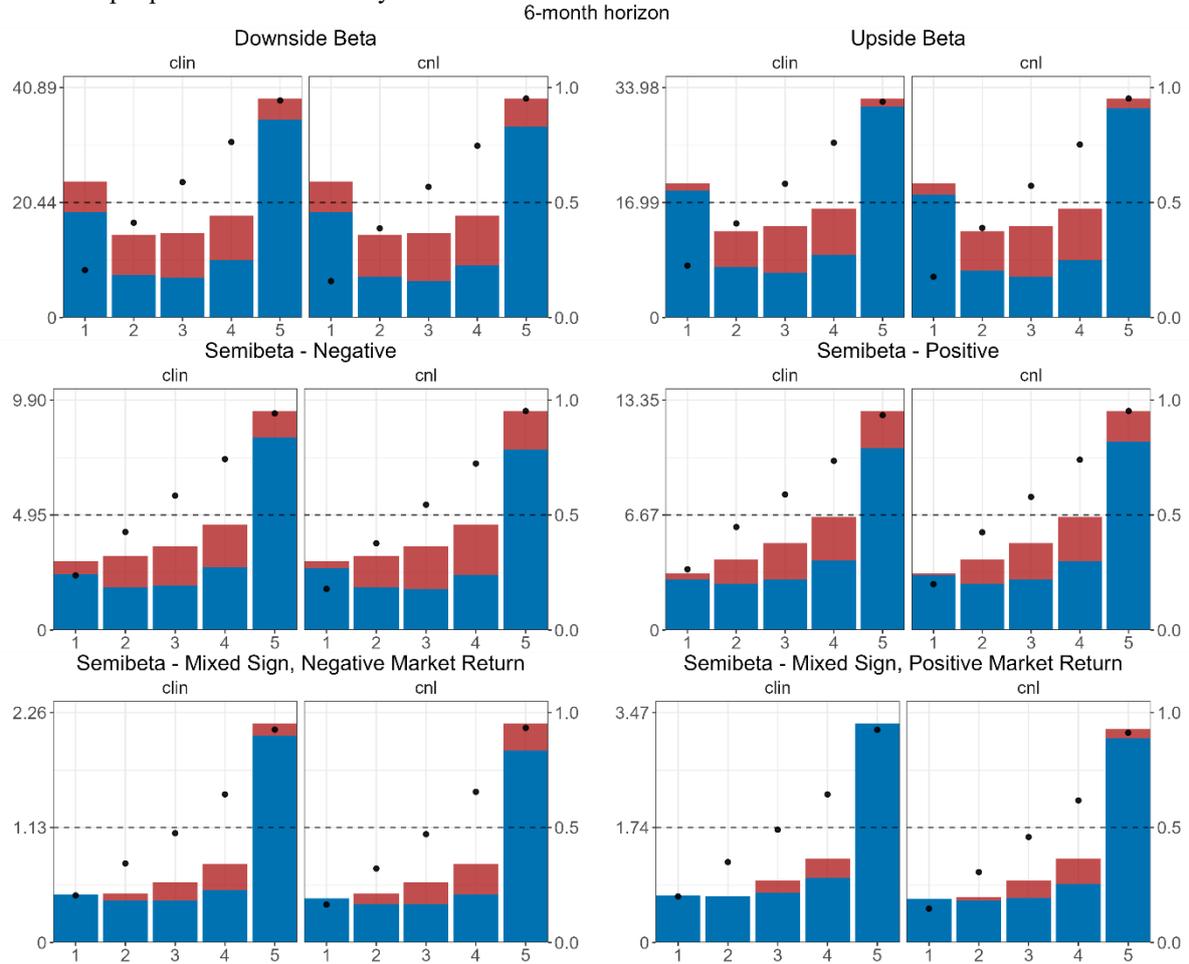

Figure 9 Average Predictive Performance of Quintile Portfolios for Forecast Combinations: 12-Month Horizon. This figure displays the time series averages of monthly mean squared errors (MSE) for quintile portfolios for $h = 12$. Portfolios are formed monthly by sorting stocks into quintiles according to realised beta values (1-low beta, 5-high beta). The equally weighted MSE is constructed for each quintile, for the benchmark (red bars) and the conditional beta forecasts (blue bars). The figure also reports the fraction of stocks within each portfolio for which the difference between realized and forecast portfolio betas is positive (dots, right-hand axis). The benchmark is the $h$-lagged realized beta measure for the respective asymmetric beta. The conditional beta forecasts are based on the equally weighted forecast combinations of the linear models (clin) and the non-linear models (cnl). The out-of-sample period is from January 1990 to December 2024.

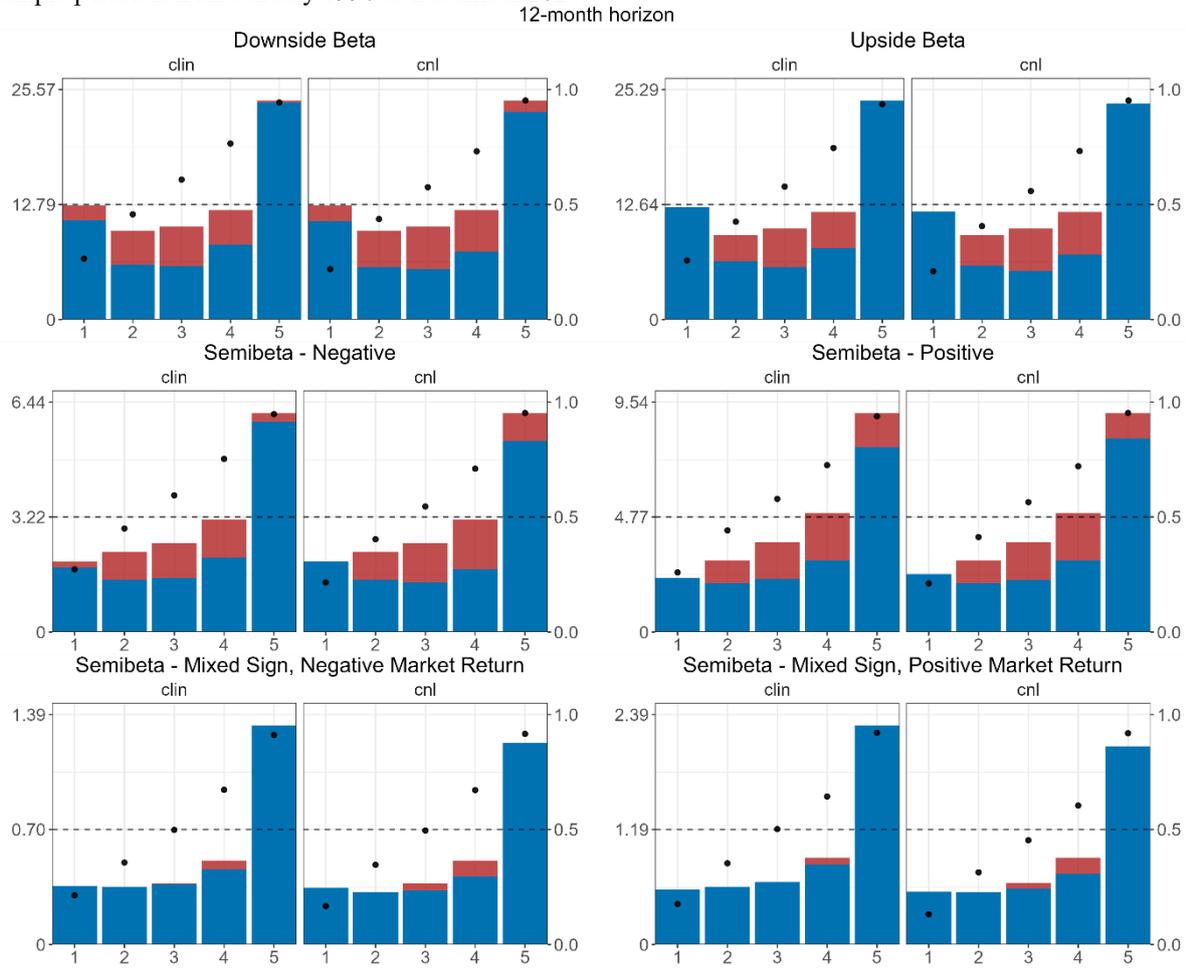

Figure 10 Average Variable Importance: 1-Month Horizon.
This figure displays the variable importance for each of the six groups of predictors. The variable importance in each period is computed according to the permutation feature importance. Variable importance is averaged throughout the out-of-sample period and is normalized for each model to sum to 100 (darker colours indicate greater variable importance). The conditional beta forecasts are based on principal component analysis (pca), partial least squares (pls), elastic net (en), random forests (rf), gradient boosting (gb) and feed-forward neural network (ffnn). The out-of-sample period is from January 1990 to December 2024.

1-month horizon

| | Downside Beta | | | | | | Semibeta - Negative | | | | | | Semibeta - Mixed Sign, Negative Market Return | | | | | |
|---|---|---|---|---|---|---|---|---|---|---|---|---|---|---|---|---|---|---|
| | pca | pls | en | gb | rf | ffnn | pca | pls | en | gb | rf | ffnn | pca | pls | en | gb | rf | ffnn |
| Value vs Growth | 1.55 | 2.96 | 2.52 | 2.84 | 11.22 | 5.98 | 1.41 | 3.51 | 3.80 | 3.41 | 11.90 | 6.62 | 1.15 | 2.16 | 3.27 | 3.53 | 8.21 | 5.65 |
| Trading Frictions | 94.60 | 89.28 | 92.54 | 77.87 | 47.12 | 75.33 | 93.92 | 85.70 | 88.78 | 74.12 | 43.02 | 71.22 | 93.80 | 87.55 | 89.04 | 55.25 | 39.92 | 71.47 |
| Profitability | 0.18 | 0.59 | 0.34 | 1.57 | 4.47 | 2.60 | 0.21 | 0.71 | 0.44 | 1.92 | 4.79 | 2.74 | 0.18 | 0.55 | 0.42 | 3.81 | 5.75 | 3.19 |
| Momentum | 2.08 | 4.45 | 3.15 | 5.34 | 10.66 | 6.93 | 2.71 | 6.47 | 4.78 | 7.20 | 12.43 | 8.65 | 2.85 | 6.77 | 4.87 | 7.61 | 11.21 | 9.35 |
| Investment | 0.11 | 0.82 | 0.45 | 2.25 | 8.54 | 3.52 | 0.21 | 1.04 | 0.69 | 2.42 | 9.20 | 3.97 | 0.09 | 0.87 | 0.73 | 1.64 | 7.06 | 4.24 |
| Intangibles | 1.48 | 1.91 | 1.01 | 10.14 | 18.00 | 5.63 | 1.55 | 2.56 | 1.50 | 10.94 | 18.66 | 6.81 | 1.92 | 2.10 | 1.67 | 28.16 | 27.84 | 6.10 |

| | Upside Beta | | | | | | Semibeta - Positive | | | | | | Semibeta - Mixed Sign, Positive Market Return | | | | | |
|---|---|---|---|---|---|---|---|---|---|---|---|---|---|---|---|---|---|---|
| | pca | pls | en | gb | rf | ffnn | pca | pls | en | gb | rf | ffnn | pca | pls | en | gb | rf | ffnn |
| Value vs Growth | 2.04 | 3.35 | 3.54 | 4.08 | 11.54 | 7.32 | 1.27 | 1.59 | 1.78 | 3.07 | 9.58 | 4.78 | 5.26 | 5.70 | 9.60 | 5.73 | 11.40 | 9.20 |
| Trading Frictions | 94.97 | 87.11 | 91.03 | 72.93 | 43.94 | 67.31 | 94.68 | 88.98 | 91.65 | 60.04 | 38.84 | 74.41 | 88.62 | 82.57 | 80.02 | 42.73 | 30.91 | 57.83 |
| Profitability | 0.11 | 0.59 | 0.39 | 2.59 | 5.06 | 3.33 | 0.06 | 0.48 | 0.37 | 4.72 | 6.15 | 2.49 | -0.57 | 0.64 | 0.62 | 3.11 | 5.72 | 5.32 |
| Momentum | 1.69 | 5.89 | 3.46 | 6.74 | 11.93 | 9.82 | 2.27 | 6.64 | 4.33 | 9.67 | 13.20 | 9.69 | 2.47 | 5.36 | 4.27 | 7.81 | 12.14 | 10.39 |
| Investment | -0.08 | 0.97 | 0.59 | 2.33 | 8.69 | 5.03 | -0.01 | 0.73 | 0.65 | 1.19 | 7.28 | 3.58 | 0.78 | 1.74 | 1.90 | 2.83 | 9.81 | 7.14 |
| Intangibles | 1.26 | 2.08 | 0.99 | 11.33 | 18.84 | 7.19 | 1.72 | 1.58 | 1.22 | 21.31 | 24.95 | 5.06 | 3.44 | 3.99 | 3.59 | 37.79 | 30.03 | 10.13 |

Figure 11 Cumulative Difference of Forecast Errors Over Time of CAPM Beta for Forecast Combinations.
This figure displays the cumulative differences in squared forecast errors of CAPM beta for the benchmark relative to the conditional beta forecasting models across horizons. The benchmark is the $h$-lagged realized CAPM beta. The conditional beta forecasts are based on equally weighted forecast combinations of the linear models (clin) and the non-linear models (cnl). Estimates of CAPM beta are derived as a combination of forecasts of downside and upside betas (green line), and as a combination of forecasts of four semibetas (blue line). The shaded regions depict NBER-dated recessions. Higher values indicate improved predictive performance from the benchmark. The out-of-sample period is from January 1990 to December 2024.

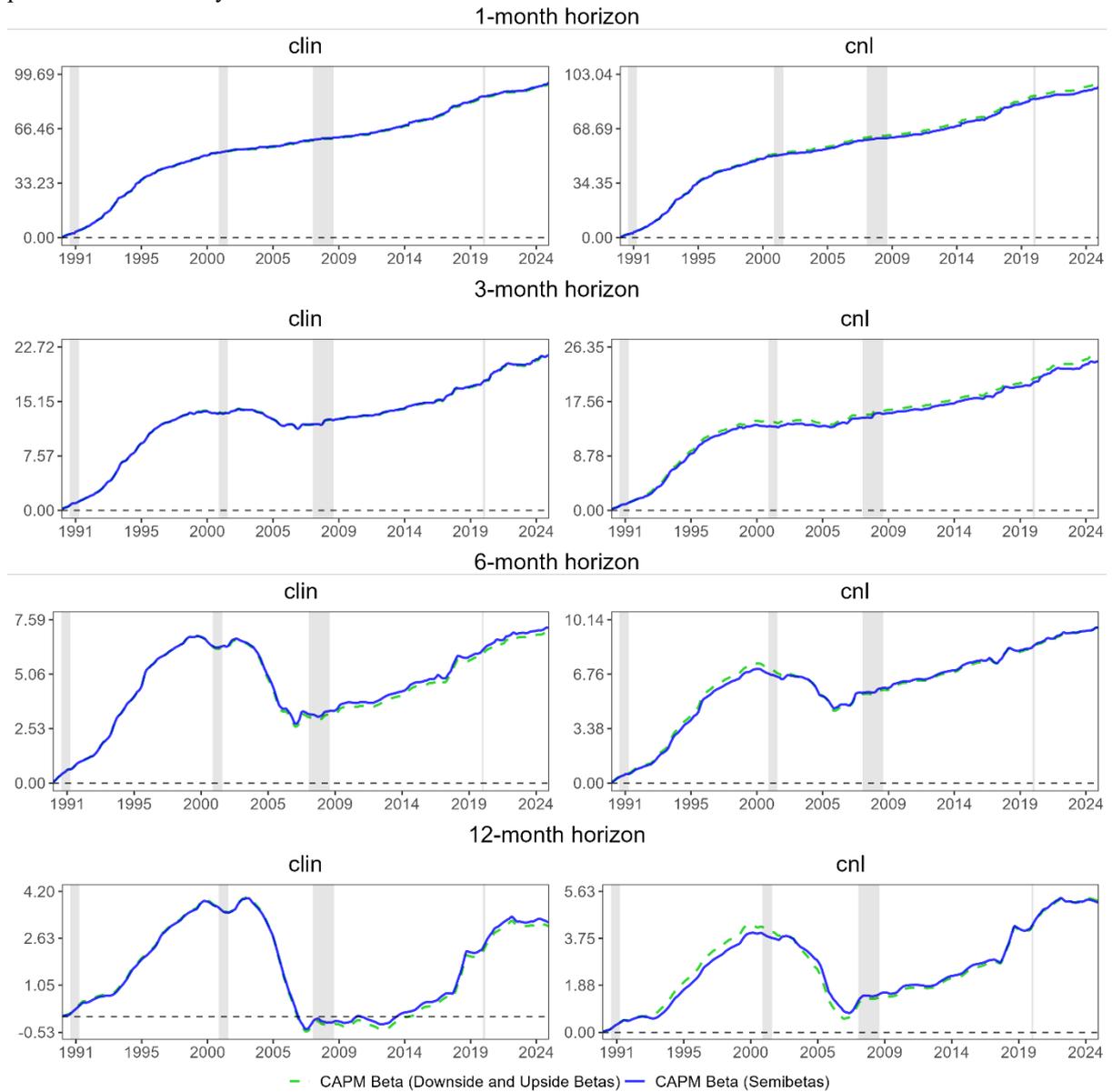

Figure 12 Average Predictive Performance of Quintile Portfolios of CAPM Beta for Forecast Combinations.
This figure displays the time series averages of monthly mean squared errors (MSE) for quintile portfolios across horizons. Portfolios are formed monthly by sorting stocks into quintiles according to realised beta values (1-low beta, 5-high beta). The equally weighted MSE is constructed for each quintile, for the benchmark (red bars) and the conditional beta forecasts (blue bars). The figure also reports the fraction of stocks within each portfolio for which the difference between realized and forecast portfolio betas is positive (dots, right-hand axis). The benchmark is the $h$-lagged realized CAPM beta. The conditional beta forecasts are based on the equally weighted forecast combinations of the linear models (clin) and the non-linear models (cnl). Estimates of CAPM beta are derived as a combination of forecasts of downside and upside betas, and as a combination of forecasts of four semibetas. The out-of-sample period is from January 1990 to December 2024.

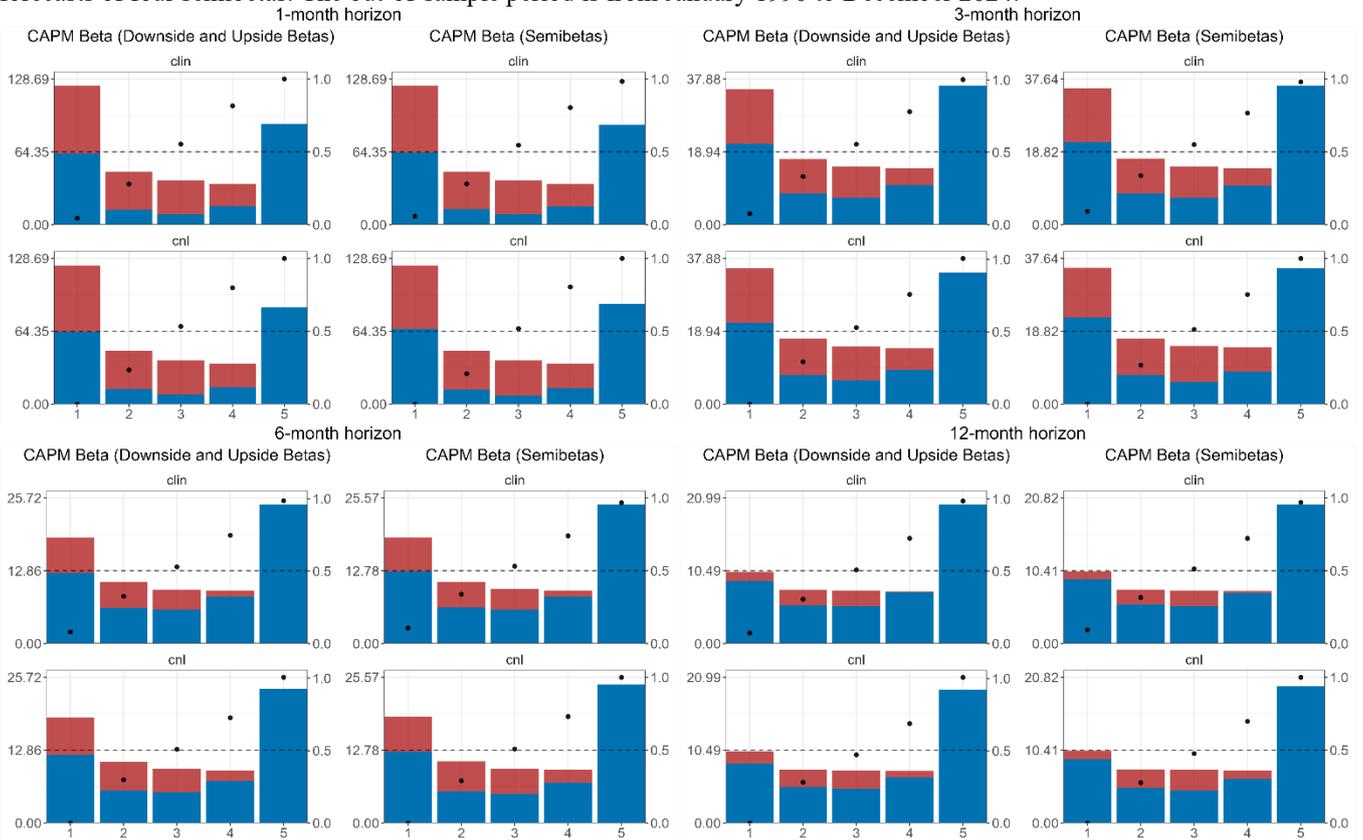

Figure 13 Market-Neutral Minimum-Variance Portfolio: 1-Month Horizon.
This figure displays the distribution of monthly ex-post realized betas for market-neutral minimum-variance portfolios for $h = 1$. Portfolios are constructed based on the $h$-lagged historical CAPM beta benchmark (grey), and conditional CAPM betas reconstructed from forecasts of downside and upside betas (green), and from forecasts of the four semibetas (blue). Dashed vertical lines and annotations indicate the modes of each distribution. The conditional asymmetric beta forecasts are based on equally weighted forecast combinations of the linear models (clin) and the non-linear models (cnl). The out-of-sample period is from January 1990 to December 2024.

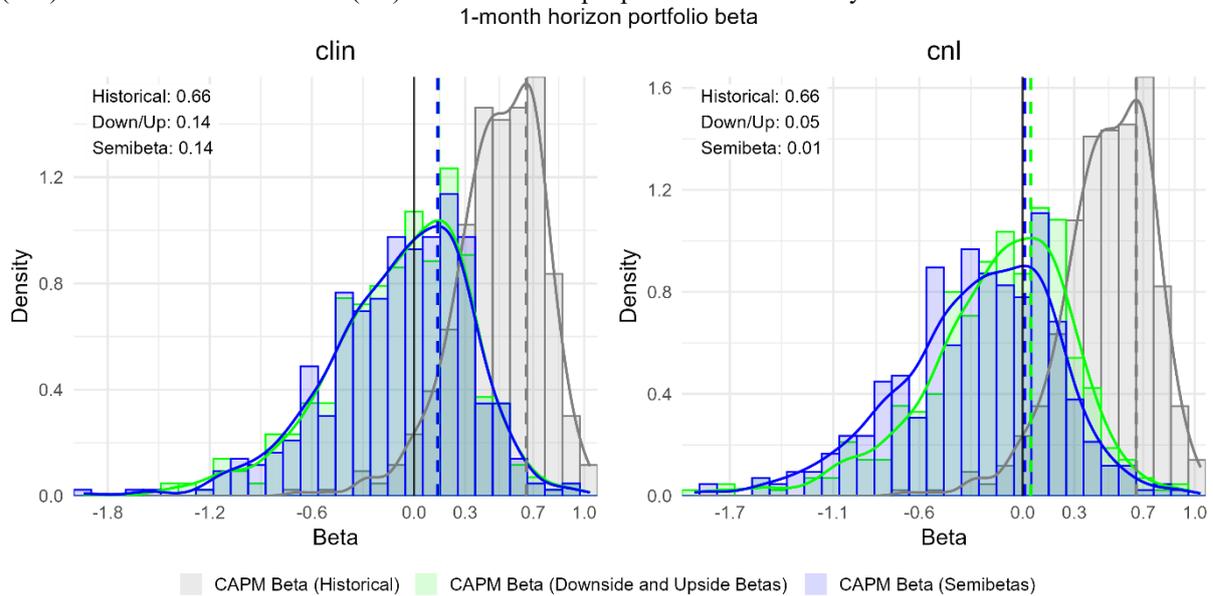

Figure 14 Market-Neutral Minimum-Variance Portfolio: 3-Month Horizon.
This figure displays the distribution of monthly ex-post realized betas for market-neutral minimum-variance portfolios for $h = 3$. Portfolios are constructed based on the $h$-lagged historical CAPM beta benchmark (grey), and conditional CAPM betas reconstructed from forecasts of downside and upside betas (green), and from forecasts of the four semibetas (blue). Dashed vertical lines and annotations indicate the modes of each distribution. The conditional asymmetric beta forecasts are based on equally weighted forecast combinations of the linear models (clin) and the non-linear models (cnl). The out-of-sample period is from January 1990 to December 2024.

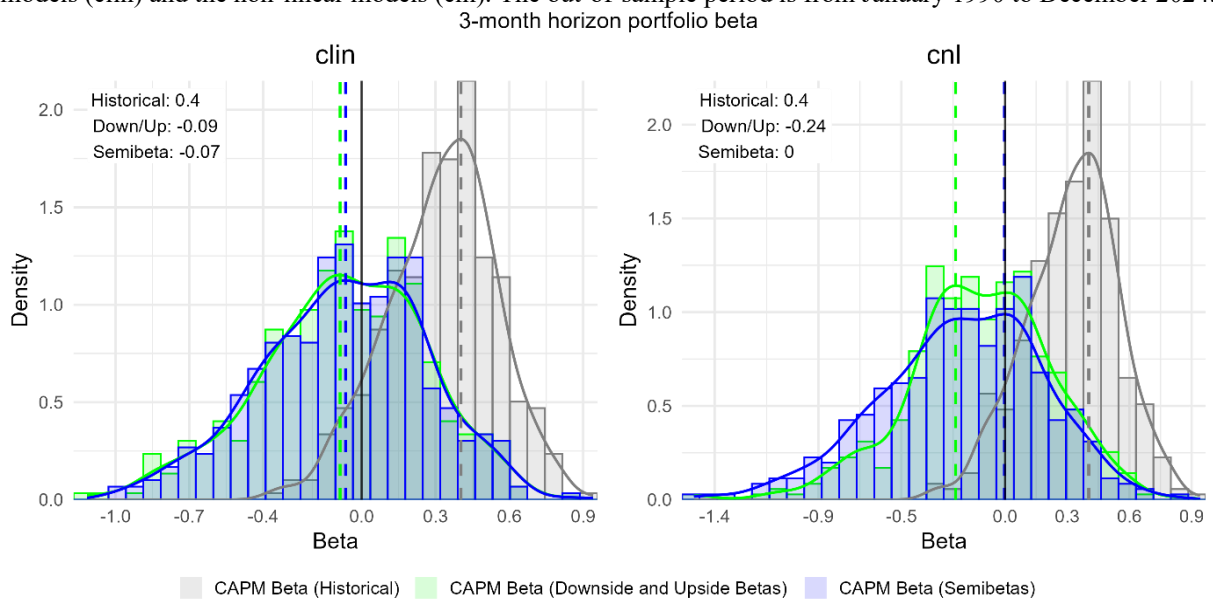

Figure 15 Market-Neutral Minimum-Variance Portfolio: 6-Month Horizon.
This figure displays the distribution of monthly ex-post realized betas for market-neutral minimum-variance portfolios for $h = 6$. Portfolios are constructed based on the $h$-lagged historical CAPM beta benchmark (grey), and conditional CAPM betas reconstructed from forecasts of downside and upside betas (green), and from forecasts of the four semibetas (blue). Dashed vertical lines and annotations indicate the modes of each distribution. The conditional asymmetric beta forecasts are based on equally weighted forecast combinations of the linear models (clin) and the non-linear models (cnl). The out-of-sample period is from January 1990 to December 2024.

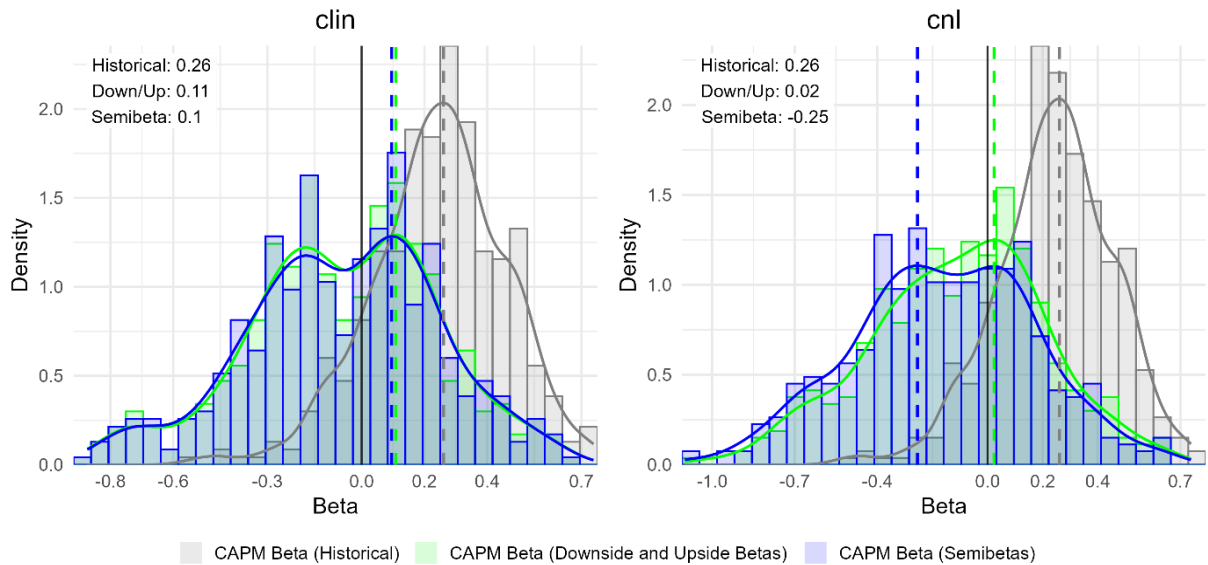

Figure 16 Market-Neutral Minimum-Variance Portfolio: 12-Month Horizon.
This figure displays the distribution of monthly ex-post realized betas for market-neutral minimum-variance portfolios for $h = 12$. Portfolios are constructed based on the $h$-lagged historical CAPM beta benchmark (grey), and conditional CAPM betas reconstructed from forecasts of downside and upside betas (green), and from forecasts of the four semibetas (blue). Dashed vertical lines and annotations indicate the modes of each distribution. The conditional asymmetric beta forecasts are based on equally weighted forecast combinations of the linear models (clin) and the non-linear models (cnl). The out-of-sample period is from January 1990 to December 2024.

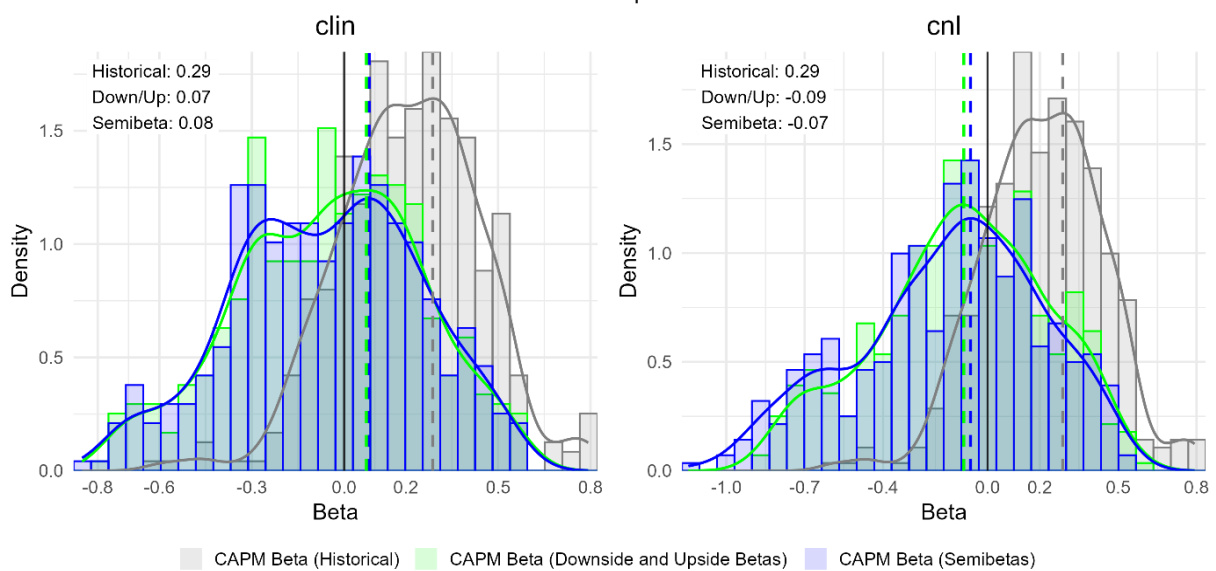

# Appendix 1: Supplementary Tables and Figures

Table A1 Predictive Time Series Performance of Asymmetric Betas Across Forecasting Horizons.
This table reports the out-of-sample $R^2$ of the alternative prediction models across horizons ($h = 1, 3, 6, 12$ months). The forecasting performance is evaluated using the time-series (TS) $R^2$ calculated as the average of firm-level time-series errors. The benchmark is the $h$-lagged realized beta measure for the respective asymmetric beta. The conditional beta forecasts are based on principal component analysis (pca), partial least squares (pls), elastic net (en), random forests (rf), gradient boosting (gb) and feed-forward neural network (ffnn). Performance is also reported for the equally weighted forecast combinations of the three linear models (clin) and the three non-linear models (cnl). This table also reports the statistical significance of the forecasts according to the Clark and West (2007) test for equal predictive accuracy. The statistical significance of the alternative beta estimators from the benchmark is denoted by: *, **, and *** for significance at the 10%, 5%, and 1% level, respectively. The out-of-sample period is from January 1990 to December 2024.

| | Downside Beta | | | | | Upside Beta | | | |
|---|---|---|---|---|---|---|---|---|---|
| | h=1 | h=3 | h=6 | h=12 | | h=1 | h=3 | h=6 | h=12 |
| pca | 45.75*** | 37.11*** | 29.31*** | 20.96*** | pca | 44.24*** | 34.31*** | 23.06*** | 14.61*** |
| pls | 46.30*** | 38.08*** | 30.71*** | 22.41*** | pls | 45.21*** | 35.92*** | 25.06*** | 17.06*** |
| en | 46.36*** | 38.36*** | 31.10*** | 23.26*** | en | 45.24*** | 36.00*** | 25.63*** | 18.00*** |
| clin | 46.35*** | 38.35*** | 31.09*** | 23.35*** | clin | 45.23*** | 36.04*** | 25.48*** | 17.91*** |
| gb | 46.26*** | 38.61*** | 31.62*** | 23.68*** | gb | 45.14*** | 36.91*** | 25.76*** | 17.63*** |
| rf | 46.37*** | 39.68*** | 33.81*** | 26.61*** | rf | 45.41*** | 37.63*** | 27.85*** | 21.00*** |
| ffnn | 46.41*** | 38.26*** | 30.76*** | 21.95*** | ffnn | 45.42*** | 36.04*** | 25.02*** | 16.74*** |
| cnl | 46.72*** | 39.87*** | 33.58*** | 26.34*** | cnl | 45.79*** | 37.97*** | 27.82*** | 21.00*** |
| | Semibeta - Negative | | | | | Semibeta - Positive | | | |
| | h=1 | h=3 | h=6 | h=12 | | h=1 | h=3 | h=6 | h=12 |
| pca | 45.14*** | 34.78*** | 25.58*** | 18.38*** | pca | 42.41*** | 37.70*** | 25.47*** | 23.65*** |
| pls | 46.11*** | 36.09*** | 26.96*** | 19.38*** | pls | 43.42*** | 39.04*** | 26.92*** | 24.72*** |
| en | 46.13*** | 36.41*** | 27.44*** | 20.38*** | en | 43.22*** | 39.12*** | 26.52*** | 25.54*** |
| clin | 46.11*** | 36.32*** | 27.40*** | 20.49*** | clin | 43.30*** | 39.07*** | 27.00*** | 25.72*** |
| gb | 43.98*** | 34.00*** | 28.12*** | 23.45*** | gb | 39.88*** | 32.02*** | 19.73*** | 16.65*** |
| rf | 45.69*** | 36.76*** | 30.31*** | 25.39*** | rf | 41.60*** | 36.58*** | 24.03*** | 24.63*** |
| ffnn | 46.24*** | 36.32*** | 27.17*** | 19.63*** | ffnn | 43.63*** | 38.75*** | 24.02*** | 18.98*** |
| cnl | 46.05*** | 37.53*** | 30.69*** | 25.73*** | cnl | 42.88*** | 37.54*** | 24.98*** | 23.69*** |
| | Semibeta - Mixed Sign, Negative Market Return | | | | | Semibeta - Mixed Sign, Positive Market Return | | | |
| | h=1 | h=3 | h=6 | h=12 | | h=1 | h=3 | h=6 | h=12 |
| pca | 43.70*** | 27.64*** | 12.07*** | -14.72*** | pca | 36.70*** | 21.32*** | 1.80*** | -15.83*** |
| pls | 44.68*** | 30.18*** | 15.08*** | -9.72*** | pls | 37.78*** | 23.90*** | 5.33*** | -12.22*** |
| en | 44.89*** | 30.91*** | 16.18*** | -8.32*** | en | 38.10*** | 24.51*** | 5.98*** | -11.55*** |
| clin | 44.74*** | 30.29*** | 15.64*** | -9.36*** | clin | 37.84*** | 24.02*** | 5.56*** | -11.82*** |
| gb | 43.11*** | 33.47*** | 19.05*** | -2.77*** | gb | 38.42*** | 23.96*** | 10.01*** | -4.65*** |
| rf | 45.52*** | 34.10*** | 19.86*** | -2.73*** | rf | 39.32*** | 28.64*** | 10.64*** | -3.05*** |
| ffnn | 44.97*** | 30.42*** | 20.80*** | -8.34*** | ffnn | 38.44*** | 26.85*** | 8.46*** | -10.22*** |
| cnl | 45.46*** | 34.28*** | 21.86*** | -0.88*** | cnl | 40.07*** | 28.79*** | 12.45*** | -1.89*** |

Table A2 Predictive Cross-Sectional Performance of Asymmetric Betas Across Forecasting Horizons.

This table reports the out-of-sample $R^2$ of the alternative prediction models across horizons ($h = 1, 3, 6, 12$ months). The forecasting performance is evaluated using the cross-sectional (CS) $R^2$ computed as the average of the monthly cross-sectional errors. The benchmark is the $h$-lagged realized beta measure for the respective asymmetric beta. The conditional beta forecasts are based on principal component analysis (pca), partial least squares (pls), elastic net (en), random forests (rf), gradient boosting (gb) and feed-forward neural network (ffnn). Performance is also reported for the equally weighted forecast combinations of the three linear models (clin) and the three non-linear models (cnl). This table also reports the statistical significance of the forecasts according to the Clark and West (2007) test for equal predictive accuracy. The statistical significance of the alternative beta estimators from the benchmark is denoted by: *, **, and *** for significance at the 10%, 5%, and 1% level, respectively. The out-of-sample period is from January 1990 to December 2024.

| | Downside Beta | | | | | Upside Beta | | | |
|---|---|---|---|---|---|---|---|---|---|
| | h=1 | h=3 | h=6 | h=12 | | h=1 | h=3 | h=6 | h=12 |
| pca | 45.02*** | 35.42*** | 26.88*** | 16.85*** | pca | 43.34*** | 32.01*** | 19.65*** | 10.50*** |
| pls | 45.73*** | 36.72*** | 28.88*** | 19.16*** | pls | 44.53*** | 33.98*** | 22.11*** | 13.60*** |
| en | 45.78*** | 36.88*** | 29.04*** | 19.89*** | en | 44.56*** | 33.96*** | 22.65*** | 14.58*** |
| clin | 45.76*** | 36.91*** | 29.08*** | 19.94*** | clin | 44.53*** | 34.04*** | 22.50*** | 14.45*** |
| gb | 45.71*** | 37.23*** | 29.62*** | 20.29*** | gb | 44.62*** | 35.29*** | 23.20*** | 14.44*** |
| rf | 45.85*** | 38.55*** | 32.27*** | 23.96*** | rf | 44.89*** | 36.26*** | 25.75*** | 18.51*** |
| ffnn | 45.89*** | 36.97*** | 28.95*** | 18.89*** | ffnn | 44.77*** | 34.13*** | 22.07*** | 13.20*** |
| cnl | 46.24*** | 38.77*** | 31.96*** | 23.54*** | cnl | 45.29*** | 36.50*** | 25.48*** | 18.17*** |
| | Semibeta - Negative | | | | | Semibeta - Positive | | | |
| | h=1 | h=3 | h=6 | h=12 | | h=1 | h=3 | h=6 | h=12 |
| pca | 45.11*** | 35.18*** | 25.69*** | 17.73*** | pca | 42.28*** | 37.11*** | 25.49*** | 23.63*** |
| pls | 46.14*** | 36.52*** | 27.25*** | 19.03*** | pls | 43.32*** | 38.39*** | 26.90*** | 24.70*** |
| en | 46.19*** | 36.84*** | 27.64*** | 20.02*** | en | 43.10*** | 38.44*** | 26.44*** | 25.00*** |
| clin | 46.15*** | 36.77*** | 27.67*** | 20.17*** | clin | 43.22*** | 38.48*** | 27.06*** | 25.68*** |
| gb | 43.82*** | 33.87*** | 27.86*** | 22.40*** | gb | 39.08*** | 31.00*** | 19.70*** | 16.21*** |
| rf | 45.66*** | 37.01*** | 30.39*** | 25.07*** | rf | 41.31*** | 36.05*** | 24.46*** | 24.50*** |
| ffnn | 46.28*** | 36.71*** | 27.11*** | 18.76*** | ffnn | 43.49*** | 38.03*** | 23.68*** | 18.33*** |
| cnl | 46.03*** | 37.80*** | 30.77*** | 25.18*** | cnl | 42.55*** | 36.93*** | 25.13*** | 23.45*** |
| | Semibeta - Mixed Sign, Negative Market Return | | | | | Semibeta - Mixed Sign, Positive Market Return | | | |
| | h=1 | h=3 | h=6 | h=12 | | h=1 | h=3 | h=6 | h=12 |
| pca | 42.71*** | 25.57*** | 8.93*** | -15.92*** | pca | 36.19*** | 20.20*** | -1.24*** | -19.28* |
| pls | 44.01*** | 28.68*** | 13.04*** | -10.56*** | pls | 37.49*** | 23.29*** | 3.34*** | -13.72 |
| en | 44.07*** | 28.98*** | 13.24*** | -10.65*** | en | 37.52*** | 23.16*** | 2.65*** | -15.04 |
| clin | 43.97*** | 28.58*** | 13.16*** | -10.52*** | clin | 37.43*** | 23.13*** | 3.03*** | -14.29 |
| gb | 41.93*** | 31.73*** | 17.34*** | -5.92*** | gb | 37.98*** | 23.13*** | 8.22*** | -8.88* |
| rf | 44.72*** | 32.59*** | 17.66*** | -4.22*** | rf | 38.95*** | 27.90*** | 7.65*** | -5.77* |
| ffnn | 44.56*** | 30.05*** | 18.53*** | -6.78*** | ffnn | 38.02*** | 25.92*** | 7.02*** | -10.35* |
| cnl | 44.72*** | 33.08*** | 19.97*** | -1.67*** | cnl | 39.60*** | 28.06*** | 10.64*** | -3.92* |

Table A3 Predictive Time Series Performance of Reconstructed CAPM Betas Across Forecasting Horizons.

This table reports the out-of-sample $R^2$ of the alternative prediction models across horizons ($h = 1, 3, 6, 12$ months). The forecasting performance is evaluated using the time-series (TS) $R^2$ calculated as the average of firm-level time-series errors. The benchmark is the $h$-lagged realized CAPM beta. The conditional beta forecasts are based on principal component analysis (pca), partial least squares (pls), elastic net (en), random forests (rf), gradient boosting (gb) and feed-forward neural network (ffnn). Performance is also reported for the equally weighted forecast combinations of the three linear models (clin) and the three non-linear models (cnl). This table also reports the statistical significance of the forecasts according to the Clark and West (2007) test for equal predictive accuracy. The statistical significance of the alternative beta estimators from the benchmark is denoted by: *, **, and *** for significance at the 10%, 5%, and 1% level, respectively. The out-of-sample period is from January 1990 to December 2024.

| | CAPM Beta (Downside and Upside Betas) | | | |
|---|---|---|---|---|
| | h=1 | h=3 | h=6 | h=12 |
| pca | 36.82*** | 24.13*** | 13.11*** | 7.65*** |
| pls | 38.51*** | 26.55*** | 15.84*** | 10.21*** |
| en | 38.64*** | 26.84*** | 16.45*** | 11.41*** |
| clin | 38.55*** | 26.71*** | 16.23*** | 11.25*** |
| gb | 38.81*** | 28.61*** | 18.13*** | 13.32*** |
| rf | 39.10*** | 29.65*** | 20.52*** | 15.74*** |
| ffnn | 38.88*** | 27.01*** | 16.21*** | 10.18*** |
| cnl | 39.73*** | 29.96*** | 20.24*** | 15.64*** |
| | CAPM Beta (Semibetas) | | | |
| | h=1 | h=3 | h=6 | h=12 |
| pca | 37.18*** | 24.27*** | 14.00*** | 8.08*** |
| pls | 38.95*** | 26.75*** | 16.01*** | 10.01*** |
| en | 38.92*** | 27.34*** | 16.57*** | 11.75*** |
| clin | 38.89*** | 26.96*** | 16.73*** | 11.59*** |
| gb | 36.17*** | 25.98*** | 18.61*** | 12.92*** |
| rf | 38.13*** | 28.28*** | 19.16*** | 15.06*** |
| ffnn | 38.99*** | 26.86*** | 15.11*** | 7.74*** |
| cnl | 38.92*** | 29.06*** | 20.08*** | 15.32*** |

Table A4 Predictive Cross-Sectional Performance of Reconstructed CAPM Betas Across Forecasting Horizons.

This table reports the out-of-sample $R^2$ of the alternative prediction models across horizons ($h = 1, 3, 6, 12$ months). The forecasting performance is evaluated using the cross-sectional (CS) $R^2$ computed as the average of the monthly cross-sectional errors. The benchmark is the $h$-lagged realized CAPM beta. The conditional beta forecasts are based on principal component analysis (pca), partial least squares (pls), elastic net (en), random forests (rf), gradient boosting (gb) and feed-forward neural network (ffnn). Performance is also reported for the equally weighted forecast combinations of the three linear models (clin) and the three non-linear models (cnl). This table also reports the statistical significance of the forecasts according to the Clark and West (2007) test for equal predictive accuracy. The statistical significance of the alternative beta estimators from the benchmark is denoted by: *, **, and *** for significance at the 10%, 5%, and 1% level, respectively. The out-of-sample period is from January 1990 to December 2024.

| | CAPM Beta (Downside and Upside Betas) | | | |
|---|---|---|---|---|
| | h=1 | h=3 | h=6 | h=12 |
| pca | 34.87*** | 20.61*** | 8.91*** | 2.71*** |
| pls | 37.01*** | 23.64*** | 12.48*** | 6.21*** |
| en | 37.13*** | 23.73*** | 12.81*** | 7.35*** |
| clin | 36.98*** | 23.64*** | 12.64*** | 7.11*** |
| gb | 37.56*** | 26.00*** | 14.90*** | 9.48*** |
| rf | 37.82*** | 27.44*** | 17.94*** | 12.74*** |
| ffnn | 37.48*** | 24.18*** | 12.85*** | 6.31*** |
| cnl | 38.52*** | 27.65*** | 17.40*** | 12.31*** |
| | CAPM Beta (Semibetas) | | | |
| | h=1 | h=3 | h=6 | h=12 |
| pca | 35.11*** | 20.52*** | 9.33*** | 2.85*** |
| pls | 37.43*** | 23.81*** | 12.69*** | 6.17*** |
| en | 37.35*** | 24.29*** | 12.88*** | 7.62*** |
| clin | 37.26*** | 23.84*** | 13.00*** | 7.41*** |
| gb | 34.28*** | 23.12*** | 15.87*** | 9.46*** |
| rf | 36.64*** | 25.98*** | 16.60*** | 12.21*** |
| ffnn | 37.57*** | 24.00*** | 11.68*** | 3.87*** |
| cnl | 37.44*** | 26.58*** | 17.39*** | 12.15*** |

Table A5 Share Price Prediction for Individual Models Using 1-Month Horizon Betas.

This table reports the out-of-sample panel $R^2$ for observed prices compared to intrinsic values derived from discounted cash flow models with alternative discount rates. Intrinsic values are calculated using a one-year holding period with monthly cash flows based on 12-month rolling averages of ordinary dividends. Terminal values assume annual long-term growth rates from 0% to 2%. Conditional betas are converted to discount rates using a risk-free rate of zero and market risk premiums from 8% to 12%. The discount rates vary based on the model used to estimate the reconstructed CAPM beta for $h = 1$. The benchmark is the respective $h$-lagged realized CAPM beta. The conditional beta forecasts are based on principal component analysis (pca), partial least squares (pls), elastic net (en), random forests (rf), gradient boosting (gb) and feed-forward neural network (ffnn). Estimates of CAPM beta are derived as a combination of forecasts of downside and upside betas, and as a combination of forecasts of four semibetas. The out-of-sample period is from January 1990 to December 2024.

| Growth Rate | 0% | | | 1% | | | 2% | | |
|---|---|---|---|---|---|---|---|---|---|
| Market Premium | 8% | 10% | 12% | 8% | 10% | 12% | 8% | 10% | 12% |
| CAPM Beta (Downside and Upside Betas) | | | | | | | | | |
| pca | 20.88 | 12.01 | 6.93 | 33.30 | 20.61 | 11.86 | 21.76 | 26.47 | 16.37 |
| pls | 20.10 | 11.68 | 6.84 | 28.09 | 18.42 | 10.82 | 17.01 | 20.00 | 12.96 |
| en | 19.94 | 11.27 | 6.32 | 30.80 | 19.06 | 10.90 | 25.24 | 23.15 | 14.49 |
| gb | 20.57 | 11.62 | 6.52 | 31.22 | 19.71 | 11.40 | -8.02 | 13.35 | 10.75 |
| rf | 20.32 | 11.31 | 6.20 | 33.42 | 19.87 | 11.18 | 20.51 | 26.87 | 16.14 |
| ffnn | 21.36 | 12.36 | 7.19 | 34.49 | 21.26 | 12.40 | 26.41 | 26.71 | 17.51 |
| CAPM Beta (Semibetas) | | | | | | | | | |
| pca | 21.19 | 12.27 | 7.15 | 32.51 | 20.44 | 12.03 | 25.95 | 24.90 | 16.17 |
| pls | 20.98 | 12.23 | 7.19 | 31.75 | 20.14 | 11.92 | 21.37 | 23.46 | 15.77 |
| en | 20.57 | 11.69 | 6.62 | 32.82 | 20.17 | 11.51 | 25.96 | 24.39 | 15.84 |
| gb | 19.22 | 10.14 | 5.08 | 33.50 | 19.21 | 9.99 | 31.80 | 25.59 | 15.66 |
| rf | 19.59 | 10.51 | 5.42 | 34.08 | 19.79 | 10.57 | 32.24 | 27.44 | 16.36 |
| ffnn | 21.27 | 12.22 | 7.04 | 34.79 | 21.38 | 12.34 | 29.64 | 27.96 | 17.54 |

Table A6 Share Price Prediction for Individual Models Using 3-Month Horizon Betas.

This table reports the out-of-sample panel $R^2$ for observed prices compared to intrinsic values derived from discounted cash flow models with alternative discount rates. Intrinsic values are calculated using a one-year holding period with monthly cash flows based on 12-month rolling averages of ordinary dividends. Terminal values assume annual long-term growth rates from 0% to 2%. Conditional betas are converted to discount rates using a risk-free rate of zero and market risk premiums from 8% to 12%. The discount rates vary based on the model used to estimate the reconstructed CAPM beta for $h = 1$. The benchmark is the respective $h$-lagged realized CAPM beta. The conditional beta forecasts are based on principal component analysis (pca), partial least squares (pls), elastic net (en), random forests (rf), gradient boosting (gb) and feed-forward neural network (ffnn). Estimates of CAPM beta are derived as a combination of forecasts of downside and upside betas, and as a combination of forecasts of four semibetas. The out-of-sample period is from January 1990 to December 2024.

| Growth Rate | 0% | | | 1% | | | 2% | | |
|---|---|---|---|---|---|---|---|---|---|
| Market Premium | 8% | 10% | 12% | 8% | 10% | 12% | 8% | 10% | 12% |
| CAPM Beta (Downside and Upside Betas) | | | | | | | | | |
| pca | 14.48 | 7.61 | 3.92 | 33.50 | 18.09 | 10.44 | 31.44 | 27.92 | 16.69 |
| pls | 14.39 | 7.67 | 4.04 | 31.69 | 17.33 | 10.20 | 26.90 | 24.80 | 15.23 |
| en | 13.91 | 7.05 | 3.40 | 32.87 | 17.35 | 9.76 | 33.46 | 27.25 | 15.85 |
| gb | 14.17 | 7.26 | 3.57 | 33.12 | 17.73 | 10.07 | 11.08 | 20.38 | 15.09 |
| rf | 14.12 | 7.14 | 3.44 | 33.98 | 18.02 | 10.12 | 28.80 | 27.50 | 16.59 |
| ffnn | 14.97 | 7.94 | 4.16 | 34.83 | 18.89 | 10.92 | 35.64 | 29.59 | 17.73 |
| CAPM Beta (Semibetas) | | | | | | | | | |
| pca | 14.59 | 7.70 | 3.99 | 33.31 | 18.12 | 10.49 | 32.34 | 26.02 | 16.20 |
| pls | 14.55 | 7.71 | 4.03 | 32.97 | 17.94 | 10.43 | 30.66 | 25.76 | 16.15 |
| en | 14.15 | 7.24 | 3.55 | 33.44 | 17.80 | 10.05 | 34.57 | 26.97 | 16.44 |
| gb | 13.26 | 6.29 | 2.64 | 32.90 | 16.56 | 8.59 | 36.14 | 26.41 | 14.60 |
| rf | 13.60 | 6.62 | 2.94 | 33.86 | 17.53 | 9.54 | 30.52 | 27.30 | 16.15 |
| ffnn | 14.58 | 7.56 | 3.80 | 34.61 | 18.52 | 10.54 | 36.68 | 29.44 | 17.46 |

Table A7 Share Price Prediction for Individual Models Using 6-Month Horizon Betas.

This table reports the out-of-sample panel $R^2$ for observed prices compared to intrinsic values derived from discounted cash flow models with alternative discount rates. Intrinsic values are calculated using a one-year holding period with monthly cash flows based on 12-month rolling averages of ordinary dividends. Terminal values assume annual long-term growth rates from 0% to 2%. Conditional betas are converted to discount rates using a risk-free rate of zero and market risk premiums from 8% to 12%. The discount rates vary based on the model used to estimate the reconstructed CAPM beta for $h = 1$. The benchmark is the respective $h$-lagged realized CAPM beta. The conditional beta forecasts are based on principal component analysis (pca), partial least squares (pls), elastic net (en), random forests (rf), gradient boosting (gb) and feed-forward neural network (ffnn). Estimates of CAPM beta are derived as a combination of forecasts of downside and upside betas, and as a combination of forecasts of four semibetas. The out-of-sample period is from January 1990 to December 2024.

| Growth Rate | 0% | | | 1% | | | 2% | | |
|---|---|---|---|---|---|---|---|---|---|
| Market Premium | 8% | 10% | 12% | 8% | 10% | 12% | 8% | 10% | 12% |
| CAPM Beta (Downside and Upside Betas) | | | | | | | | | |
| pca | 7.11 | 3.03 | 1.02 | 28.05 | 14.20 | 7.19 | 39.51 | 34.18 | 16.26 |
| pls | 7.06 | 3.07 | 1.10 | 26.70 | 13.84 | 7.14 | 35.01 | 31.29 | 14.73 |
| en | 6.52 | 2.45 | 0.49 | 27.45 | 13.58 | 6.63 | 39.83 | 33.27 | 15.38 |
| gb | 6.47 | 2.49 | 0.57 | 23.90 | 12.13 | 6.21 | 14.15 | 23.48 | 9.74 |
| rf | 6.56 | 2.42 | 0.43 | 28.10 | 13.89 | 6.72 | 33.28 | 33.73 | 15.60 |
| ffnn | 7.36 | 3.13 | 1.04 | 29.26 | 14.71 | 7.42 | 42.48 | 35.34 | 17.02 |
| CAPM Beta (Semibetas) | | | | | | | | | |
| pca | 7.19 | 3.12 | 1.12 | 27.39 | 14.07 | 7.23 | 35.30 | 31.29 | 15.52 |
| pls | 7.19 | 3.14 | 1.14 | 27.27 | 14.03 | 7.25 | 36.23 | 31.94 | 15.47 |
| en | 6.56 | 2.47 | 0.50 | 27.78 | 13.78 | 6.72 | 36.60 | 33.03 | 15.64 |
| gb | 6.36 | 2.27 | 0.31 | 27.44 | 13.53 | 6.49 | 26.08 | 29.77 | 14.51 |
| rf | 6.34 | 2.21 | 0.23 | 28.11 | 13.70 | 6.50 | 33.85 | 32.45 | 15.02 |
| ffnn | 6.82 | 2.67 | 0.66 | 28.16 | 14.00 | 6.88 | 40.92 | 33.48 | 15.84 |

Table A8 Share Price Prediction for Individual Models Using 12-Month Horizon Betas.

This table reports the out-of-sample panel $R^2$ for observed prices compared to intrinsic values derived from discounted cash flow models with alternative discount rates. Intrinsic values are calculated using a one-year holding period with monthly cash flows based on 12-month rolling averages of ordinary dividends. Terminal values assume annual long-term growth rates from 0% to 2%. Conditional betas are converted to discount rates using a risk-free rate of zero and market risk premiums from 8% to 12%. The discount rates vary based on the model used to estimate the reconstructed CAPM beta for $h = 1$. The benchmark is the respective $h$-lagged realized CAPM beta. The conditional beta forecasts are based on principal component analysis (pca), partial least squares (pls), elastic net (en), random forests (rf), gradient boosting (gb) and feed-forward neural network (ffnn). Estimates of CAPM beta are derived as a combination of forecasts of downside and upside betas, and as a combination of forecasts of four semibetas. The out-of-sample period is from January 1990 to December 2024.

| Growth Rate | 0% | | | 1% | | | 2% | | |
|---|---|---|---|---|---|---|---|---|---|
| Market Premium | 8% | 10% | 12% | 8% | 10% | 12% | 8% | 10% | 12% |
| CAPM Beta (Downside and Upside Betas) | | | | | | | | | |
| pca | 2.23 | 0.27 | -0.59 | 23.97 | 11.50 | 5.26 | 40.92 | 30.12 | 16.28 |
| pls | 2.10 | 0.23 | -0.59 | 22.63 | 11.00 | 5.09 | 36.18 | 26.80 | 14.95 |
| en | 1.52 | -0.41 | -1.22 | 23.48 | 10.87 | 4.65 | 41.64 | 29.70 | 15.60 |
| gb | 1.59 | -0.27 | -1.06 | 21.85 | 10.47 | 4.61 | 22.58 | 23.75 | 14.15 |
| rf | 1.62 | -0.39 | -1.23 | 24.27 | 11.14 | 4.72 | 41.18 | 30.27 | 16.10 |
| ffnn | 2.36 | 0.26 | -0.66 | 24.96 | 11.86 | 5.41 | 42.54 | 31.27 | 16.85 |
| CAPM Beta (Semibetas) | | | | | | | | | |
| pca | 2.20 | 0.22 | -0.64 | 23.82 | 11.44 | 5.24 | 38.53 | 29.11 | 15.91 |
| pls | 2.16 | 0.24 | -0.60 | 23.17 | 11.17 | 5.15 | 37.67 | 27.65 | 15.38 |
| en | 1.60 | -0.36 | -1.17 | 23.84 | 11.03 | 4.75 | 42.59 | 29.83 | 15.86 |
| gb | 1.56 | -0.34 | -1.14 | 22.00 | 10.47 | 4.55 | 31.74 | 23.86 | 13.65 |
| rf | 1.58 | -0.44 | -1.27 | 24.42 | 11.17 | 4.72 | 35.45 | 30.75 | 16.21 |
| ffnn | 1.61 | -0.37 | -1.20 | 23.09 | 10.79 | 4.61 | 42.01 | 29.32 | 15.39 |

Table A9 Share Price Prediction for Individual Models Using Betas Across Horizons.

This table reports the out-of-sample panel $R^2$ for observed prices compared to intrinsic values derived from discounted cash flow models with alternative discount rates. Intrinsic values are calculated using a one-year holding period with monthly cash flows based on 12-month rolling averages of ordinary dividends. Terminal values assume annual long-term growth rates from 0% to 2%. Conditional betas are converted to discount rates using a risk-free rate of zero and market risk premiums from 8% to 12%. The discount rates vary based on the model used to estimate the reconstructed CAPM beta for $h = 1$. The benchmark is the respective $h$-lagged realized CAPM beta. The conditional beta forecasts are based on principal component analysis (pca), partial least squares (pls), elastic net (en), random forests (rf), gradient boosting (gb) and feed-forward neural network (ffnn). Estimates of CAPM beta are derived as a combination of forecasts of downside and upside betas, and as a combination of forecasts of four semibetas. The out-of-sample period is from January 1990 to December 2024.

| Growth Rate | 0% | | | 1% | | | 2% | | |
|---|---|---|---|---|---|---|---|---|---|
| Market Premium | 8% | 10% | 12% | 8% | 10% | 12% | 8% | 10% | 12% |
| CAPM Beta (Downside and Upside Betas) | | | | | | | | | |
| pca | 5.69 | 2.60 | 1.09 | 35.49 | 18.68 | 10.36 | 60.15 | 48.22 | 28.20 |
| pls | 5.57 | 2.57 | 1.09 | 34.36 | 18.24 | 10.20 | 54.86 | 44.68 | 27.08 |
| en | 5.02 | 1.95 | 0.49 | 35.12 | 18.13 | 9.80 | 61.69 | 48.11 | 27.79 |
| gb | 5.08 | 2.07 | 0.63 | 33.47 | 17.76 | 9.74 | 40.54 | 41.72 | 26.08 |
| rf | 5.10 | 1.96 | 0.47 | 35.79 | 18.37 | 9.85 | 61.64 | 48.60 | 28.23 |
| ffnn | 5.83 | 2.60 | 1.04 | 36.38 | 19.04 | 10.51 | 63.05 | 49.56 | 28.86 |
| CAPM Beta (Semibetas) | | | | | | | | | |
| pca | 5.66 | 2.56 | 1.05 | 35.40 | 18.65 | 10.34 | 58.79 | 47.09 | 27.91 |
| pls | 5.63 | 2.58 | 1.09 | 34.86 | 18.40 | 10.26 | 56.66 | 46.08 | 27.48 |
| en | 5.10 | 2.00 | 0.53 | 35.43 | 18.28 | 9.88 | 61.73 | 48.28 | 28.02 |
| gb | 5.06 | 2.01 | 0.56 | 33.83 | 17.76 | 9.69 | 49.15 | 40.88 | 24.98 |
| rf | 5.07 | 1.92 | 0.43 | 35.92 | 18.40 | 9.86 | 57.63 | 49.05 | 28.32 |
| ffnn | 5.10 | 1.99 | 0.51 | 34.80 | 18.06 | 9.76 | 61.23 | 47.02 | 27.40 |

Figure A1 Average Predictive Performance of Quintile Portfolios for Individual Models: 1-Month Horizon.
This figure displays the time series averages of monthly mean squared errors (MSE) for quintile portfolios for $h = 1$. Portfolios are formed monthly by sorting stocks into quintiles according to realised beta values (1-low beta, 5-high beta). The equally weighted MSE is constructed for each quintile, for the benchmark (red bars) and the conditional beta forecasts (blue bars). The figure also reports the fraction of stocks within each portfolio for which the difference between realized and forecast portfolio betas is positive (dots, right-hand axis). The benchmark is the $h$-lagged realized beta measure for the respective asymmetric beta. The conditional beta forecasts are based on principal component analysis (pca), partial least squares (pls), elastic net (en), random forests (rf), gradient boosting (gb) and feed-forward neural network (ffnn). The out-of-sample period is from January 1990 to December 2024.

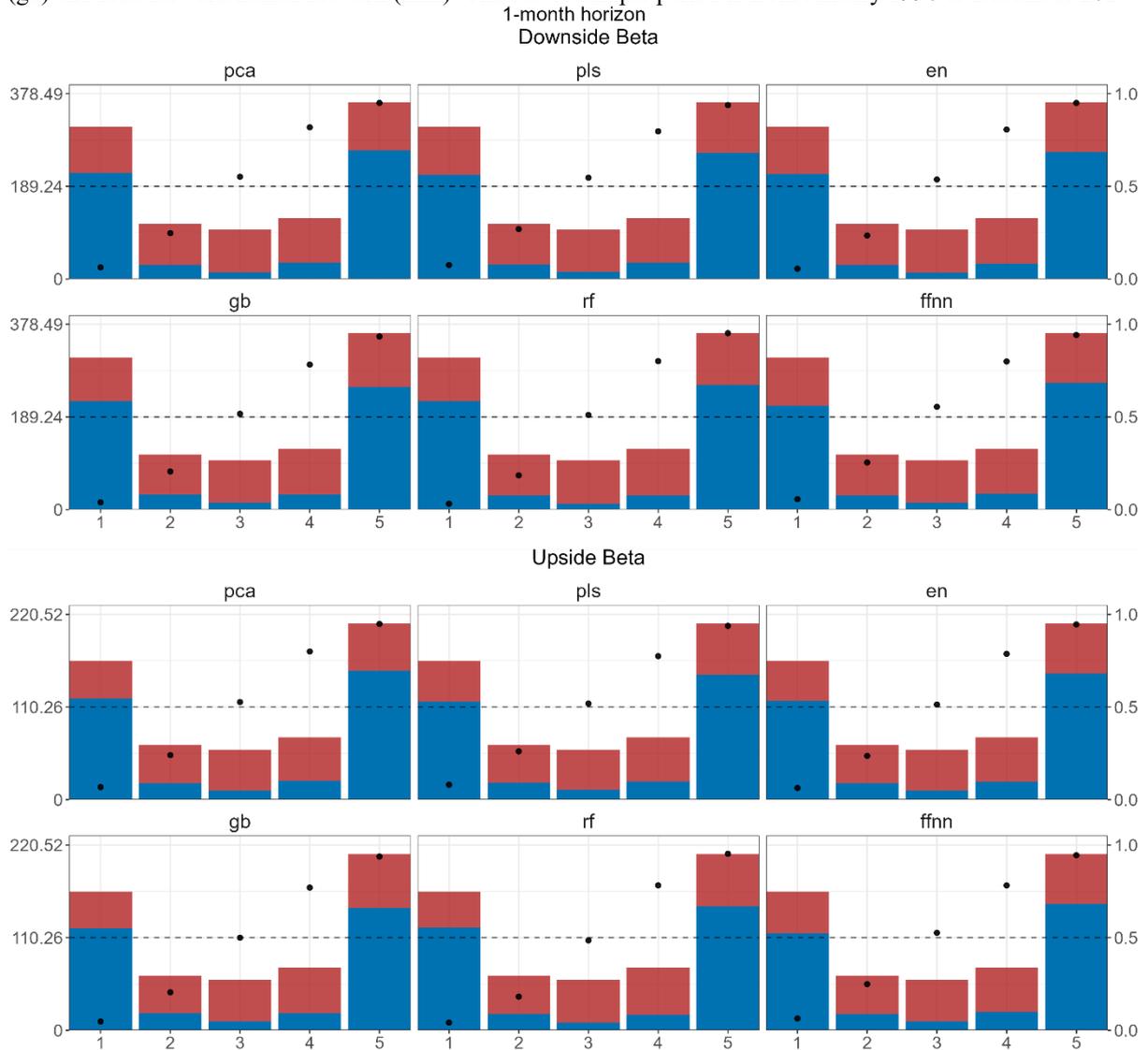

Figure A1 Average Predictive Performance of Quintile Portfolios for Individual Models: 1-Month Horizon. (Continued.)

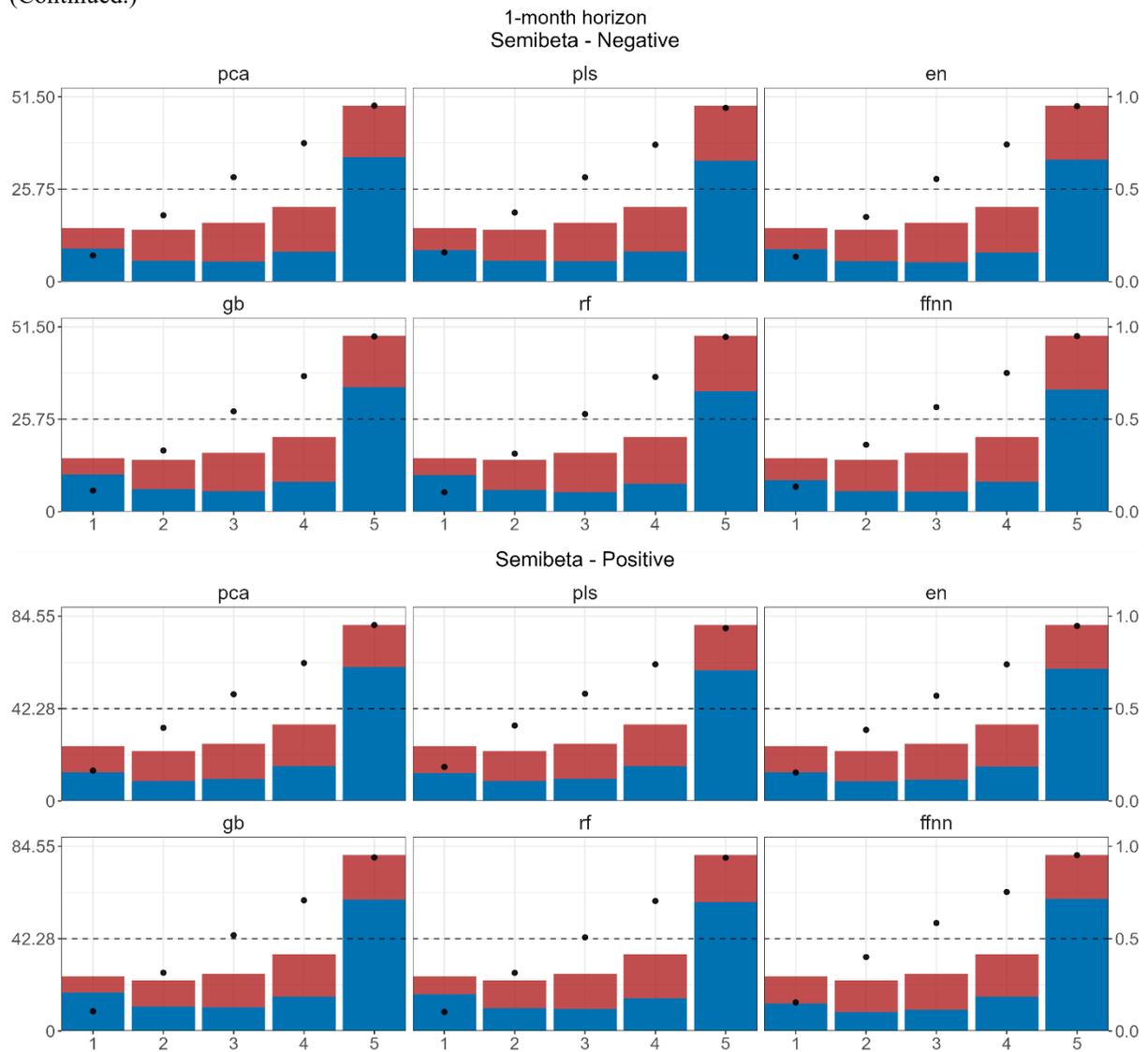

Figure A1 Average Predictive Performance of Quintile Portfolios for Individual Models: 1-Month Horizon. (Continued.)

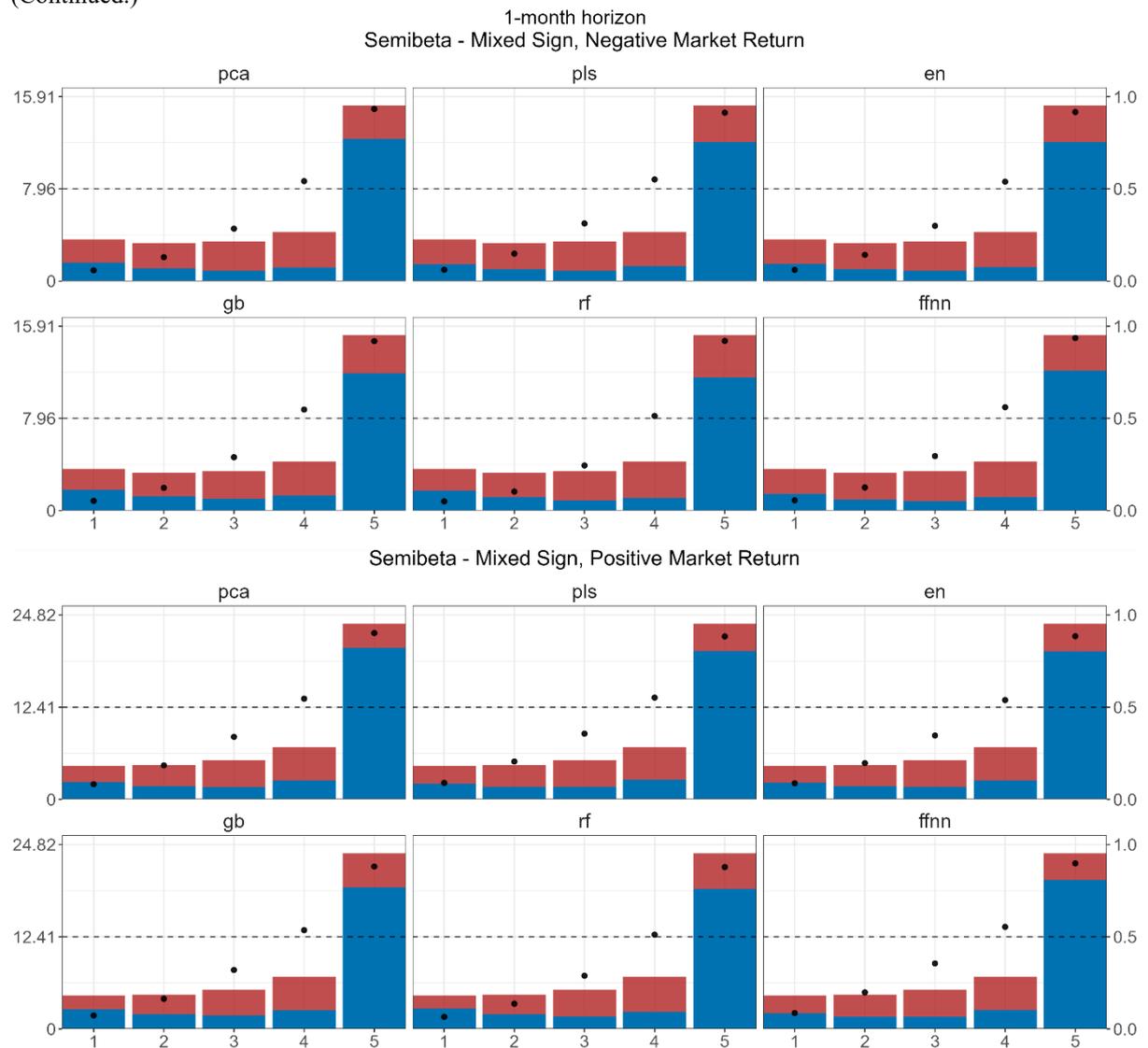

Figure A2 Average Predictive Performance of Quintile Portfolios for Individual Models: 3-Month Horizon.
This figure displays the time series averages of monthly mean squared errors (MSE) for quintile portfolios for $h = 3$. Portfolios are formed monthly by sorting stocks into quintiles according to realised beta values (1-low beta, 5-high beta). The equally weighted MSE is constructed for each quintile, for the benchmark (red bars) and the conditional beta forecasts (blue bars). The figure also reports the fraction of stocks within each portfolio for which the difference between realized and forecast portfolio betas is positive (dots, right-hand axis). The benchmark is the $h$-lagged realized beta measure for the respective asymmetric beta. The conditional beta forecasts are based on principal component analysis (pca), partial least squares (pls), elastic net (en), random forests (rf), gradient boosting (gb) and feed-forward neural network (ffnn). The out-of-sample period is from January 1990 to December 2024.

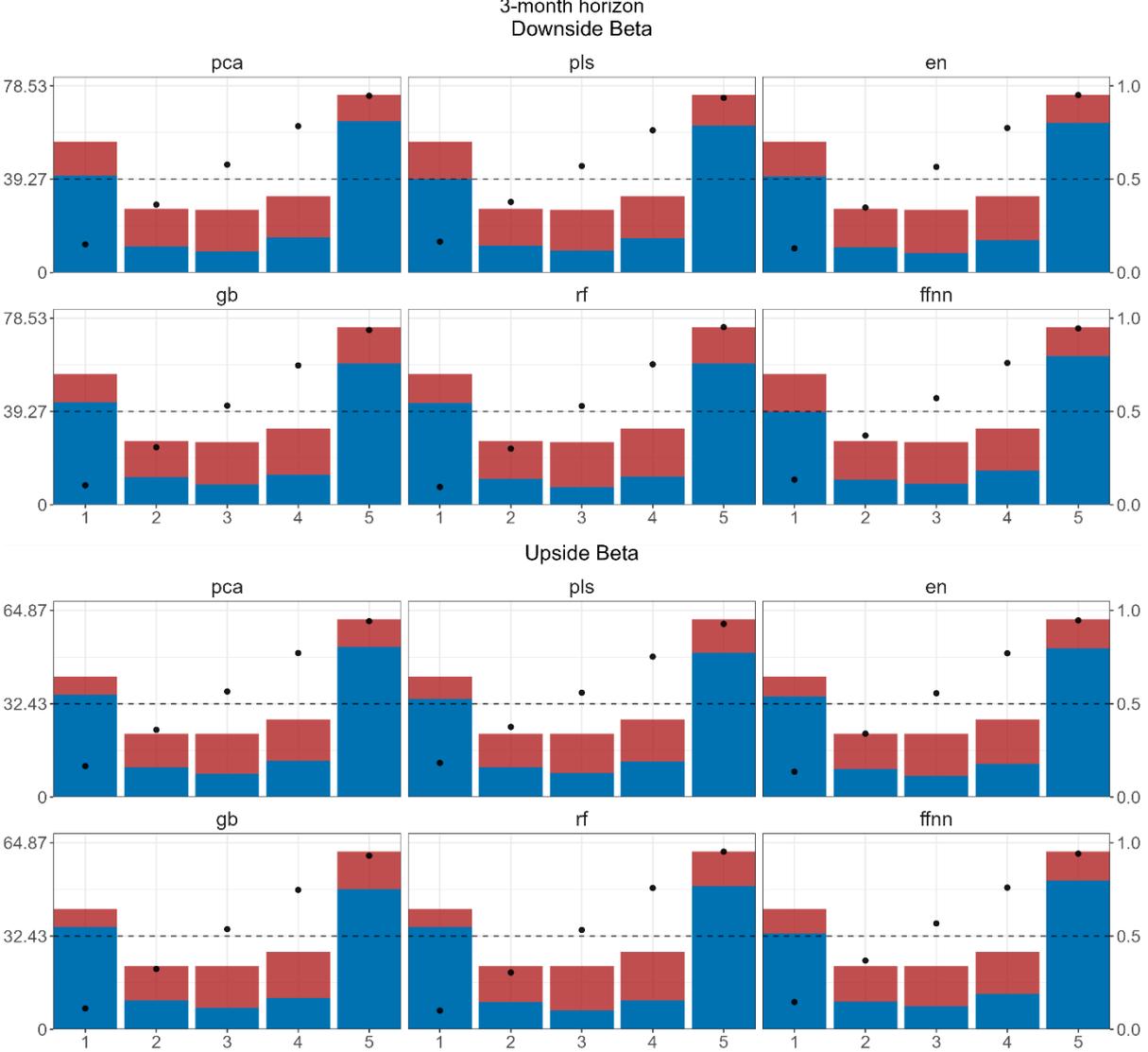

Figure A2 Average Predictive Performance of Quintile Portfolios for Individual Models: 3-Month Horizon. (Continued.)

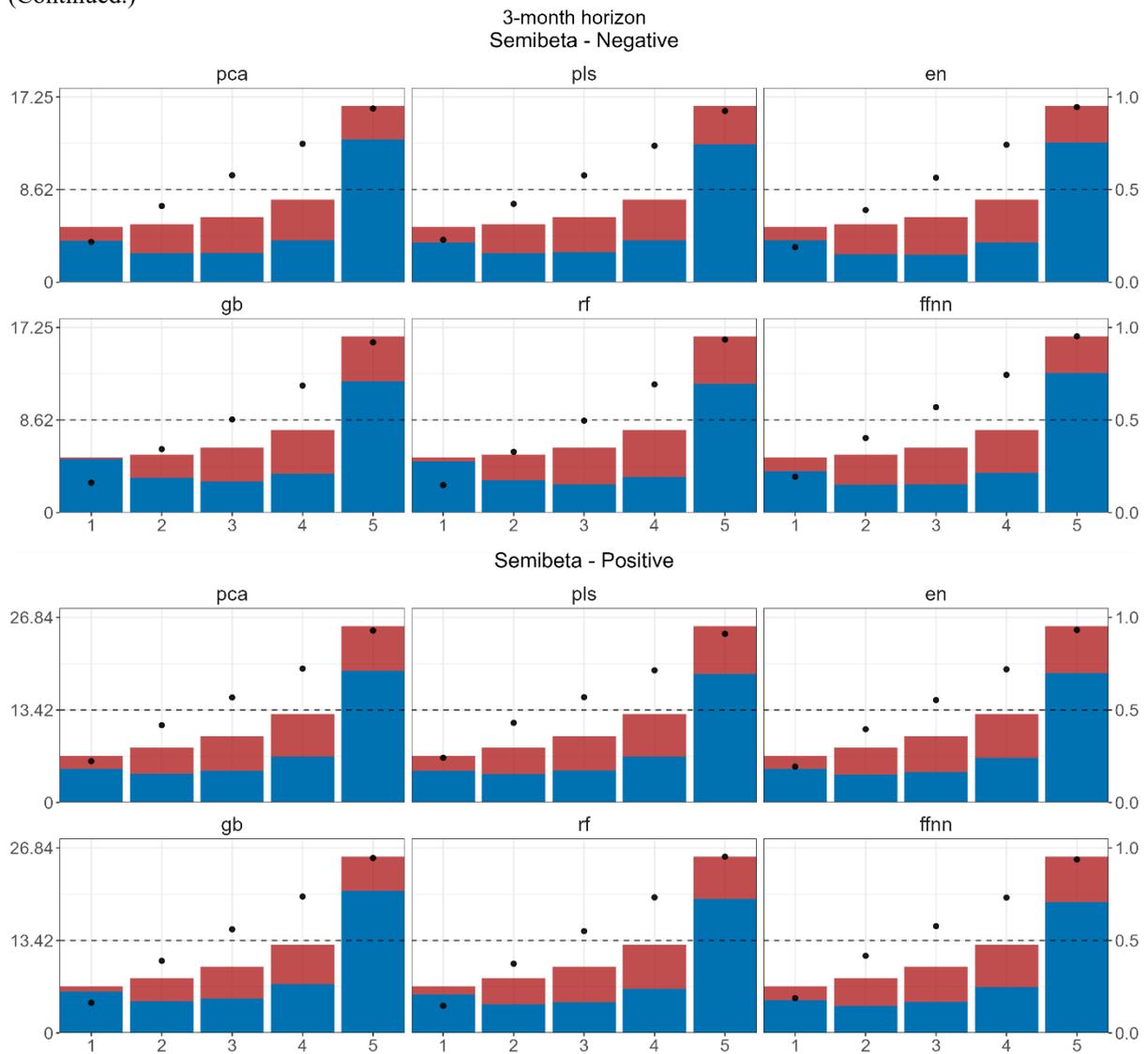

Figure A2 Average Predictive Performance of Quintile Portfolios for Individual Models: 3-Month Horizon. (Continued.)

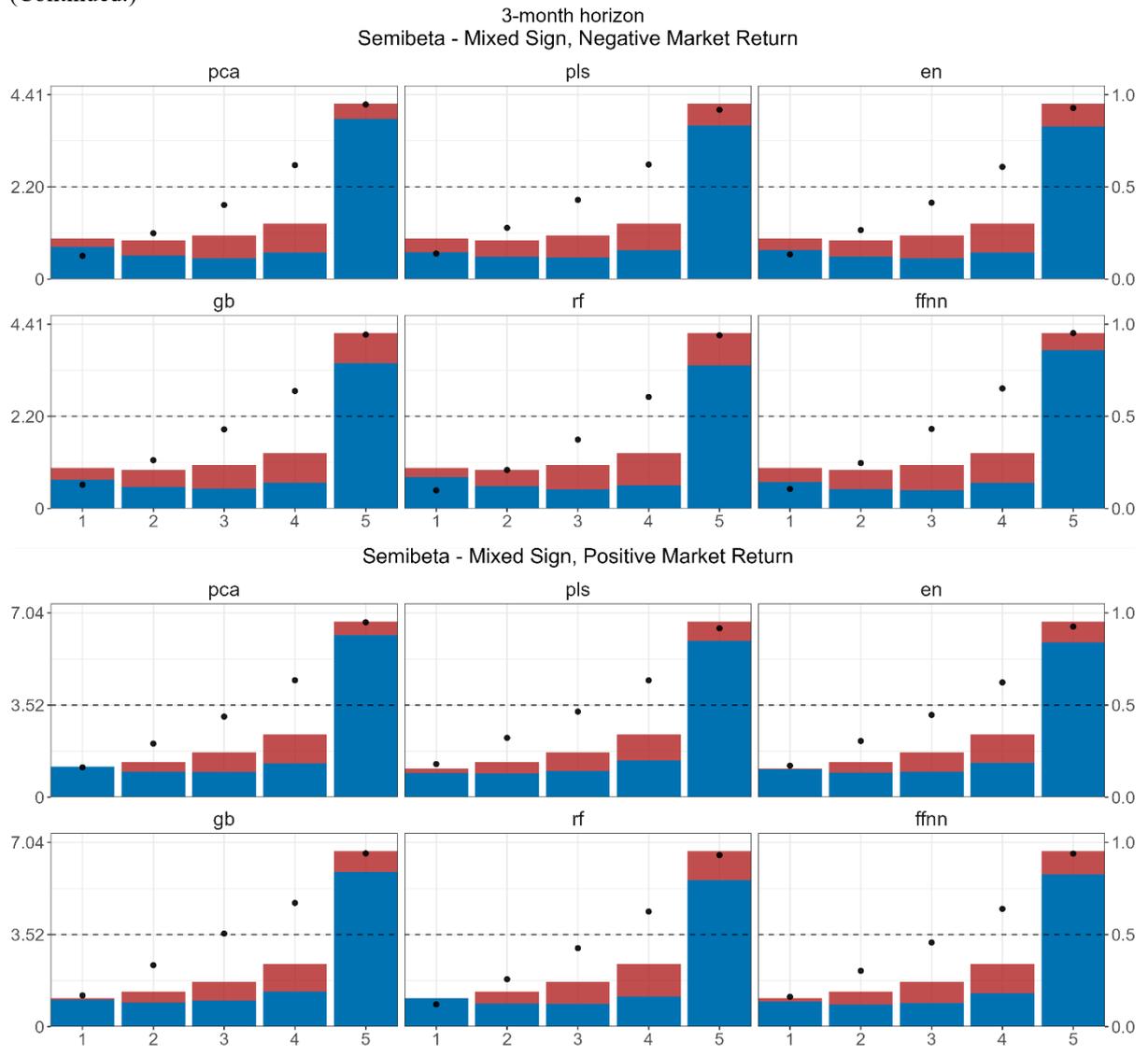

Figure A3 Average Predictive Performance of Quintile Portfolios for Individual Models: 6-Month Horizon. This figure displays the time series averages of monthly mean squared errors (MSE) for quintile portfolios for $h = 6$. Portfolios are formed monthly by sorting stocks into quintiles according to realised beta values (1-low beta, 5-high beta). The equally weighted MSE is constructed for each quintile, for the benchmark (red bars) and the conditional beta forecasts (blue bars). The figure also reports the fraction of stocks within each portfolio for which the difference between realized and forecast portfolio betas is positive (dots, right-hand axis). The benchmark is the $h$-lagged realized beta measure for the respective asymmetric beta. The conditional beta forecasts are based on principal component analysis (pca), partial least squares (pls), elastic net (en), random forests (rf), gradient boosting (gb) and feed-forward neural network (ffnn). The out-of-sample period is from January 1990 to December 2024.

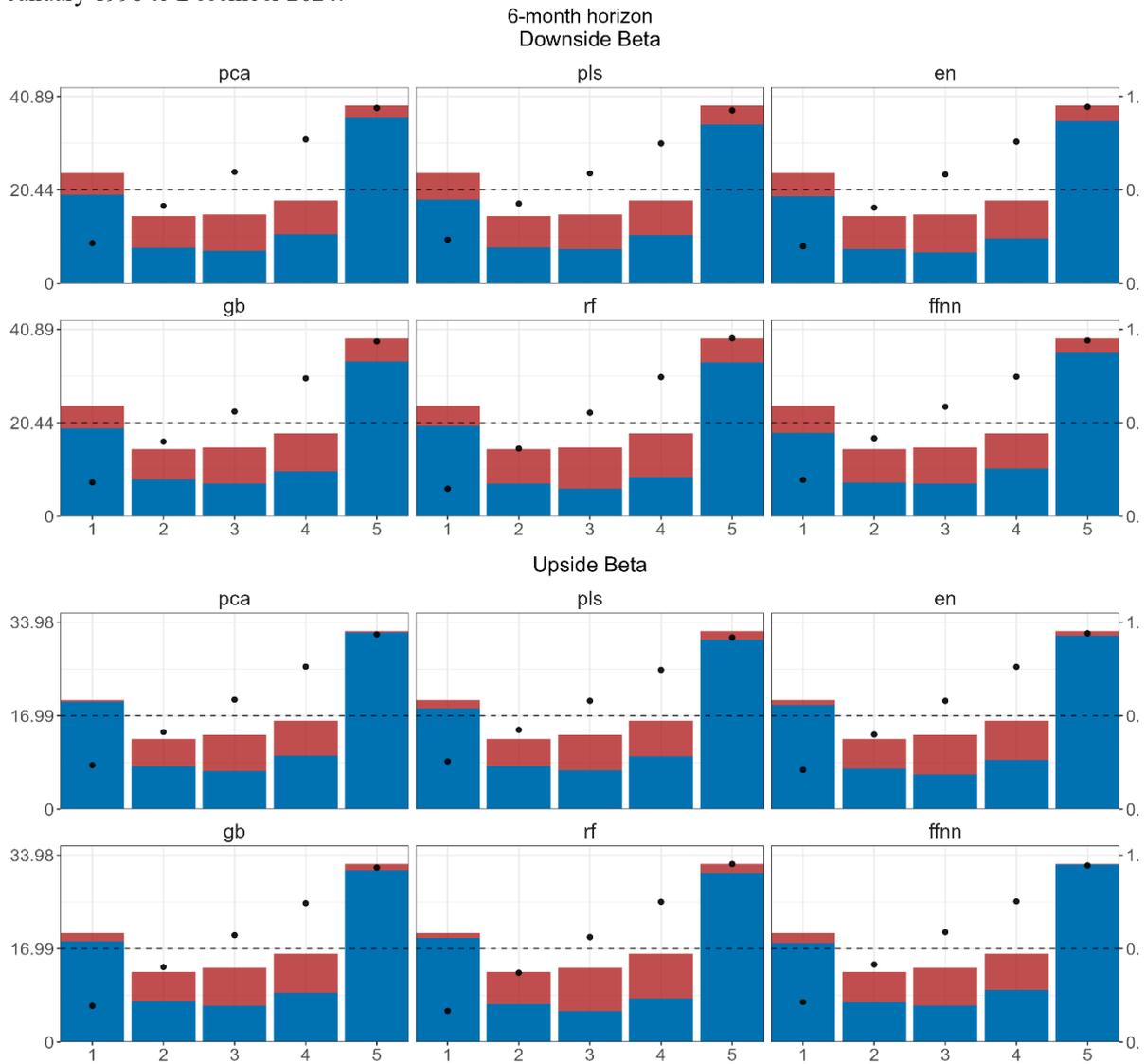

Figure A3 Average Predictive Performance of Quintile Portfolios for Individual Models: 6-Month Horizon. (Continued.)

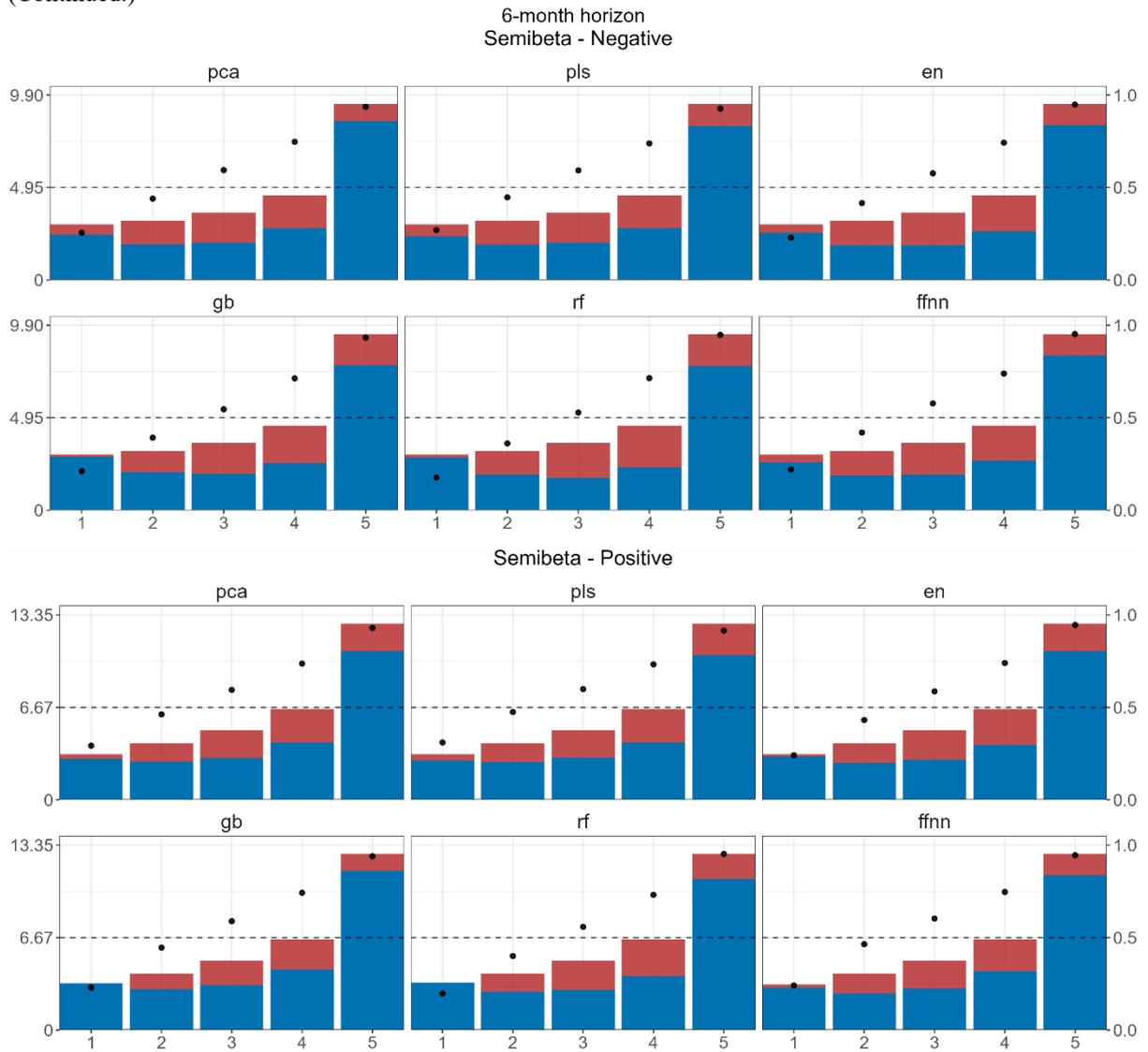

Figure A3 Average Predictive Performance of Quintile Portfolios for Individual Models: 6-Month Horizon. (Continued.)

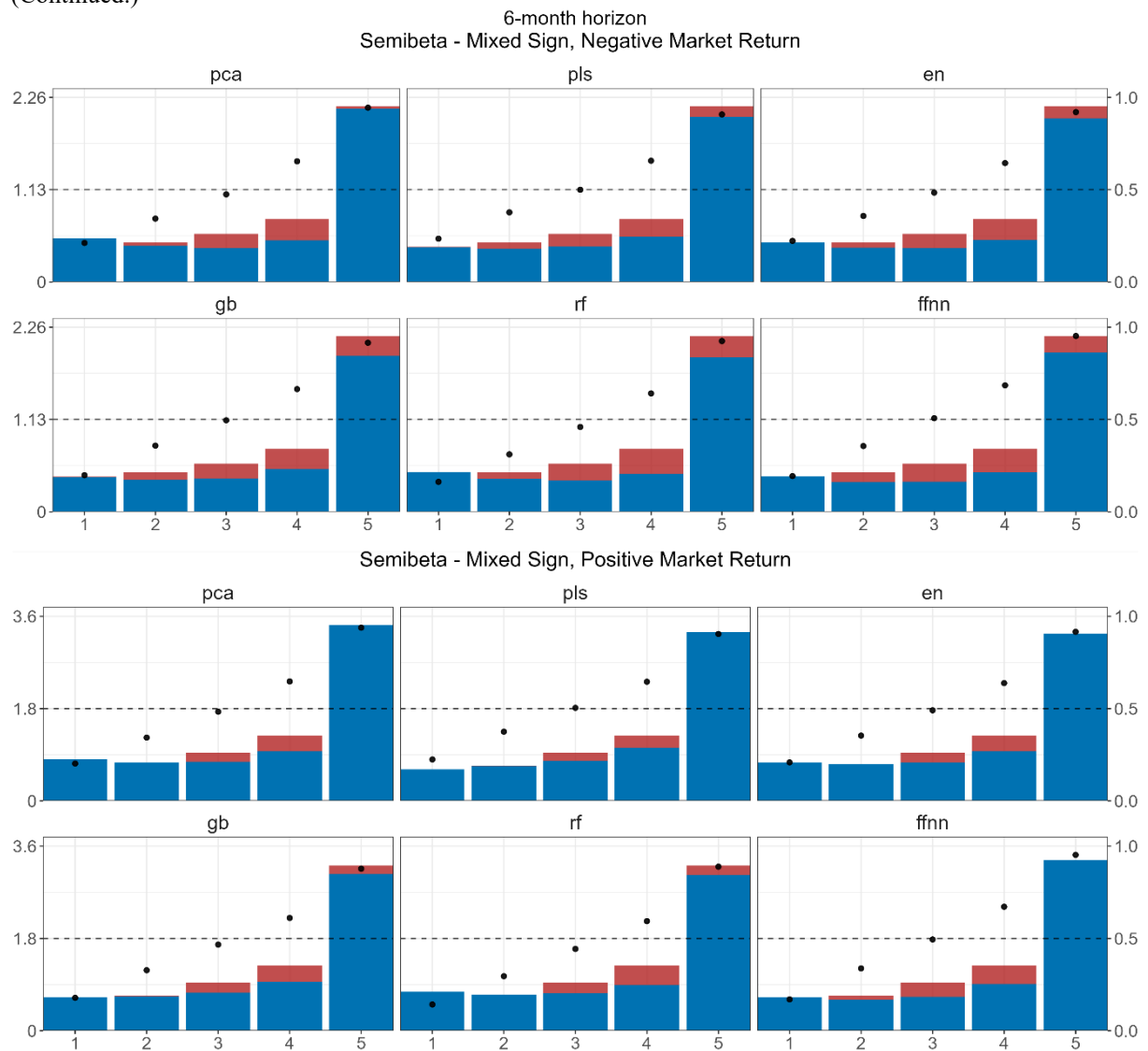

Figure A4 Average Predictive Performance of Quintile Portfolios for Individual Models: 12-Month Horizon.
This figure displays the time series averages of monthly mean squared errors (MSE) for quintile portfolios for $h = 12$. Portfolios are formed monthly by sorting stocks into quintiles according to realised beta values (1-low beta, 5-high beta). The equally weighted MSE is constructed for each quintile, for the benchmark (red bars) and the conditional beta forecasts (blue bars). The figure also reports the fraction of stocks within each portfolio for which the difference between realized and forecast portfolio betas is positive (dots, right-hand axis). The benchmark is the $h$-lagged realized beta measure for the respective asymmetric beta. The conditional beta forecasts are based on principal component analysis (pca), partial least squares (pls), elastic net (en), random forests (rf), gradient boosting (gb) and feed-forward neural network (ffnn). The out-of-sample period is from January 1990 to December 2024.

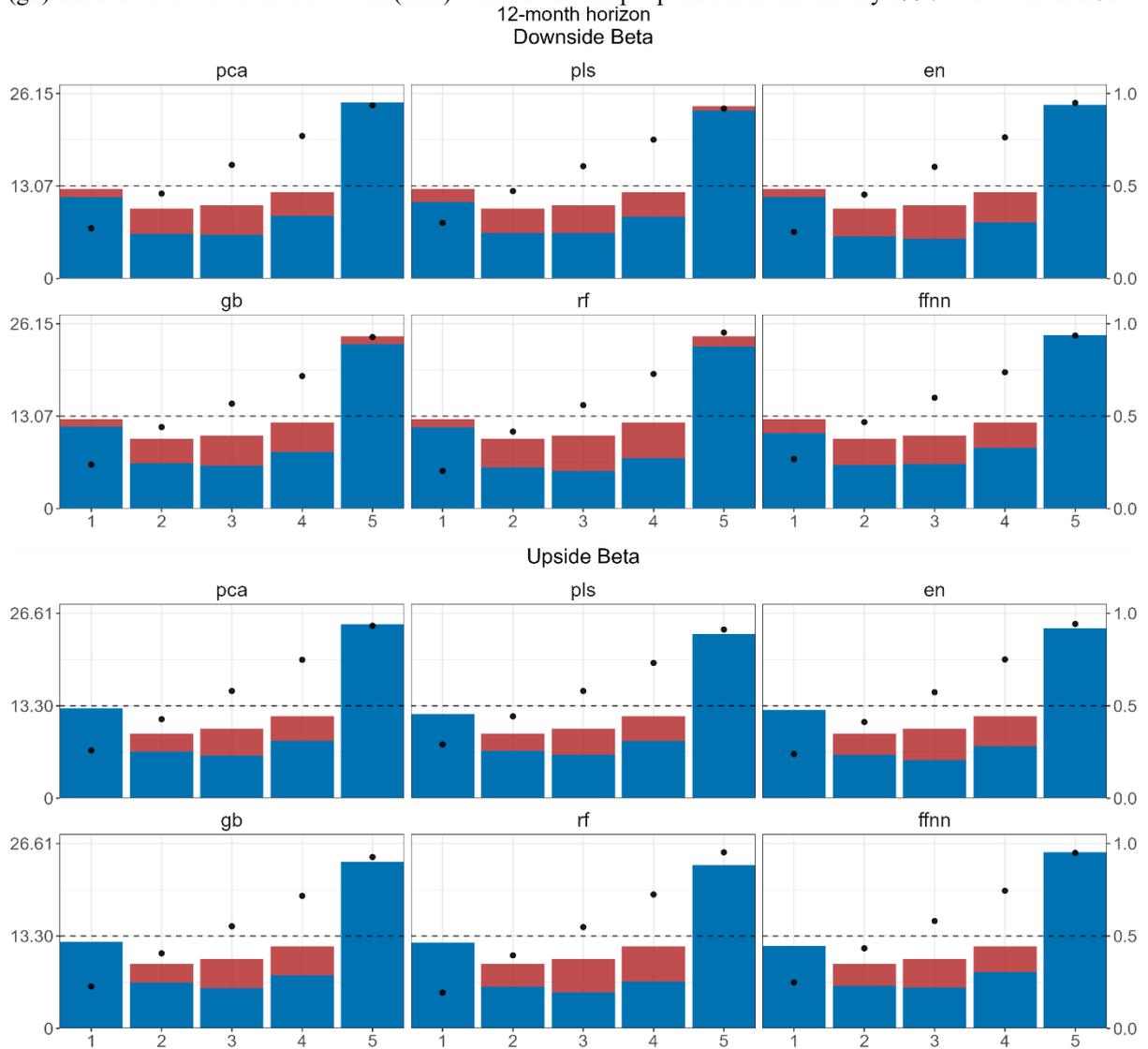

Figure A4 Average Predictive Performance of Quintile Portfolios for Individual Models: 12-Month Horizon. (Continued.)

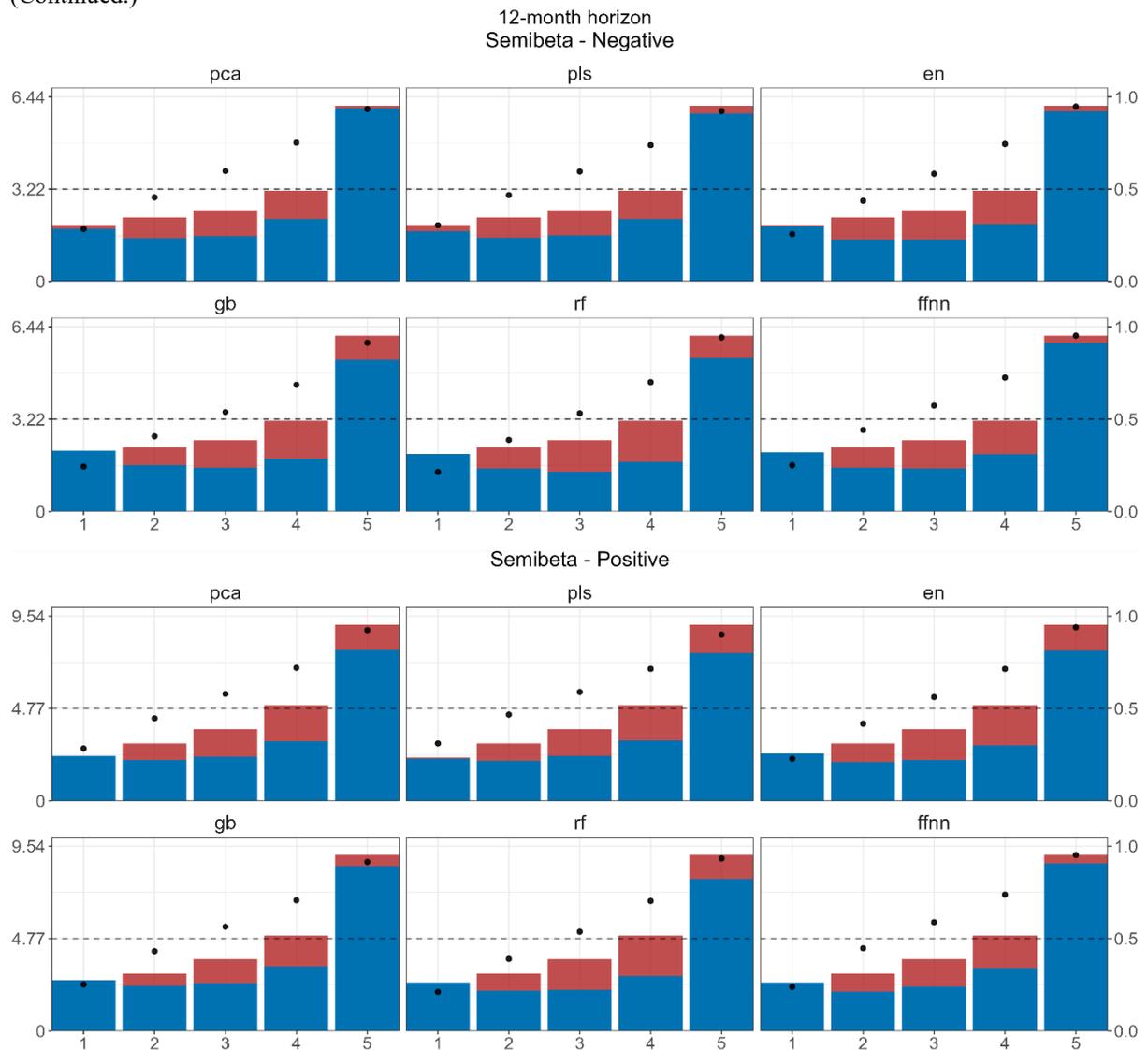

Figure A4 Average Predictive Performance of Quintile Portfolios for Individual Models: 12-Month Horizon. (Continued.)

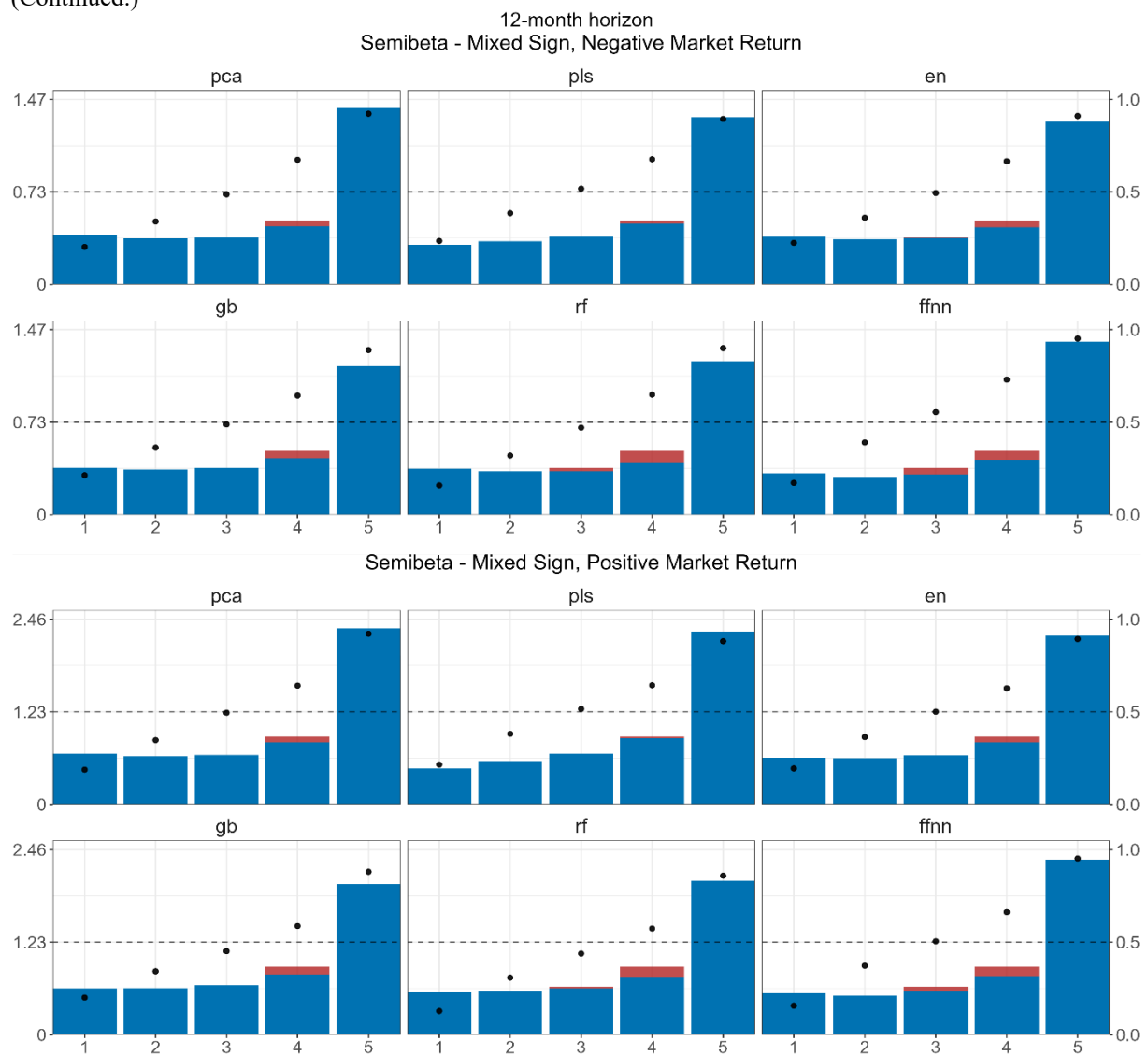

Figure A5 Average Variable Importance: 3-Month Horizon.
This figure displays the variable importance for each of the six groups of predictors. The variable importance in each period is computed according to the permutation feature importance. Variable importance is averaged throughout the out-of-sample period and is normalized for each model to sum to 100. The conditional beta forecasts are based on principal component analysis (pca), partial least squares (pls), elastic net (en), random forests (rf), gradient boosting (gb) and feed-forward neural network (ffnn). The out-of-sample period is from January 1990 to December 2024.

3-month horizon

| | Downside Beta | | | | | | Semibeta - Negative | | | | | | Semibeta - Mixed Sign, Negative Market Return | | | | | |
|---|---|---|---|---|---|---|---|---|---|---|---|---|---|---|---|---|---|---|
| | pca | pls | en | gb | rf | ffnn | pca | pls | en | gb | rf | ffnn | pca | pls | en | gb | rf | ffnn |
| Value vs Growth | 1.17 | 2.83 | 2.89 | 3.15 | 10.03 | 5.81 | 1.17 | 3.31 | 3.84 | 3.79 | 11.20 | 6.98 | 0.84 | 1.89 | 2.42 | 2.55 | 7.90 | 4.97 |
| Trading Frictions | 95.79 | 90.52 | 92.43 | 77.09 | 53.40 | 77.77 | 95.66 | 89.34 | 90.86 | 74.90 | 48.27 | 72.73 | 95.83 | 91.33 | 91.49 | 64.27 | 48.36 | 76.67 |
| Profitability | 0.13 | 0.55 | 0.39 | 1.69 | 4.00 | 2.18 | 0.18 | 0.64 | 0.42 | 1.81 | 4.50 | 2.74 | 0.15 | 0.45 | 0.41 | 2.20 | 4.11 | 2.76 |
| Momentum | 1.76 | 3.60 | 2.58 | 4.39 | 7.85 | 5.89 | 1.68 | 3.65 | 2.86 | 5.12 | 8.95 | 6.49 | 1.55 | 3.64 | 3.37 | 4.33 | 7.74 | 6.38 |
| Investment | 0.03 | 0.73 | 0.50 | 3.24 | 7.99 | 3.12 | 0.18 | 0.94 | 0.58 | 3.39 | 9.29 | 4.08 | 0.11 | 0.68 | 0.63 | 2.10 | 7.26 | 3.66 |
| Intangibles | 1.12 | 1.78 | 1.21 | 10.44 | 16.72 | 5.23 | 1.14 | 2.12 | 1.44 | 10.98 | 17.79 | 6.97 | 1.51 | 2.01 | 1.68 | 24.54 | 24.62 | 5.56 |

| | Upside Beta | | | | | | Semibeta - Positive | | | | | | Semibeta - Mixed Sign, Positive Market Return | | | | | |
|---|---|---|---|---|---|---|---|---|---|---|---|---|---|---|---|---|---|---|
| | pca | pls | en | gb | rf | ffnn | pca | pls | en | gb | rf | ffnn | pca | pls | en | gb | rf | ffnn |
| Value vs Growth | 1.32 | 2.76 | 2.24 | 3.01 | 11.13 | 6.18 | 0.88 | 1.46 | 1.59 | 2.57 | 8.88 | 4.63 | 5.19 | 4.73 | 6.83 | 5.27 | 11.21 | 8.94 |
| Trading Frictions | 95.64 | 89.49 | 92.91 | 76.58 | 47.95 | 73.00 | 95.41 | 90.68 | 92.90 | 69.52 | 49.16 | 75.87 | 90.35 | 85.96 | 84.78 | 50.25 | 35.91 | 62.53 |
| Profitability | 0.09 | 0.53 | 0.36 | 2.06 | 4.60 | 2.84 | 0.04 | 0.45 | 0.35 | 2.64 | 4.33 | 2.61 | -0.38 | 0.59 | 0.65 | 3.29 | 5.17 | 5.43 |
| Momentum | 1.93 | 4.60 | 3.03 | 4.49 | 9.32 | 7.59 | 2.34 | 5.23 | 3.45 | 4.83 | 8.34 | 8.46 | 0.90 | 2.89 | 2.45 | 4.54 | 8.26 | 7.50 |
| Investment | -0.10 | 0.77 | 0.44 | 2.75 | 8.87 | 4.22 | -0.02 | 0.60 | 0.54 | 1.94 | 7.40 | 3.65 | 0.26 | 1.81 | 1.83 | 3.30 | 10.08 | 6.52 |
| Intangibles | 1.12 | 1.84 | 1.03 | 11.11 | 18.13 | 6.18 | 1.35 | 1.58 | 1.17 | 18.50 | 21.89 | 4.79 | 3.69 | 4.02 | 3.46 | 33.35 | 29.39 | 9.08 |

Figure A6 Average Variable Importance: 6-Month Horizon.
This figure displays the variable importance for each of the six groups of predictors. The variable importance in each period is computed according to the permutation feature importance. Variable importance is averaged throughout the out-of-sample period and is normalized for each model to sum to 100. The conditional beta forecasts are based on principal component analysis (pca), partial least squares (pls), elastic net (en), random forests (rf), gradient boosting (gb) and feed-forward neural network (ffnn). The out-of-sample period is from January 1990 to December 2024.

6-month horizon

|  | Downside Beta | | | | | | Semibeta - Negative | | | | | | Semibeta - Mixed Sign, Negative Market Return | | | | | |
| --- | --- | --- | --- | --- | --- | --- | --- | --- | --- | --- | --- | --- | --- | --- | --- | --- | --- | --- |
| Value vs Growth | 1.15 | 2.63 | 3.10 | 4.32 | 9.64 | 6.36 | 1.11 | 3.01 | 3.83 | 4.35 | 10.54 | 7.37 | 0.93 | 2.23 | 2.40 | 3.12 | 8.03 | 5.19 |
| Trading Frictions | 95.78 | 90.99 | 91.90 | 72.49 | 56.15 | 75.95 | 95.91 | 90.19 | 90.94 | 72.07 | 52.21 | 72.47 | 95.54 | 90.59 | 91.89 | 68.42 | 52.10 | 77.29 |
| Profitability | 0.13 | 0.54 | 0.44 | 1.97 | 3.66 | 2.46 | 0.17 | 0.66 | 0.54 | 2.19 | 4.22 | 2.85 | 0.19 | 0.52 | 0.53 | 1.91 | 3.96 | 2.70 |
| Momentum | 1.94 | 3.31 | 2.65 | 4.16 | 6.23 | 5.89 | 1.79 | 3.01 | 2.45 | 4.05 | 6.77 | 5.82 | 1.75 | 3.46 | 2.56 | 3.04 | 5.57 | 5.69 |
| Investment | -0.02 | 0.73 | 0.63 | 3.69 | 7.75 | 3.48 | 0.09 | 0.93 | 0.72 | 4.17 | 8.79 | 4.24 | 0.14 | 0.89 | 0.74 | 1.86 | 7.07 | 3.68 |
| Intangibles | 1.01 | 1.81 | 1.29 | 13.38 | 16.56 | 5.87 | 0.93 | 2.19 | 1.52 | 13.17 | 17.48 | 7.26 | 1.46 | 2.31 | 1.88 | 21.66 | 23.26 | 5.45 |

|  | Upside Beta | | | | | | Semibeta - Positive | | | | | | Semibeta - Mixed Sign, Positive Market Return | | | | | |
| --- | --- | --- | --- | --- | --- | --- | --- | --- | --- | --- | --- | --- | --- | --- | --- | --- | --- | --- |
| Value vs Growth | 1.26 | 2.35 | 2.49 | 4.66 | 10.79 | 6.41 | 0.88 | 1.62 | 1.99 | 2.94 | 9.00 | 4.80 | 4.98 | 4.46 | 7.03 | 5.80 | 11.04 | 8.59 |
| Trading Frictions | 95.54 | 90.44 | 92.03 | 71.62 | 51.20 | 73.15 | 95.17 | 90.55 | 91.87 | 71.93 | 53.15 | 76.11 | 92.08 | 86.53 | 85.72 | 49.05 | 37.70 | 64.80 |
| Profitability | 0.12 | 0.54 | 0.49 | 1.96 | 4.12 | 2.72 | 0.05 | 0.52 | 0.49 | 1.84 | 3.84 | 2.59 | -0.57 | 0.59 | 0.59 | 3.02 | 4.77 | 5.16 |
| Momentum | 2.15 | 4.07 | 3.12 | 4.32 | 7.25 | 7.19 | 2.57 | 4.94 | 3.57 | 4.08 | 6.49 | 7.99 | 0.45 | 2.33 | 1.85 | 3.19 | 6.14 | 6.64 |
| Investment | -0.14 | 0.74 | 0.56 | 3.83 | 8.72 | 4.23 | 0.01 | 0.59 | 0.61 | 2.62 | 7.69 | 3.54 | -0.46 | 1.67 | 1.76 | 4.32 | 10.14 | 6.09 |
| Intangibles | 1.08 | 1.87 | 1.31 | 13.61 | 17.92 | 6.31 | 1.33 | 1.77 | 1.46 | 16.57 | 19.83 | 4.97 | 3.52 | 4.42 | 3.04 | 34.63 | 30.20 | 8.72 |
|  | pca | pls | en | gb | rf | ffnn | pca | pls | en | gb | rf | ffnn | pca | pls | en | gb | rf | ffnn |

Figure A7 Average Variable Importance: 12-Month Horizon.

This figure displays the variable importance for each of the six groups of predictors. The variable importance in each period is computed according to the permutation feature importance. Variable importance is averaged throughout the out-of-sample period and is normalized for each model to sum to 100. The conditional beta forecasts are based on principal component analysis (pca), partial least squares (pls), elastic net (en), random forests (rf), gradient boosting (gb) and feed-forward neural network (ffnn). The out-of-sample period is from January 1990 to December 2024.

### 12-month horizon

| | Downside Beta | | | | | | Semibeta - Negative | | | | | | Semibeta - Mixed Sign, Negative Market Return | | | | | |
|---|---|---|---|---|---|---|---|---|---|---|---|---|---|---|---|---|---|---|
| | pca | pls | en | gb | rf | ffnn | pca | pls | en | gb | rf | ffnn | pca | pls | en | gb | rf | ffnn |
| Value vs Growth | 1.28 | 2.68 | 3.58 | 5.77 | 9.46 | 7.58 | 1.25 | 2.91 | 4.76 | 7.10 | 10.43 | 8.61 | 1.01 | 2.87 | 3.40 | 4.87 | 8.80 | 6.21 |
| Trading Frictions | 95.29 | 90.19 | 90.84 | 66.60 | 58.73 | 72.19 | 95.37 | 89.45 | 88.53 | 60.78 | 55.40 | 67.83 | 95.66 | 88.77 | 90.09 | 63.15 | 53.17 | 72.96 |
| Profitability | 0.16 | 0.66 | 0.64 | 2.48 | 3.34 | 2.85 | 0.15 | 0.78 | 0.81 | 3.25 | 3.85 | 3.43 | 0.15 | 0.65 | 0.61 | 2.83 | 3.89 | 3.36 |
| Momentum | 2.20 | 3.42 | 2.53 | 3.56 | 4.66 | 5.94 | 2.08 | 3.15 | 2.80 | 3.76 | 4.89 | 5.91 | 2.01 | 3.52 | 2.56 | 2.73 | 3.99 | 6.05 |
| Investment | 0.00 | 0.76 | 0.75 | 5.14 | 7.38 | 4.19 | 0.12 | 0.98 | 1.00 | 6.60 | 8.19 | 5.12 | -0.48 | 0.93 | 0.93 | 3.16 | 7.01 | 4.39 |
| Intangibles | 1.07 | 2.29 | 1.67 | 16.45 | 16.44 | 7.26 | 1.03 | 2.72 | 2.10 | 18.51 | 17.23 | 9.10 | 1.66 | 3.26 | 2.41 | 23.26 | 23.14 | 7.03 |

| | Upside Beta | | | | | | Semibeta - Positive | | | | | | Semibeta - Mixed Sign, Positive Market Return | | | | | |
|---|---|---|---|---|---|---|---|---|---|---|---|---|---|---|---|---|---|---|
| | pca | pls | en | gb | rf | ffnn | pca | pls | en | gb | rf | ffnn | pca | pls | en | gb | rf | ffnn |
| Value vs Growth | 1.36 | 2.53 | 2.90 | 6.18 | 10.01 | 7.63 | 1.06 | 2.13 | 2.36 | 4.91 | 9.23 | 5.61 | 4.91 | 4.65 | 7.89 | 6.47 | 10.90 | 9.46 |
| Trading Frictions | 95.23 | 89.81 | 91.29 | 64.36 | 55.84 | 70.52 | 94.61 | 89.41 | 90.82 | 67.51 | 56.45 | 73.90 | 90.69 | 85.30 | 82.53 | 46.91 | 38.80 | 61.06 |
| Profitability | 0.15 | 0.71 | 0.63 | 2.72 | 3.63 | 3.13 | 0.10 | 0.72 | 0.56 | 2.51 | 3.83 | 3.05 | -0.19 | 0.61 | 0.68 | 3.66 | 4.99 | 5.76 |
| Momentum | 2.30 | 4.03 | 3.00 | 3.97 | 5.23 | 6.71 | 2.80 | 4.82 | 3.76 | 4.07 | 5.04 | 7.45 | 0.70 | 2.17 | 1.78 | 1.95 | 4.06 | 7.18 |
| Investment | -0.15 | 0.71 | 0.62 | 5.65 | 8.05 | 4.68 | -0.01 | 0.71 | 0.74 | 4.32 | 7.65 | 3.95 | 0.24 | 1.82 | 2.63 | 5.42 | 9.44 | 7.10 |
| Intangibles | 1.10 | 2.21 | 1.55 | 17.13 | 17.24 | 7.34 | 1.45 | 2.21 | 1.77 | 16.67 | 17.81 | 6.04 | 3.66 | 5.45 | 4.48 | 35.59 | 31.80 | 9.44 |

Figure A8 Cumulative Difference of Forecast Errors Over Time of CAPM Beta for Individual Models: 1-Month Horizon.

This figure displays the cumulative differences in squared forecast errors of CAPM beta for the benchmark relative to the conditional beta forecasting models for $h = 1$. The benchmark is the $h$-lagged realized CAPM beta. The conditional beta forecasts are based on principal component analysis (pca), partial least squares (pls), elastic net (en), random forests (rf), gradient boosting (gb) and feed-forward neural network (ffnn). Estimates of CAPM beta are derived as a combination of forecasts of downside and upside betas (green line), and as a combination of forecasts of four semibetas (blue line). The shaded regions depict NBER-dated recessions. Higher values indicate improved predictive performance from the benchmark. The out-of-sample period is from January 1990 to December 2024.

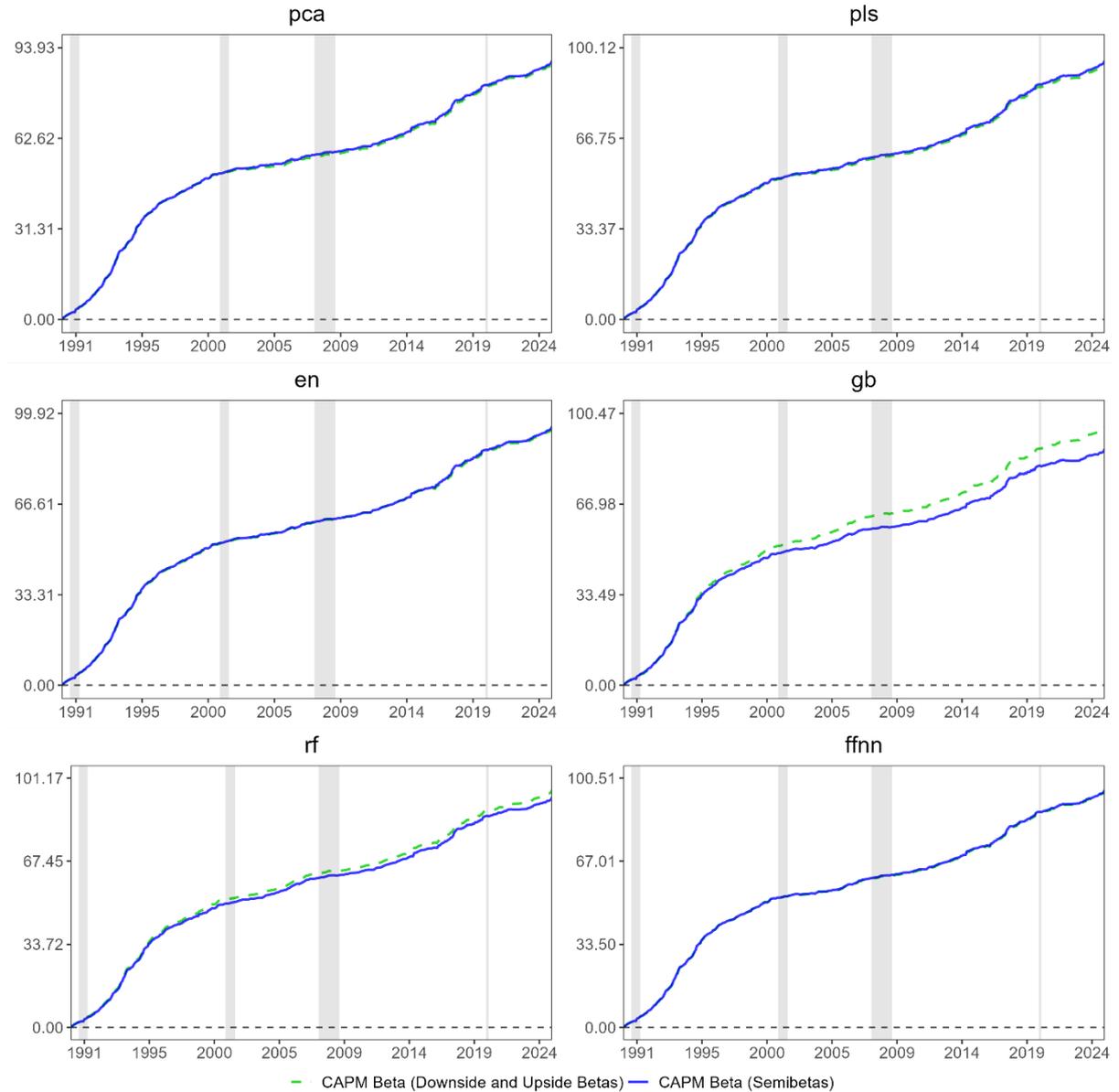

Figure A9 Cumulative Difference of Forecast Errors Over Time of CAPM Beta for Individual Models: 3-Month Horizon.
This figure displays the cumulative differences in squared forecast errors of CAPM beta for the benchmark relative to the conditional beta forecasting models for $h = 3$. The benchmark is the $h$-lagged realized CAPM beta. The conditional beta forecasts are based on principal component analysis (pca), partial least squares (pls), elastic net (en), random forests (rf), gradient boosting (gb) and feed-forward neural network (ffnn). Estimates of CAPM beta are derived as a combination of forecasts of downside and upside betas (green line), and as a combination of forecasts of four semibetas (blue line). The shaded regions depict NBER-dated recessions. Higher values indicate improved predictive performance from the benchmark. The out-of-sample period is from January 1990 to December 2024.

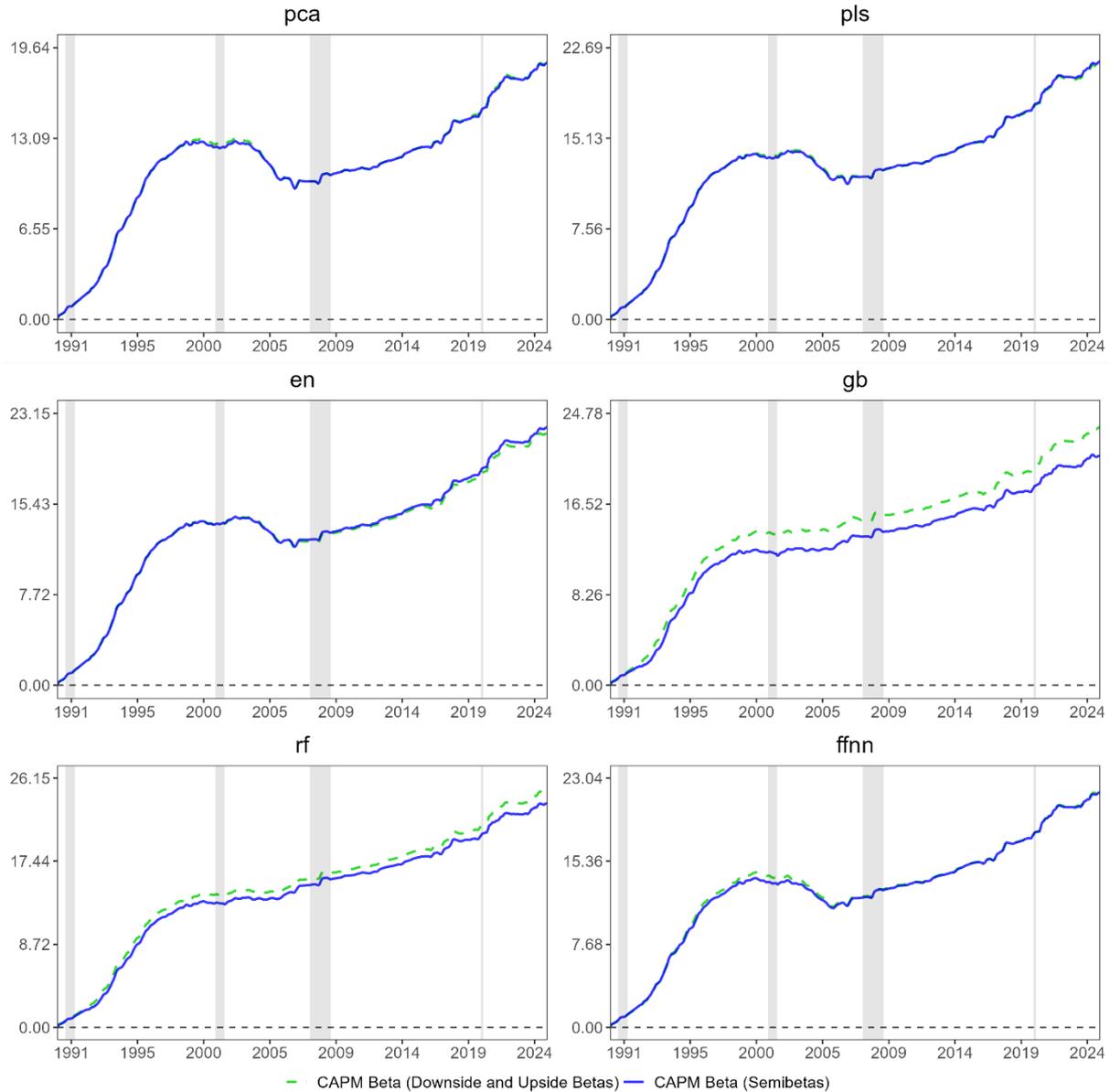

Figure A10 Cumulative Difference of Forecast Errors Over Time of CAPM Beta for Individual Models: 6-Month Horizon.
This figure displays the cumulative differences in squared forecast errors of CAPM beta for the benchmark relative to the conditional beta forecasting models for $h = 6$. The benchmark is the $h$-lagged realized CAPM beta. The conditional beta forecasts are based on principal component analysis (pca), partial least squares (pls), elastic net (en), random forests (rf), gradient boosting (gb) and feed-forward neural network (ffnn). Estimates of CAPM beta are derived as a combination of forecasts of downside and upside betas (green line), and as a combination of forecasts of four semibetas (blue line). The shaded regions depict NBER-dated recessions. Higher values indicate improved predictive performance from the benchmark. The out-of-sample period is from January 1990 to December 2024.

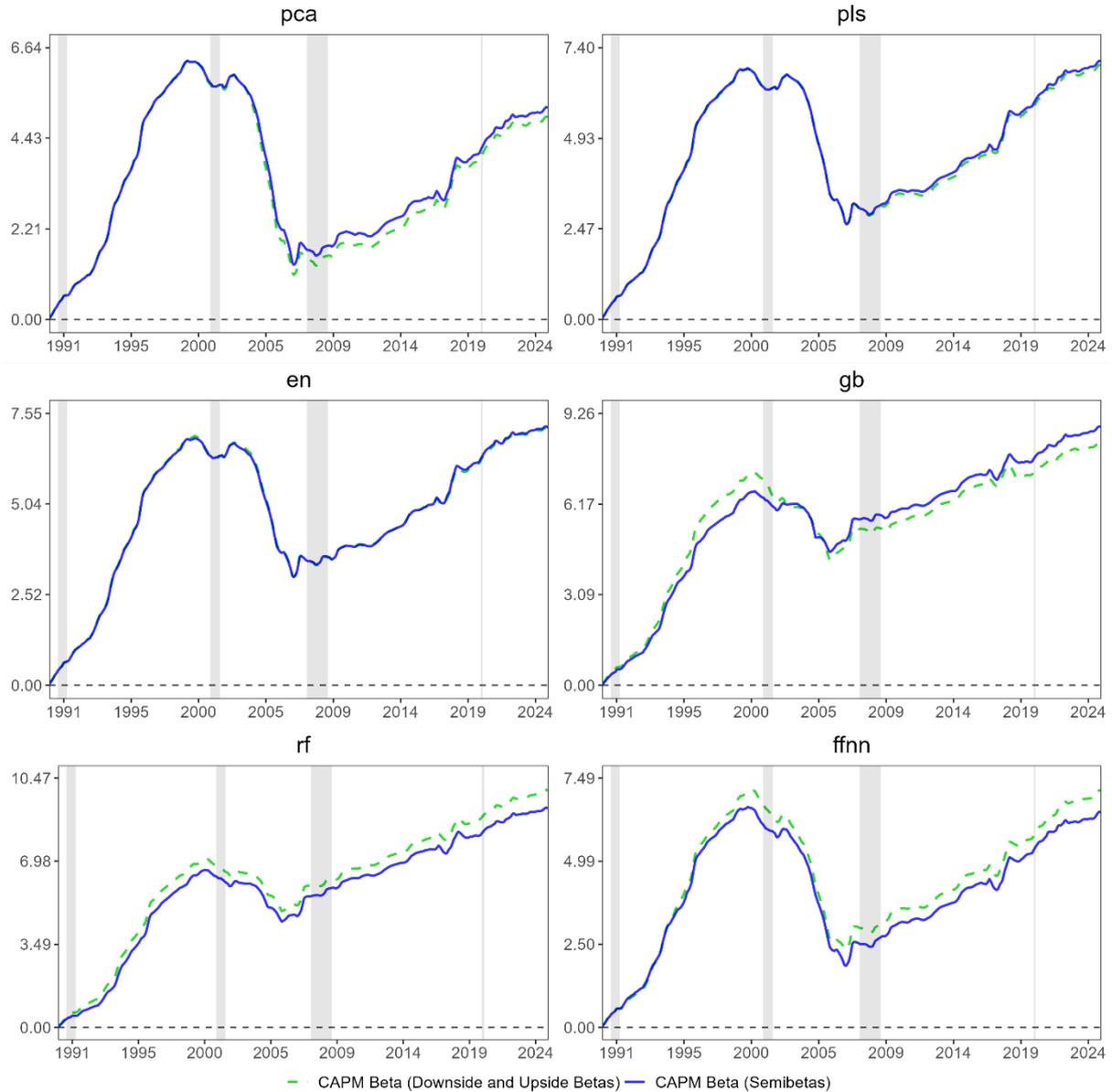

Figure A11 Cumulative Difference of Forecast Errors Over Time of CAPM Beta for Individual Models: 12-Month Horizon.
This figure displays the cumulative differences in squared forecast errors of CAPM beta for the benchmark relative to the conditional beta forecasting models for $h = 12$. The benchmark is the $h$-lagged realized CAPM beta. The conditional beta forecasts are based on principal component analysis (pca), partial least squares (pls), elastic net (en), random forests (rf), gradient boosting (gb) and feed-forward neural network (ffnn). Estimates of CAPM beta are derived as a combination of forecasts of downside and upside betas (green line), and as a combination of forecasts of four semibetas (blue line). The shaded regions depict NBER-dated recessions. Higher values indicate improved predictive performance from the benchmark. The out-of-sample period is from January 1990 to December 2024.

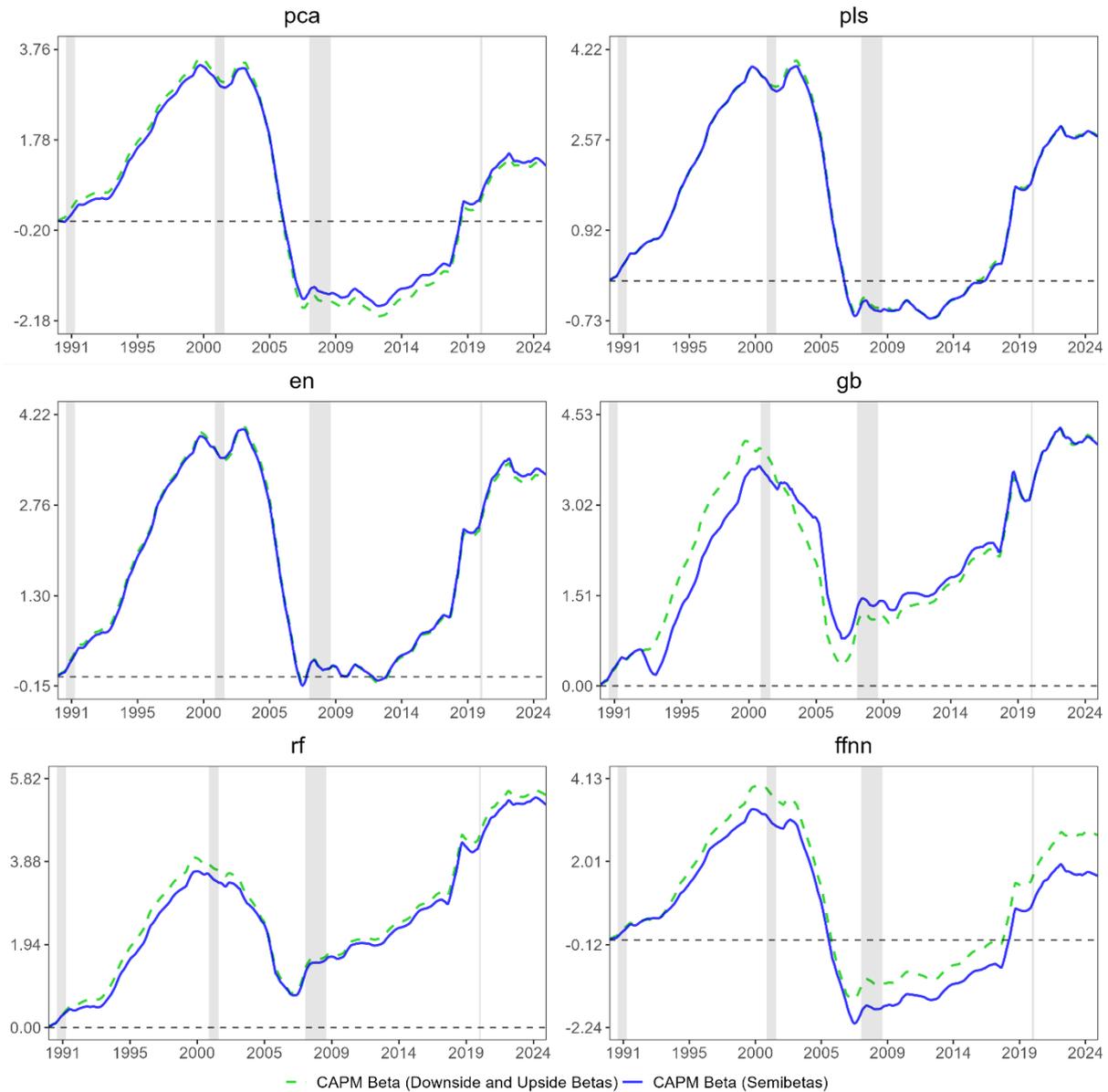

Figure A12 Average Predictive Performance of Quintile Portfolios of CAPM Beta for Individual Models: 1-Month Horizon.
This figure displays the time series averages of monthly mean squared errors (MSE) for quintile portfolios for $h = 1$. Portfolios are formed monthly by sorting stocks into quintiles according to realised beta values (1-low beta, 5-high beta). The equally weighted MSE is constructed for each quintile, for the benchmark (red bars) and the conditional beta forecasts (blue bars). The figure also reports the fraction of stocks within each portfolio for which the difference between realized and forecast portfolio betas is positive (dots, right-hand axis). The benchmark is the $h$-lagged realized CAPM beta. The conditional beta forecasts are based on principal component analysis (pca), partial least squares (pls), elastic net (en), random forests (rf), gradient boosting (gb) and feed-forward neural network (ffnn). Estimates of CAPM beta are derived as a combination of forecasts of downside and upside betas, and as a combination of forecasts of four semibetas. The out-of-sample period is from January 1990 to December 2024.

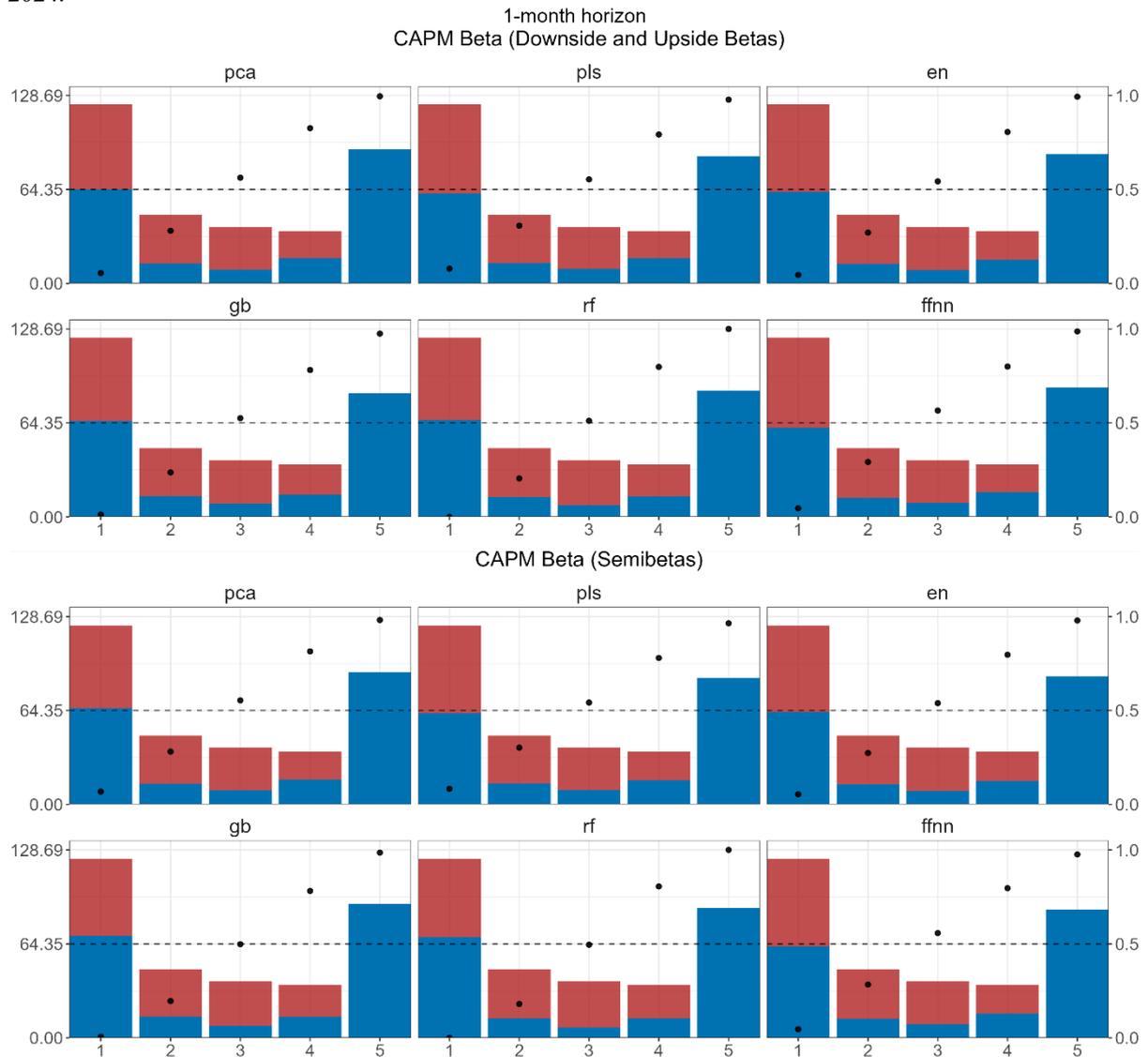

Figure A13 Average Predictive Performance of Quintile Portfolios of CAPM Beta for Individual Models: 3-Month Horizon.
This figure displays the time series averages of monthly mean squared errors (MSE) for quintile portfolios for $h = 3$. Portfolios are formed monthly by sorting stocks into quintiles according to realised beta values (1-low beta, 5-high beta). The equally weighted MSE is constructed for each quintile, for the benchmark (red bars) and the conditional beta forecasts (blue bars). The figure also reports the fraction of stocks within each portfolio for which the difference between realized and forecast portfolio betas is positive (dots, right-hand axis). The benchmark is the $h$-lagged realized CAPM beta. The conditional beta forecasts are based on principal component analysis (pca), partial least squares (pls), elastic net (en), random forests (rf), gradient boosting (gb) and feed-forward neural network (ffnn). Estimates of CAPM beta are derived as a combination of forecasts of downside and upside betas, and as a combination of forecasts of four semibetas. The out-of-sample period is from January 1990 to December 2024.

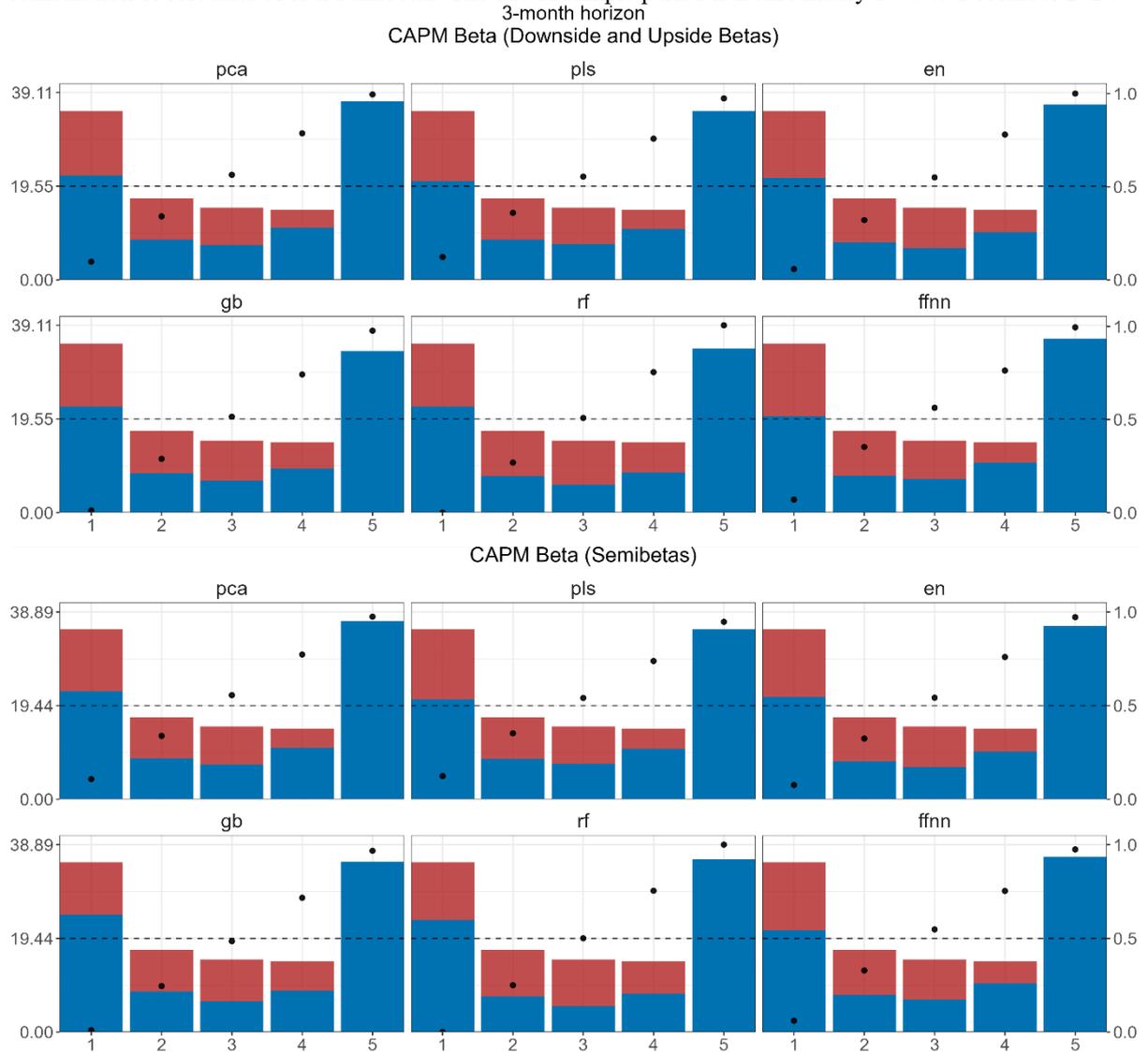

Figure A14 Average Predictive Performance of Quintile Portfolios of CAPM Beta for Individual Models: 6-Month Horizon.
This figure displays the time series averages of monthly mean squared errors (MSE) for quintile portfolios for $h = 6$. Portfolios are formed monthly by sorting stocks into quintiles according to realised beta values (1-low beta, 5-high beta). The equally weighted MSE is constructed for each quintile, for the benchmark (red bars) and the conditional beta forecasts (blue bars). The figure also reports the fraction of stocks within each portfolio for which the difference between realized and forecast portfolio betas is positive (dots, right-hand axis). The benchmark is the $h$-lagged realized CAPM beta. The conditional beta forecasts are based on principal component analysis (pca), partial least squares (pls), elastic net (en), random forests (rf), gradient boosting (gb) and feed-forward neural network (ffnn). Estimates of CAPM beta are derived as a combination of forecasts of downside and upside betas, and as a combination of forecasts of four semibetas. The out-of-sample period is from January 1990 to December 2024.

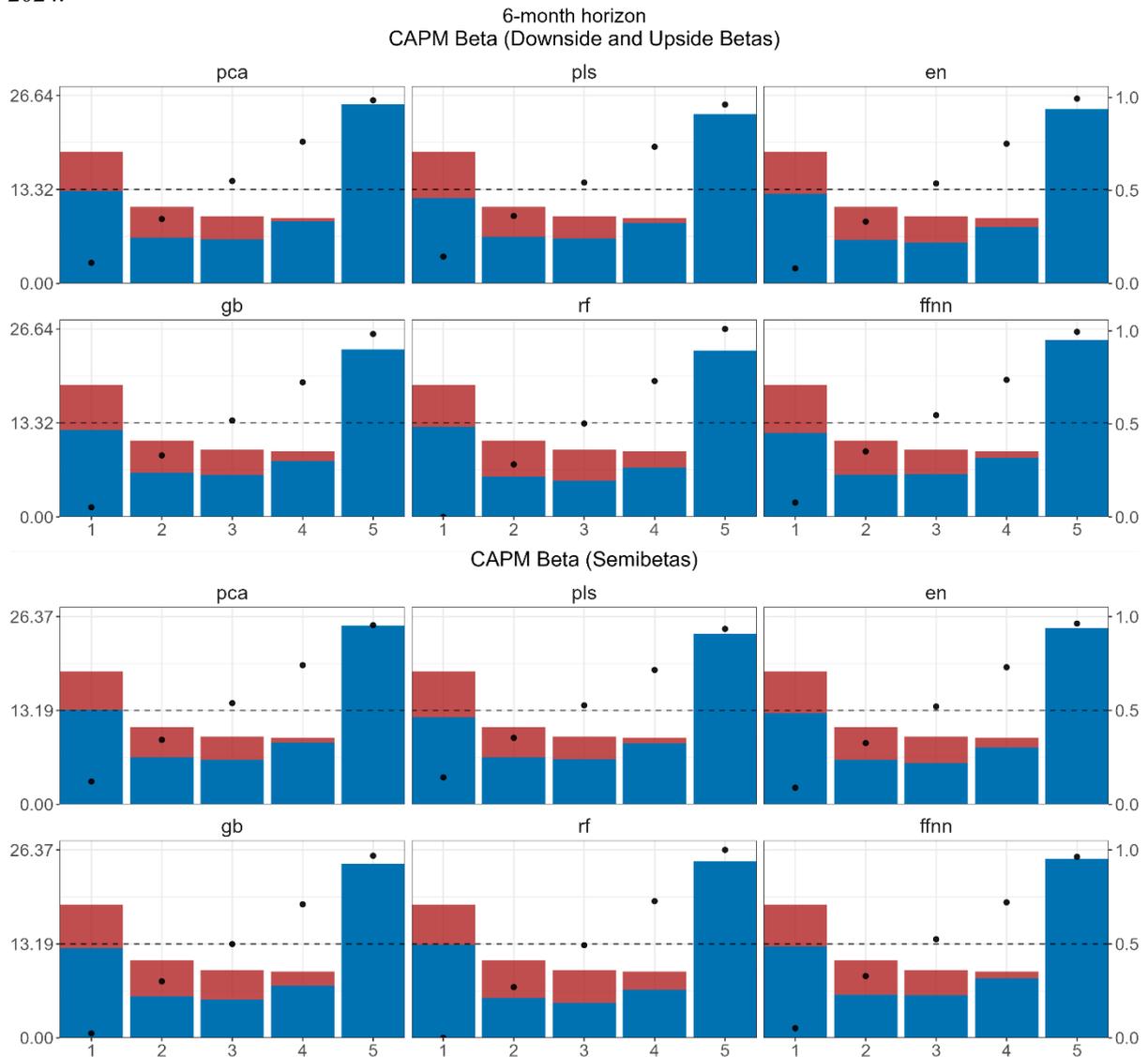

Figure A15 Average Predictive Performance of Quintile Portfolios of CAPM Beta for Individual Models: 12-Month Horizon.

This figure displays the time series averages of monthly mean squared errors (MSE) for quintile portfolios for $h = 12$. Portfolios are formed monthly by sorting stocks into quintiles according to realised beta values (1-low beta, 5-high beta). The equally weighted MSE is constructed for each quintile, for the benchmark (red bars) and the conditional beta forecasts (blue bars). The figure also reports the fraction of stocks within each portfolio for which the difference between realized and forecast portfolio betas is positive (dots, right-hand axis). The benchmark is the $h$-lagged realized CAPM beta. The conditional beta forecasts are based on principal component analysis (pca), partial least squares (pls), elastic net (en), random forests (rf), gradient boosting (gb) and feed-forward neural network (ffnn). Estimates of CAPM beta are derived as a combination of forecasts of downside and upside betas, and as a combination of forecasts of four semibetas. The out-of-sample period is from January 1990 to December 2024.

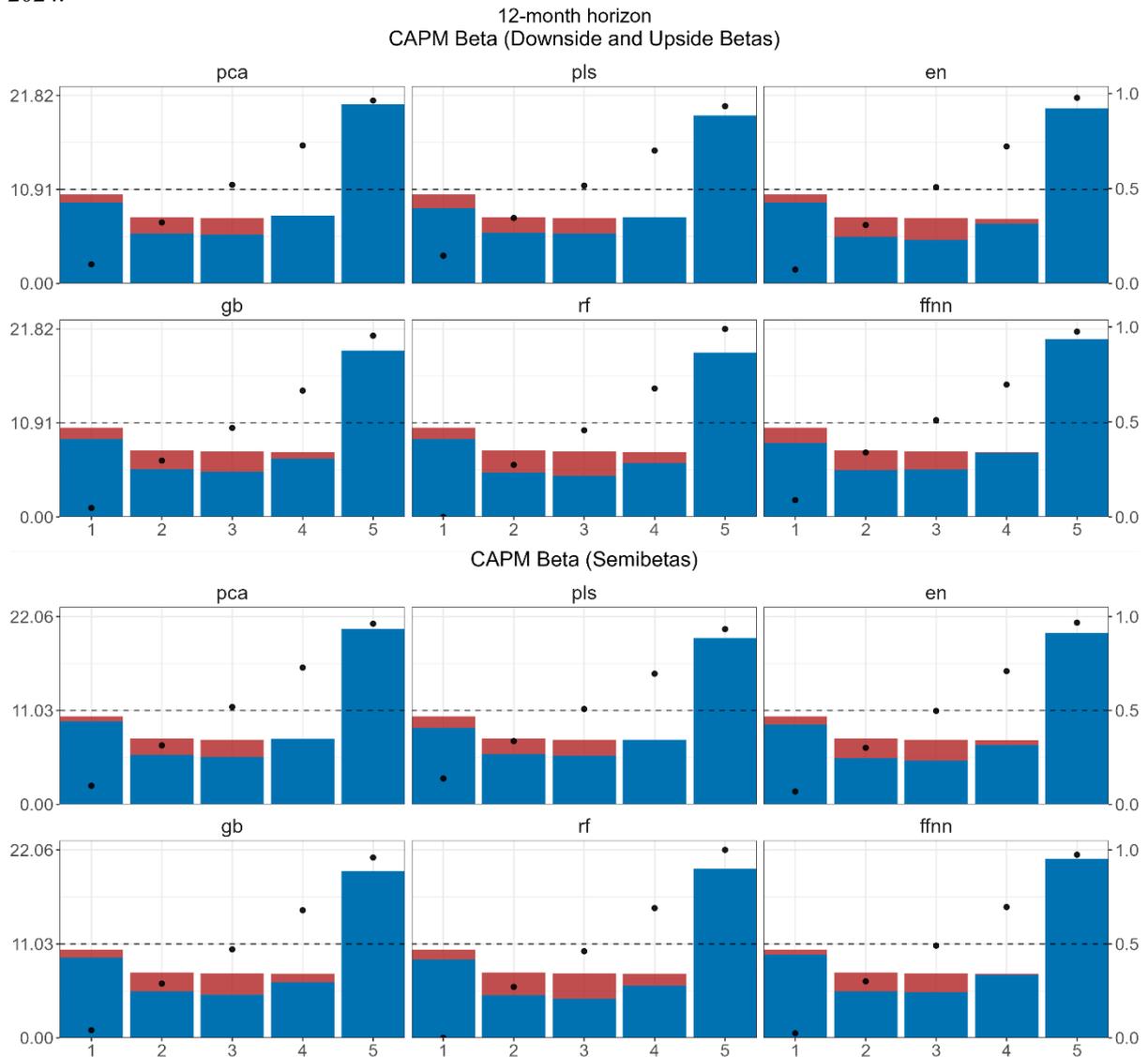

Figure A16 Market-Neutral Minimum-Variance Portfolio Based on Individual Model Forecasts: 1-Month Horizon. This figure displays the distribution of monthly ex-post realized betas for market-neutral minimum-variance portfolios for $h = 1$. Portfolios are constructed based on the $h$-lagged historical CAPM beta benchmark (grey), and conditional CAPM betas reconstructed from forecasts of downside and upside betas (green), and from forecasts of the four semibetas (blue). Dashed vertical lines and annotations indicate the modes of each distribution. The conditional beta forecasts are based on principal component analysis (pca), partial least squares (pls), elastic net (en), random forests (rf), gradient boosting (gb) and feed-forward neural network (ffnn). The out-of-sample period is from January 1990 to December 2024.

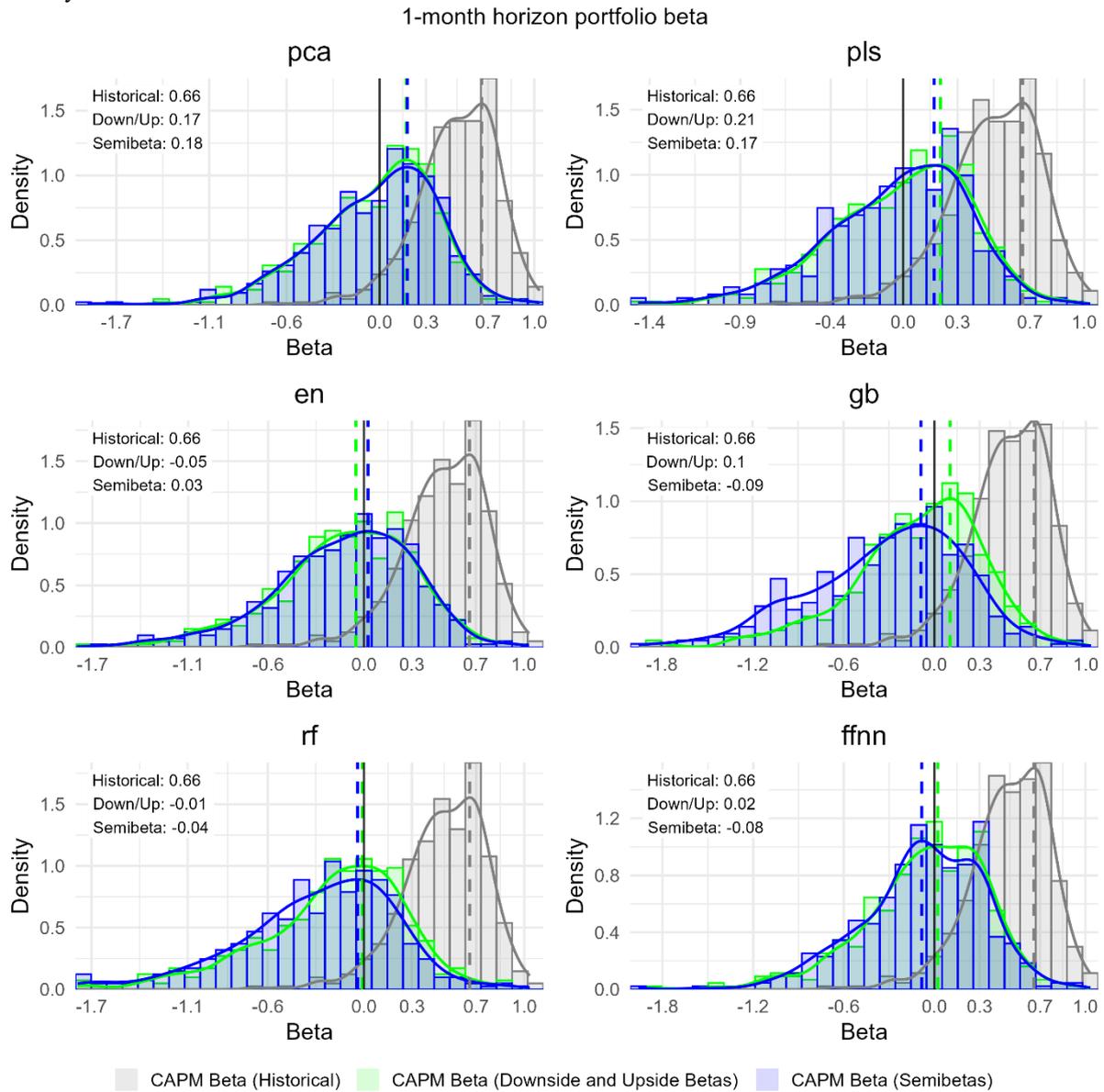

Figure A17 Market-Neutral Minimum-Variance Portfolio Based on Individual Model Forecasts: 3-Month Horizon. This figure displays the distribution of monthly ex-post realized betas for market-neutral minimum-variance portfolios for $h = 3$. Portfolios are constructed based on the $h$-lagged historical CAPM beta benchmark (grey), and conditional CAPM betas reconstructed from forecasts of downside and upside betas (green), and from forecasts of the four semibetas (blue). Dashed vertical lines and annotations indicate the modes of each distribution. The conditional beta forecasts are based on principal component analysis (pca), partial least squares (pls), elastic net (en), random forests (rf), gradient boosting (gb) and feed-forward neural network (ffnn). The out-of-sample period is from January 1990 to December 2024.

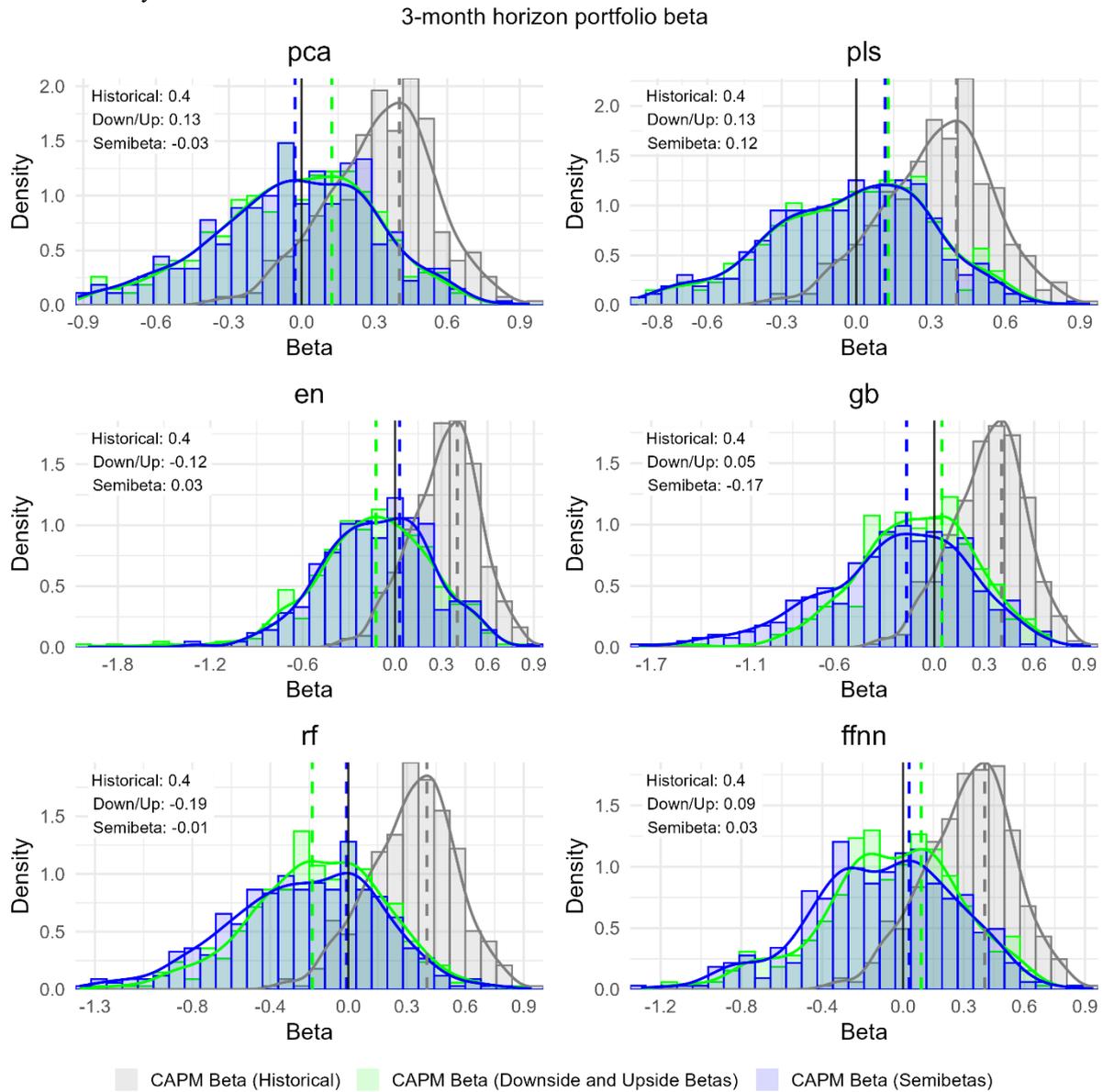

Figure A18 Market-Neutral Minimum-Variance Portfolio based on Individual Model Forecasts: 6-Month Horizon. This figure displays the distribution of monthly ex-post realized betas for market-neutral minimum-variance portfolios for $h = 6$. Portfolios are constructed based on the $h$-lagged historical CAPM beta benchmark (grey), and conditional CAPM betas reconstructed from forecasts of downside and upside betas (green), and from forecasts of the four semibetas (blue). Dashed vertical lines and annotations indicate the modes of each distribution. The conditional beta forecasts are based on principal component analysis (pca), partial least squares (pls), elastic net (en), random forests (rf), gradient boosting (gb) and feed-forward neural network (ffnn). The out-of-sample period is from January 1990 to December 2024.

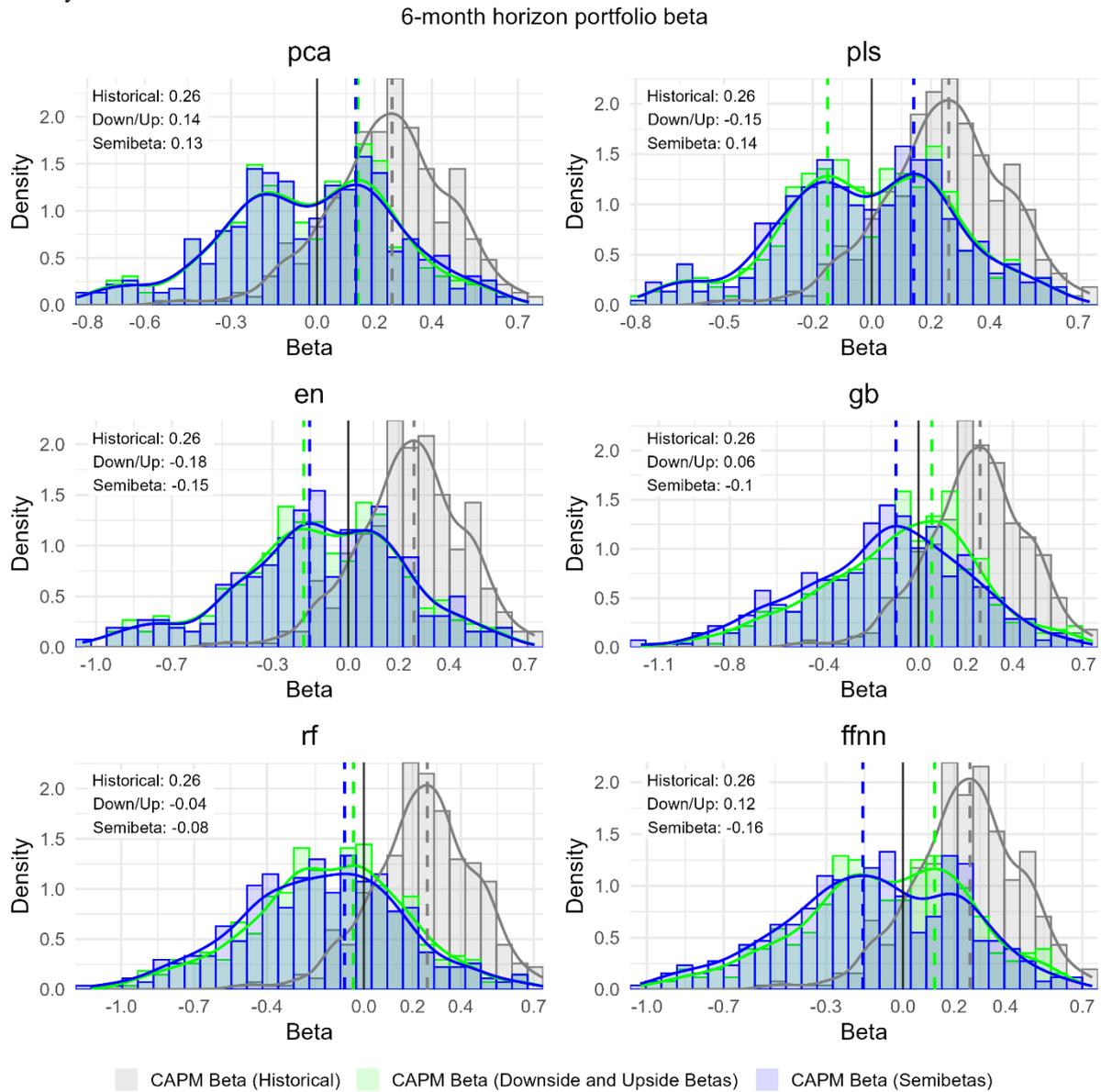

Figure A19 Market-Neutral Minimum-Variance Portfolio based on Individual Model Forecasts: 12-Month Horizon. This figure displays the distribution of monthly ex-post realized betas for market-neutral minimum-variance portfolios for $h = 12$. Portfolios are constructed based on the $h$-lagged historical CAPM beta benchmark (grey), and conditional CAPM betas reconstructed from forecasts of downside and upside betas (green), and from forecasts of the four semibetas (blue). Dashed vertical lines and annotations indicate the modes of each distribution. The conditional beta forecasts are based on principal component analysis (pca), partial least squares (pls), elastic net (en), random forests (rf), gradient boosting (gb) and feed-forward neural network (ffnn). The out-of-sample period is from January 1990 to December 2024.

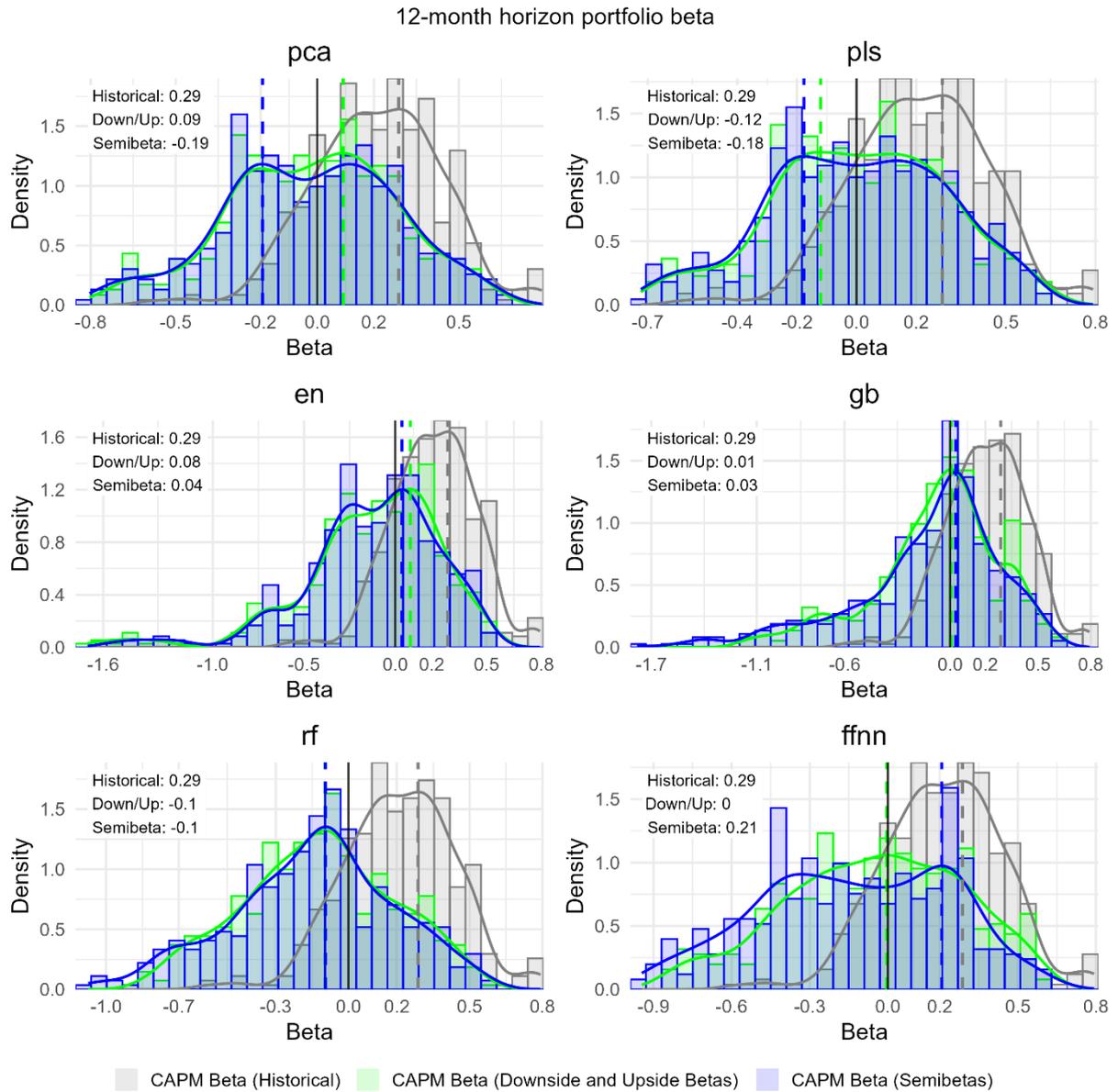

# Appendix 2: Information on the Predictor Set

Table B1 Variable Description.
This table reports the firm-specific characteristics that comprise the predictor set. Of the 214 predictors, 207 are from the Open Source Asset Pricing database by Chen and Zimmermann (2022), the remaining 7 predictors are lagged realized beta estimators of the CAPM beta, the upside and downside betas and the four semibetas, based on daily data (predictor # 173-179). The second column provides the acronyms of the variables in Open Source Asset Pricing database, the third column gives a short description of the variables. The predictors are categorised according to the six groups from Hou et al. (2020): Intangibles, Investment, Momentum, Profitability, Trading Frictions, and Value vs Growth. More information on the characteristics and their construction can be found on the Open Source Asset Pricing website: https://www.openassetpricing.com/. The full sample period is from January 1980 to December 2024. All predictors have been lagged by $h$ months.

| #  | Acronym | Description |
|---|---|---|
| A. Predictors in the Momentum Group | | |
| 1  | AnnouncementReturn | Earnings announcement return |
| 2  | CustomerMomentum | Customer momentum |
| 3  | EarningsStreak | Earnings surprise streak |
| 4  | EarningsSurprise | Earnings surprise |
| 5  | EarnSupBig | Earnings surprise of big firms |
| 6  | FirmAgeMom | Firm age momentum |
| 7  | High52 | 52-week high |
| 8  | IndMom | Industry momentum |
| 9  | IndRetBig | Industry return of big firms |
| 10 | IntMom | Intermediate momentum |
| 11 | iomom_cust | Customers momentum |
| 12 | iomom_supp | Suppliers momentum |
| 13 | LRreversal | Long-term reversal |
| 14 | Mom12m | Momentum (12 months) |
| 15 | Mom12mOffSeason | Momentum without the seasonal part |
| 16 | Mom6m | Momentum (6 months) |
| 17 | Mom6mJunk | Junk stock momentum |
| 18 | MomOffSeason | Off-season long-term reversal |
| 19 | MomOffSeason06YrPlus | Off-season reversal years 6 to 10 |
| 20 | MomOffSeason11YrPlus | Off-season reversal years 11 to 15 |
| 21 | MomOffSeason16YrPlus | Off-season reversal years 16 to 20 |
| 22 | MomRev | Momentum and long-term reversal |
| 23 | MomSeason | Return seasonality years 2 to 5 |
| 24 | MomSeason06YrPlus | Return seasonality years 6 to 10 |
| 25 | MomSeason11YrPlus | Return seasonality years 11 to 15 |
| 26 | MomSeason16YrPlus | Return seasonality years 16 to 20 |
| 27 | MomSeasonShort | Return seasonality last year |
| 28 | MomVol | Momentum in high-volume stocks |
| 29 | MRreversal | Momentum reversal |
| 30 | NumEarnIncrease | Earnings streak length |
| 31 | ResidualMomentum | Momentum based on FF3 residuals |
| 32 | retConglomerate | Conglomerate return |
| 33 | REV6 | Earnings forecast revisions |
| 34 | RevenueSurprise | Revenue surprise |
| 35 | streversal | Short-term reversal |
| 36 | TrendFactor | Trend factor |

B. Predictors in the Value vs Growth Group

| | | |
|---|---|---|
| 37 | AM | Total assets to market |
| 38 | BM | Book to market using most recent ME |
| 39 | BMdec | Book to market using December ME |
| 40 | BookLeverage | Book leverage (annual) |
| 41 | BPEBM | Leverage component of BM |
| 42 | Cash | Cash to assets |
| 43 | CF | Cash flow to market |
| 44 | cfp | Operating cash flows to price |
| 45 | DivInit | Dividend initiation |
| 46 | DivOmit | Dividend omission |
| 47 | DivSeason | Dividend seasonality |
| 48 | DivYieldST | Predicted dividend yield next month |
| 49 | EBM | Enterprise component of BM |
| 50 | EntMult | Enterprise multiple |
| 51 | EP | Earnings-to-price ratio |
| 52 | EquityDuration | Equity duration |
| 53 | FEPS | Analyst earnings per share |
| 54 | IntanBM | Intangible return using BM |
| 55 | IntanCFP | Intangible return using CF-to-P |
| 56 | IntanEP | Intangible return using EP |
| 57 | IntanSP | Intangible return using Sale-to-P |
| 58 | Size | Log size |
| 59 | MeanRankRevGrowth | Revenue growth rank |
| 60 | NetDebtPrice | Net debt to price |
| 61 | NetPayoutYield | Net payout yield |
| 62 | PayoutYield | Payout yield |
| 63 | sfe | Earnings forecast to price |
| 64 | SP | Sales to price |

| | | C. Predictors in the Investment Group | |
|---|---|---|---|
| 65 | AbnormalAccruals | Abnormal accruals |
| 66 | Accruals | Accruals |
| 67 | AccrualsBM | Book to market and accruals |
| 68 | AssetGrowth | Asset growth |
| 69 | BrandInvest | Brand capital investment |
| 70 | ChInv | Change in capital investment |
| 71 | ChInvIA | Change in capital investment (industry adjusted) |
| 72 | ChNNCOA | Change in net noncurrent operating assets |
| 73 | CompEquIss | Composite equity issuance |
| 74 | CompositeDebtIssuance | Composite debt issuance |
| 75 | DebtIssuance | Debt issuance |
| 76 | DelCOA | Change in current operating assets |
| 77 | DelCOL | Change in current operating liabilities |
| 78 | DelEqu | Change in equity to assets |
| 79 | DelFINL | Change in financial liabilities |
| 80 | DelLTI | Change in long-term investment |
| 81 | DelNetFin | Change in net financial assets |
| 82 | dNoa | Change in net operating assets |
| 83 | grcapx | Change in capex (2 years) |
| 84 | grcapx3y | Change in capex (3 years) |
| 85 | GrLTNOA | Growth in long-term operating assets |
| 86 | Investment | Investment to revenue |
| 87 | InvestPPEInv | Change in PPE and inventory / assets |
| 88 | InvGrowth | Inventory growth |
| 89 | NetDebtFinance | Net debt financing |
| 90 | NetEquityFinance | Net equity financing |
| 91 | NOA | Net operating assets |
| 92 | PctAcc | Percent operating accruals |
| 93 | PctTotAcc | Percent total accruals |
| 94 | ShareIss1Y | Share issuance (1 year) |
| 95 | ShareIss5Y | Share issuance (5 years) |
| 96 | ShareRepurchase | Share repurchases |
| 97 | TotalAccruals | Total accruals |
| 98 | XFIN | Net external financing |
| | | D. Predictors in the Profitability Group | |
| 99 | CBOperProf | Cash-based operating profitability |
| 100 | ChAssetTurnover | Change in asset turnover |
| 101 | ChNWC | Change in net working capital |
| 102 | CredRatDG | Credit rating downgrade |
| 103 | DelDRC | Deferred revenue |
| 104 | GP | Gross profits / total assets |
| 105 | Leverage | Market leverage |
| 106 | MS | Mohanram G-score |
| 107 | OperProf | Operating profits / book equity |
| 108 | OperProfRD | Operating profitability R&D adjusted |
| 109 | OScore | O score |
| 110 | PS | Piotroski F-score |
| 111 | roaq | Return on assets (quarterly) |
| 112 | RoE | Net income / book equity |
| 113 | Tax | Taxable income to income |

E. Predictors in the Intangibles Group

| | | |
|---|---|---|
| 114 | Activism1 | Takeover vulnerability |
| 115 | Activism2 | Active shareholders |
| 116 | AdExp | Advertising expense |
| 117 | AgeIPO | IPO and age |
| 118 | AnalystRevision | EPS forecast revision |
| 119 | AnalystValue | Analyst value |
| 120 | AOP | Analyst optimism |
| 121 | CashProd | Cash productivity |
| 122 | ChangeInRecommendation | Change in recommendation |
| 123 | ChEQ | Growth in book equity |
| 124 | ChForecastAccrual | Change in forecast and accrual |
| 125 | ChNAnalyst | Decline in analyst coverage |
| 126 | ChTax | Change in taxes |
| 127 | CitationsRD | Citations to R&D expenses |
| 128 | ConsRecomm | Consensus recommendation |
| 129 | ConvDebt | Convertible debt indicator |
| 130 | DelBreadth | Breadth of ownership |
| 131 | DownRecomm | Down forecast |
| 132 | EarningsConsistency | Earnings consistency |
| 133 | EarningsForecastDisparity | Long vs. short EPS forecasts |
| 134 | ExclExp | Excluded expenses |
| 135 | fgr5yrLag | Long-term EPS forecast |
| 136 | FirmAge | Firm age based on CRSP |
| 137 | ForecastDispersion | EPS forecast dispersion |
| 138 | FR | Pension funding status |
| 139 | Frontier | Efficient frontier index |
| 140 | Governance | Governance index |
| 141 | GrAdExp | Growth in advertising expenses |
| 142 | GrSaleToGrInv | Sales growth over inventory growth |
| 143 | GrSaleToGrOverhead | Sales growth over overhead growth |
| 144 | Herf | Industry concentration (sales) |
| 145 | HerfAsset | Industry concentration (assets) |
| 146 | HerfBE | Industry concentration (equity) |
| 147 | hire | Employment growth |
| 148 | IndIPO | Initial public offerings |
| 149 | IO_ShortInterest | Institutional ownership among high short interest |
| 150 | OPLeverage | Operating leverage |
| 151 | OrderBacklog | Order backlog |
| 152 | OrderBacklogChg | Change in order backlog |
| 153 | OrgCap | Organizational capital |
| 154 | PatentsRD | Patents to R&D expenses |
| 155 | PredictedFE | Predicted analyst forecast error |
| 156 | RD | R&D over market cap |
| 157 | RDAbility | R&D ability |
| 158 | RDcap | R&D capital to assets |
| 159 | RDIPO | IPO and no R&D spending |
| 160 | RDS | Real dirty surplus |
| 161 | realestate | Real estate holdings |
| 162 | Recomm_ShortInterest | Analyst recommendation and short interest |
| 163 | RIO_Disp | Institutional ownership and forecast dispersion |
| 164 | RIO_MB | Institutional ownership and market to book |
| 165 | RIO_Turnover | Institutional ownership and turnover |
| 166 | RIO_Volatility | Institutional ownership and idiosyncratic volatility |
| 167 | sinAlgo | Sin stock (selection criteria) |
| 168 | Spinoff | Spin-offs |
| 169 | SurpriseRD | Unexpected R&D increase |
| 170 | tang | Tangibility |
| 171 | UpRecomm | Up forecast |
| 172 | VarCF | Cash flow to price variance |

F. Predictors in the Trading Frictions Group

| | | |
|---|---|---|
| 173 | capm_rl_lag | CAPM realized beta (h months) |
| 174 | down_rl_lag | Downside realized beta (h months) |
| 175 | up_rl_lag | Upside realized beta (h months) |
| 176 | neg_rl_lag | Negative realized semibeta (h months) |
| 177 | pos_rl_lag | Positive realized semibeta (h months) |
| 178 | mneg_rl_lag | Mixed sign realized semibeta, negative market return (h months) |
| 179 | mpos_rl_lag | Mixed sign realized semibeta, positive market return (h months) |
| 180 | Beta | CAPM beta |
| 181 | BetaFP | Frazzini-Pedersen beta |
| 182 | BetaLiquidityPS | Pastor-Stambaugh liquidity beta |
| 183 | BetaTailRisk | Tail risk beta |
| 184 | betaVIX | Systematic volatility |
| 185 | BidAskSpread | Bid-ask spread |
| 186 | CoskewACX | Coskewness using daily returns |
| 187 | Coskewness | Coskewness |
| 188 | DolVol | Past trading volume |
| 189 | ExchSwitch | Exchange switch |
| 190 | RealizedVol | Idiosyncratic risk (1 factor) |
| 191 | IdioVol3F | Idiosyncratic risk (3 factors) |
| 192 | IdioVolAHT | Idiosyncratic risk (AHT) |
| 193 | Illiquidity | Amihud's illiquidity |
| 194 | Price | Log price |
| 195 | MaxRet | Maximum return over month |
| 196 | OptionVolume1 | Option to stock volume |
| 197 | OptionVolume2 | Option volume to average |
| 198 | PriceDelayRsq | Price delay R2 |
| 199 | PriceDelaySlope | Price delay coefficient |
| 200 | PriceDelayTstat | Price delay SE adjusted |
| 201 | ProbInformedTrading | Probability of informed trading |
| 202 | ReturnSkew | Return skewness |
| 203 | ReturnSkew3F | Idiosyncratic skewness (3F model) |
| 204 | ShareVol | Share volume |
| 205 | ShortInterest | Short interest |
| 206 | skew1 | Volatility smirk near the money |
| 207 | SmileSlope | Put vol. minus call vol. |
| 208 | std_turn | Share turnover volatility |
| 209 | VolMkt | Volume to market equity |
| 210 | VolSD | Volume variance |
| 211 | VolumeTrend | Volume trend |
| 212 | zerotrade1m | Days with zero trades (1m) |
| 213 | zerotrade6m | Days with zero trades (6m) |
| 214 | zerotrade12m | Days with zero trades (12m) |